\documentclass{jpp-JAK}

\usepackage{bm,amsmath,amssymb,graphicx,array,url,natbib}
\usepackage[normalem]{ulem} 
\usepackage{xcolor}
\usepackage{multicol}
\usepackage{upmath}

\usepackage[
urlcolor={red!50!black}%
,citecolor={blue!80!black}%
,allbordercolors={0 0 0}%
,pdfborder={0 0 0}
,colorlinks=true%
,urlbordercolor={0 0 0}%
,hyperfootnotes=false%
,pdfborderstyle={/S/U/W 0}%
,plainpages=false%
,breaklinks=true%
]{hyperref}

\usepackage[all]{hypcap}

\usepackage{breakurl}

\usepackage{mathrsfs,pstricks}

\makeatletter

\gdef\@underjournal{%
  \vbox to 5.5\p@{\noindent
    \parbox[t]{4.5in}{\normalfont\indexsize{\itshape \
Under consideration for publication in J.\ Plasma\ Phys.
}\\[2.5\p@]
      {\ \ }}%
  \vss}%
}

\long\def\comment#1\endcomment{}

\let\marginparo\marginpar
\def\marginpar#1{\marginparo{\raggedright #1}}

\comment

\endcomment

\def\@sect
#1#2#3#4#5#6[#7]#8{%
\ifnum #2>\c@secnumdepth 
 \let \@svsec \@empty 
\else 
 \refstepcounter {#1}%
 \ifnum #2>\@ne 
  \protected@edef \@svsec {\@seccntformat {#1}\relax }%
 \else 
  \def \apphe@d {#8}%
  \protected@edef \@svsec {\@secappcntformat {#1}\relax }%
 \fi
\fi 
\@tempskipa #5\relax 
\ifdim \@tempskipa >\z@ 
 \begingroup 
 #6{\@hangfrom {\hskip #3\relax \@svsec }%
    \interlinepenalty \@M 
     #8\@@par 
   }
 \endgroup 
\csname #1mark\endcsname {#7}%
\ifnum #2=\@ne 
 \addcontentsline {toc}{#1}%
 {\ifnum #2>\c@secnumdepth \else 
   \ifappendix 
    \appendixname ~\csname the#1\endcsname:  #7%
   \else 
    \protect \numberline {\csname the#1\endcsname .}#7%
   \fi 
  \fi 
  }%
\else 
 \addcontentsline {toc}{#1}%
 {\ifnum #2>\c@secnumdepth \else 
  \protect \numberline {\csname the#1\endcsname .}#7%
  \fi 
  }%
\fi 
\else 
 \def \@svsechd {#6{\hskip #3\relax \@svsec #8}%
 \csname #1mark\endcsname {#7}%
 \addcontentsline {toc}{#1}%
 {\ifnum #2>\c@secnumdepth \else 
  \protect \numberline {\csname the#1\endcsname .}%
  \fi #7%
  }}%
\fi 
\@xsect {#5}%
}

\let\cite\citep
\let\citeo\cite
\def\cite{\ \citeo}
\newcommand\Ref{\citet} 

\comment
\def\NAT@sep{;}

\def\NAT@split#1#2#3#4#5@@{\gdef\NAT@num{#2}%
\gdef\NAT@name{#3}%
\gdef\NAT@date{#2}%
\gdef\NAT@all@names{#4}%
\ifx\NAT@num\@empty
 \gdef\NAT@num{0}%
\fi
\ifx\NAT@noname\NAT@all@names
 \gdef\NAT@all@names{#3}%
\fi
}
\endcomment

\catcode`\@=11 

\newif\ifjournal

\renewcommand*\boldmath{\protect\mathversion{bold}} 
\newcommand*\Unskip{\unskip~}	

\newcommand*\Appabbrev{appendix} 
\newcommand*\Appsabbrev{appendixes}
\newcommand*\App[1]{\Appabbrev~\seco{#1}}
\newcommand*\Apps[1]{\Appsabbrev~\seco{#1}}
\newcommand*\APP[1]{\Unskip\ref{#1}}

\newcommand*\eq[1]{\label{#1}}	
\newcommand*\Eq[1]{Eq.~(\ref{#1})}
\newcommand*\Eqs[1]{Eqs.~(\ref{#1})}
\newcommand*\EQ[1]{\Unskip\EQo{#1}}
\newcommand*\Equation[1]{Equation~(\ref{#1})}	

\newcommand*\Fig[2][]{Fig.~\seco[#1]{Fg.#2}}

\newcommand*\Footnote[1]{Footnote~\seco{#1}}	

\comment
\newcommand*\Ref[1]{Ref.~\Onlinecite{#1}}

\newcommand*\REFo[1]{{\let\ \relax\Onlinecite{#1}}} 
\endcomment

\newcommand*\Onlinecite{\onlinecite} 

\newcommand*\seco[2][]{\ref{#2}#1}	

\newcommand*\Sec[1]{Sec.~\seco{#1}}

\newcommand*\Ifarrayflag{\ifarrayflag}
\newcommand*\Ifletterflag{\ifletterflag}
\newcommand*\Ifbeginningeq{\ifbeginningeq}

\newcommand*\BE{\arrayflagfalse\beginningeqfalse\begin{equation}} 

\newcommand*\BEA{\arrayflagtrue\beginningeqfalse\begin{eqnarray}} 
\newcommand*\BA{\BEA} 
\def\BAams#1\EAams{\begin{align}#1\end{align}} 

\newcommand*\EEA{\end{eqnarray}}

\newcommand*\BAL[1][]{\letterflagtrue\st@rtarray[#1]}
\newcommand*\EAL{\end{eqnarray}\EM}

\newcommand*\EE{%
\Ifbeginningeq
	\beginningeqfalse
	\BE
\else
	\Endanequation
	\beginningeqtrue
\fi
\Ifletterflag
	\EM
\fi
}

\newcommand*\Endanequation{%
\Ifarrayflag
	\end{eqnarray}%
\else
	\end{equation}%
\fi
}

\newcommand*\BM[1][]{\begin{subequations}%
	\gdef\theletters{a}%
	\def\NP##1{##1%
		\ifxf'@label\@empty\else\@xp\ltx@label\@xp{f'@label}\fi
		\letf'@label\@empty
		\def\theequation{\theparentequation\theletters}%
		\stepcounter{equation}%
		\protected@edef\@currentlabel{\theparentequation\alph{equation}}%
		\xdef\theletters{\theletters,\alph{equation}}%
		}%
	\def\NQ##1{\xdef\theletters{\alph{equation}}%
		\NP{##1}}%
	\l@belletters{#1}%
	}

\newcommand*\EM{\end{subequations}\COMMENT
	\letterflagfalse
	}

\newcommand*\NN{\nonumber}

\newcommand*\BI{\begin{itemize}}
\newcommand*\EI{\end{itemize}}

\newcommand*\verticalbar{|}

\newcommand*\vhacek[1]{{\accent20 #1}}

\newcommand*\e{\epsilon}

\newcommand*\g{\gamma}
\newcommand*\G{\Gamma}

\renewcommand*\k{\kappa}

\renewcommand*\p{\phi} 

\renewcommand*\r{\rho}
\newcommand*\s{\sigma}
\renewcommand*\th{\theta}

\newcommand*\U{\Upsilon}
\newcommand*\w{\omega}

\newcommand*\z{\zeta}

\newcommand*\shortGreek{
	\renewcommand*\a{\alpha}
	\renewcommand*\b{\beta}
	\renewcommand*\c{\chi}
	\renewcommand*\d{\delta}
	\newcommand*\D{\Delta}
	\renewcommand*\l{\lambda}
	\renewcommand*\L{\Lambda}
	\renewcommand*\t{\tau}
	\renewcommand*\u{\upsilon}
}

\newcommand*\rcvr@tmp[1]{\edef\next{\global\let\noexpand#1\csname#1@tmp}\next}

\newcommand*\longGreek{
	\rcvr@tmp{a}
	\rcvr@tmp{b}
	\rcvr@tmp{c}
	\rcvr@tmp{d}
	\rcvr@tmp{D}
	\rcvr@tmp{l}
	\rcvr@tmp{L}
	\rcvr@tmp{t}
	\rcvr@tmp{u}
}
	
\shortGreek

\newcommand*\BIGavg[1]{\left\langle#1\right\rangle}

\renewcommand*\({\left(}
\renewcommand*\){\right)}

\renewcommand*\[{\left[}
\renewcommand*\]{\right]}

\newcommand*\?{}

\newcommand*\Ahat{{\Hat{A}}}	

\newcommand*\abso[1]{\verticalbar#1\verticalbar} 

\newcommand*\abs[1]{\mathopen{\ifBIG\left\fi 
	\verticalbar}
	#1\mathclose{\ifBIG\global\BIGfalse\right\fi
	\verticalbar}} 

\newcommand*\adj{^{\dagger}}	

\newcommand*\At[1]{{}_{\big\vert#1}} 
\newcommand*\at[1]{_{\vert#1}}

\renewcommand*\Bar[1]{{\overline{#1}}}	

\newcommand*\BL{Balescu--Lenard}

\newcommand*\Bhat{\Hat{B}}	
\newcommand*\bfit{\rmfamily\bfseries\itshape} 
\newcommand*\bdot{\boldsymbol{\cdot}}	
\newcommand*\bhat{\unit{b}}	

\newcommand*\bigast{{\mathchoice{\asterisk}
	{\displaystyle\asterisk}
	{\textstyle\asterisk}
	{\scriptstyle\asterisk}
	}}

\def\case{\frac}

\newcommand*\CLandau{C_{s\sbar}\on{f} &= 2\pi\,\Fr{q^2,m}_s(\nbar
	q^2)_\sbar\ln\Lambda_{s,\sbar} 
		\Partial{}{\vv}\bdot\Int \dd\vvbar\,\mU(\vv-\vvbar)
\NN
\\
	&\qquad\bdot
		\(\fr{1,m_{\sbar}}\Partial{}{\vvbar}-\fr{1,m_s}\Partial{}{\vv} 
	\)\!f_s(\vv) f_\sbar(\vvbar)}

\newcommand*\CE{Chapman--Enskog}

\newcommand*\Chat{{\Hat{C}}}	

\newcommand*\cE{{\mathcal{E}}} 	
\newcommand*\cEk{\cE_\vk}	

\newcommand*\ck{c_\vk}		
\newcommand*\cL{{\mathcal{L}}}	

\newcommand*\cp{\text{c.p.}}	

\newcommand*\cf{\LatinAIP{cf.}}
\newcommand*\chik{\chi_\vk}	

\newcommand*\chibar{{\Bar{\chi}}}

\newcommand*\cs{c_{\rm s}} 	

\newcommand*\conj{^\bigast}	
\newcommand*\const{\text{const}}
\newcommand*\cross{\boldsymbol{\times}} 

\newcommand*\Dperp{D_\perp}	
\newcommand*\Dn{D_n}		
\newcommand*\Dv{D_v}		

\newcommand*\DT{\Delta T}		

\newcommand*\dA{\delta \Ekern A}

 \newcommand*\Ekern{\kern-0.12em}

\newcommand*\degreeso{^\circ}		
\renewcommand*\degrees{\ifmmode\degreeso\else$\degreeso$\fi} 

\newcommand*\del{\partial}		

\newcommand*\delt{\partial_t}

\newcommand*\delvv{\partial_{\vv}}

\newcommand*\f{\delta \fkern f} \newcommand*\fkern{\kern-0.125em}

\newcommand*\diel{{\mathcal{D}}} 	
\newcommand*\dielperp{\diel_\perp} 	

\renewcommand*\div{\vgrad\bdot} 	

\newcommand*\Dirac[1]{\delta(#1)}

\newcommand*\Ext{\updown{\textrm{ext}}}	

\newcommand*\ed{\epsilon_\delta}
	
\newcommand*\ehalf{^{1/2}}		
\newcommand*\ek{\v{\epsilon}_\vk}
\newcommand*\ep{\epsilon_{\Rm p}}

\newcommand*\ETAL{\Latin{et~al.}}	

\newcommand*\etahat{{\widehat{\eta}}}

\newcommand*\ETC{\LatinAIP{etc.}}		

\newcommand*\Fhat{\Hat{F}}		

\newcommand*\fM{f_{\Rm M}}		
\newcommand*\flM{f_{\textrm{lM}}}	
\newcommand*\FP{Fokker--Planck}
\newcommand*\FPE{\FP\ equation}
\newcommand*\Fo{F_0}			
\newcommand*\Ft{\Tilde F}		
\newcommand*\fhat{{\Hat{f}}}		

\newcommand*\Ghat{{\Hat{\Gamma}}}

\newcommand*\GK{gyrokinetic}

\newcommand*\gspace{\mbox{$\Gamma$-space}}
\newcommand*\Gspace{$\Gamma$~space}

\newcommand*\gbar{{\Bar{\gamma}}}	

\newcommand*\grad{\nabla}		
\newcommand*\gradpar{\grad_\parallel}	

\renewcommand*\Hat[1]{{\widehat{#1}}}	

\newcommand*\half{\case12}		

\newcommand*\Idvbar{\Int d\vvbar}

\newcommand*\Int{\int\!\?}		
\newcommand*\I[2]{\int_{#1}^{#2}\!\?}	
\newcommand*\InT{\I0\infty}		
\newcommand*\INT{\I{-\infty}\infty}	

\newcommand*\ie{\LatinAIP{i.e.}}	
\newcommand*\inWax[1]{reprinted in \textsl{Selected Papers on Noise and Stochastic
Processes}, edited by N.~Wax (Dover, New York, 1954), pp.~#1} 

\newcommand*\K{\cK}			

\newcommand*\Kron[1]{\delta_{#1}} 	

\newcommand*\kbar{{\Bar{k}}}		
	
\newcommand*\khat{\unit{k}}		

\newcommand*\kill@period{\futurelet\nextchar\no@period}
\newcommand*\no@period{\ifx\nextchar.\skip@period\fi}	

\newcommand*\kpar{k_\parallel}		
\newcommand*\kperp{k_\perp}		
\newcommand*\kperpbar{{\Bar{k}}_\perp} 

\newcommand*\kpq[4]{#1_{\v{#2}\v{#3}\v{#4}}}

\newcommand*\Latin{\textit} 
\newcommand*\LatinAIP{\textrm}	

\newcommand*\lD{\lambda_{\Rm D}}	
\newcommand*\lDe{\lambda_{{\Rm D}e}}	

\newcommand*\lmfp{\lambda_{\textrm{mfp}}}	

\newcommand*\lhs{left-hand side}

\newcommand*\MATH{\texttt{MATHEMATICA}}

\newcommand*\Mathop[1]{\mathop{\hbox{\rm #1}}\nolimits}
 \def\M(#1,#2,#3){\kpq{M}{#1}{#2}{#3}} 

\newcommand*\Mhat{{\widehat{M}}}

\newcommand*\Max{\mathop{\operator@font max}}	

\newcommand*\Min{\mathop{\operator@font min}}	

\newcommand*\tensor{\mathsfbi}		

\newcommand*\mA{\tensor{A}}

\newcommand*\mB{\tensor{B}}
\newcommand*\mC{\tensor{C}}
\newcommand*\mD{\tensor{D}}

\newcommand*\mI{\tensor{I}}
\newcommand*\mone{\mI}			

\newcommand*\mK{\tensor{K}}		
\newcommand*\mL{\tensor{L}}

\newcommand*\mM{\tensor{M}}
\newcommand*\mOmega{\boldsymbol{\Omega}}

\newcommand*\mPi{\boldsymbol{\Pi}}

\newcommand*\mR{\tensor{R}}
\newcommand*\mS{\tensor{S}}

\newcommand*\mU{\tensor{U}}
\newcommand*\mW{\tensor{W}}

\newcommand*\m[1]{^{-#1}}		

\newcommand*\me{m_e}
\newcommand*\mi{m_i}

\newcommand*\muo{\mu}			

\newcommand*\muhat{\Hat{\mu}}

\newcommand*\NS{Navier--Stokes}

\newcommand*\nbar{{\Bar{n}}}		
\renewcommand*\ne{n_e}			
\renewcommand*\ni{n_i}			
\newcommand*\no{n_0}			

\newcommand*\nhat{\unit{n}}		

\newcommand*\nue{\nu_e}		
\newcommand*\nui{\nu_i}		

\newcommand*\nuhat{\Hat{\nu}}

\newcommand*\Omegak{\Omega_\vk}		
\newcommand*\Order[1]{O(#1)} 		
\newcommand*\OrdeR[1]{O\vlp#1\vrp} 	
\newcommand*\on[1]{[#1]}			

\newcommand*\Phat{{\Hat{\Phi}}}		

\newcommand*\psibar{{\Bar{\psi}}}

\newcommand*\Rm[1]{{\rm #1}}			

\newcommand*\re{\rho_e}
\newcommand*\ri{\rho_i}
\newcommand*\rs{\rho_{\textrm{s}}}	

\newcommand*\rhs{right-hand side}

\newcommand*\Sbar{{\Bar{S}}}

\newcommand*\Schr{Schr\"odinger}
\newcommand*\Sigmabar{\Bar{\Sigma}}	

\newcommand*\Sigmahat{\Hat{\Sigma}}

\newcommand*\sans{\LatinAIP{sans}}
\newcommand*\sbar{{\Bar{s}}}

\newcommand*\set[1]{{\let|\mid \{#1\}}}	

\newcommand*\Sstate{steady state}
\newcommand*\sstate{steady-state}

\newcommand*\Tr{^{\Rm T}} 		
\newcommand*\Trace{\Mathop{Tr}}		

\newcommand*\Te{T_e} 		
\newcommand*\Ti{T_i} 		
\newcommand*\taue{\tau_e}       
\newcommand*\taui{\tau_i}       

\renewcommand*\Tilde{\widetilde}	
\newcommand*\Tot{\updown{\textrm{tot}}}

\newcommand*\tac{\tau_{\textrm{ac}}}	

\newcommand*\tbar{{\overline t}}
\newcommand*\taubar{{\overline\tau}}

 \def\triad(#1,#2,#3){\kpq{\theta}{#1}{#2}{#3}} 

\newcommand*\third{\case13}

\newcommand*\Unsym{^{\Rm U}}	
\newcommand*\Unsymconj{^{\Rm U\bigast}}	

\newcommand*\unit[1]{{\widehat{\v{#1}}}}	

\newcommand*\up[1]{^{(#1)}}

\newcommand*\upar{u_\parallel}		
\newcommand*\uspace{\mbox{$\muo$-space}}  
\newcommand\muspace{\uspace} 	
\newcommand*\Uspace{$\muo$~space} 
\newcommand*\Muspace{\Uspace} 	

\newcommand*\Vkern{\kern-0.1em} 

\renewcommand*\v[1]{{\protect\bm{#1}}} 	

\newcommand*\vA{\v{A}} 		
\newcommand*\va{\v{a}}		
\newcommand*\vB{\v{B}} 		

\newcommand*\vE{\v{E}} 		

\newcommand*\vGamma{\boldsymbol{\Gamma}}  
\newcommand*\vP{\v{P}}		
\newcommand*\vPhat{{\Hat{\boldsymbol{\Phi}}}}	

\newcommand*\vR{\v{R}}		
\newcommand*\vRbar{\Bar{\vR}}	

\newcommand*\vbar{{\Bar{v}}}
\newcommand*\vdel{\boldsymbol{\del}}

\newcommand*\vepsilon{\v{\epsilon}}
\newcommand*\vf{\v{f}}
\newcommand*\vgrad{\v{\nabla}}
\newcommand*\vgradperp{\vgrad_{\fkern\perp}}
\newcommand*\vgradpar{\vgrad_{\fkern\parallel}}

\newcommand*\vhat{\unit{v}}

\newcommand*\vk{\v{k}}
\newcommand*\vkperp{\v{k}_\perp}	

\newcommand*\vkw{{\vk,\omega}}		

\newcommand*\vp{\v{p}}

\newcommand*\vperp{v_\perp}
\newcommand*\vpsi{\v{\psi}}		

\newcommand*\vq{\v{q}}

\newcommand*\vlp{\mathopen{\boldsymbol(}} 

\newcommand*\vrp{\mathclose{\boldsymbol)}} 

\newcommand*\vt{v_{\Rm t}}	
\newcommand*\vT{v_T}		
\newcommand*\vte{v_{{\Rm t}e}}	
\newcommand*\vti{v_{{\Rm t}i}}	
\newcommand*\vts{v_{{\Rm t}s}}	
\newcommand*\vtu{v_{{\Rm t}\mu}} 

\newcommand*\vu{\v{u}}		
\newcommand*\vuperp{\vu_\perp}	

\newcommand*\vue{\vu_e}		
\newcommand*\vui{\vu_i}		

\newcommand*\vW{\v{W}}
\newcommand*\vv{\v{v}}
\newcommand*\vvperp{\vv_\perp}
\newcommand*\vvbar{{\Bar{\vv}}}

\newcommand*\vw{\v{w}}
\newcommand*\vx{\v{x}}		

\newcommand*\vus{\vu_\bigast}	

\newcommand*\vs{\LatinAIP{vs}}
\newcommand*\via{\LatinAIP{via}}

\newcommand*\versa{\Latin{vice versa}}

\newcommand*\wc{\omega_{\Rm c}}
\newcommand*\wce{\omega_{{\Rm c}e}}
\newcommand*\wci{\omega_{{\Rm c}i}}
\newcommand*\wcs{\omega_{{\Rm c}s}}

\renewcommand*\wp{\omega_{\Rm p}}

\newcommand*\wpi{\omega_{{\Rm p}i}}
\newcommand*\wps{\omega_{{\Rm p}s}}

\newcommand*\wrt{with respect to}

\newcommand*\xspace{\mbox{$x$-space}}

\newcommand*\xt{\widetilde x}

\newcommand*\xhat{\unit{x}}

\newcommand*\Partial[1]{\Deriv\partial\fr{#1}}

\newcommand*\PartiaL[1]{\Deriv\partial\fR{#1}}

\newcommand*\Total[1]{\Deriv{\?\dd}\fr{#1}}

\newcommand*\TotaL[1]{\Deriv{\?D}\fr{#1}}

\newcommand*\mathselect[4]{\mathchoice{#1{#3}{#4}}
	{#2{#3}{#4}}{#2{#3}{#4}}{#2{#3}{#4}}}

\newcommand*\frD{\frac} 
\newcommand*\frT[2]{#1/#2} 

\newcommand*\fr[1]{\fro#1\fro} 
 
\newcommand*\fR[1]{\left(\fr{#1}\right)}	

\newcommand*\Casefr[2]{\frac{#1}{#2}}	
\newcommand*\Case{\Casefr}

\newcommand*\Half{\Casefr12}	

\newcommand*\Third{\Casefr13}	

\newcommand*\ordspacing{%
	\normalbar
	\mathcode`|="226A
	\mathcode`+="002B
	\mathcode`-="0200
	\mathcode`*="0203
	\mathcode`=="003D
	}

\newcommand*\oF[1]{\vlp{\ordspacing #1}\vrp}

	\renewcommand*\BE{\begin{equation}}
	\renewcommand*\EE{\end{equation}}
	\renewcommand*\BEA{\begin{eqnarray}}
	\renewcommand*\BAL[1][]{\BM[#1]\BA}
	\newcommand*\BALams[1][]{\BM[#1]\BAams}
\def\BALams{\@ifnextchar[\BALams@{\BALams@[]}}
\def\BALams@[#1]#2\EALams{\BM[#1]\BAams#2\EAams\EM}
	\renewcommand*\EM{\end{subequations}}
	\newcommand*\WT{\begin{widetext}}
	\newcommand*\NT{\end{widetext}}
	\renewcommand*\etal{\ETAL}	
	\renewcommand*\etc{\ETC}	
	\renewcommand*\fkern{\!}	
	\renewcommand*\Vkern{}		
	\renewcommand*\Unskip{}		

\newcommand*\l@belletters[1]{\def\Jtemp{#1}\ifx\Jtemp\empty\else\eq{#1}\fi}

 \newcommand*\activebar{\catcode`\|=\active}

 {\activebar
 \gdef\normalbar{\activebar
 	\let|\verticalbar}%
 }

 \normalbar

\def\Derivo#1#2#3,#4\Derivo{#2{#1{#3},#1#4}}

\def\Partial#1{\Deriv\partial\fr{#1}}

\def\PartiaL#1{\Deriv\partial\fR{#1}}

\def\Total#1{\Deriv{\?d}\fr{#1}}

\def\TotaL#1{\Deriv{\?D}\fr{#1}}

\def\mathselect#1#2#3#4{\mathchoice{#1{#3}{#4}}
	{#2{#3}{#4}}{#2{#3}{#4}}{#2{#3}{#4}}}

\def\fro#1,#2\fro{{\mathselect\frac\frT{#1}{#2}}}
\let\frD\frac
\def\frT#1#2{#1/#2} 

\def\fr#1{\fro#1\fro} 

\def\fR#1{\left(\fr{#1}\right)}	

\def\choiceo#1,#2\choiceo{{#1\atop#2}}

\def\Casefr#1#2{\frac{#1}{#2}}	
\let\Case\Casefr

\def\casefr#1#2{\mathchoice{{\textstyle\frD{#1}{#2}}}%
	{{\textstyle\frD{#1}{#2}}}
	{{\scriptstyle\frD{#1}{#2}}}%
	{{\scriptscriptstyle\frD{#1}{#2}}}}
\let\case\casefr

\def\half{\casefr12}	\def\Half{\Casefr12}	
\def\third{\casefr13}	\def\Third{\Casefr13}

\def\PARENS#1{\noexpand\let
\csname#1x\endcsname
\csname#1\endcsname
\noexpand\def
\csname#1(##1){\csname#1x\endcsname{##1}}}

\renewcommand*\l@belletters[1]{\def\Jtemp{#1}\ifx\Jtemp\empty\else\eq(#1)\fi}

 \let\COMMENT\relax 

 \newif\ifarrayflag
 \newif\ifletterflag
 \newif\ifbeginningeq
 \beginningeqtrue

 \def\M(#1,#2,#3){\kpq{M}{#1}{#2}{#3}} 
 \def\Mpas(#1,#2,#3){\kpq{M\Unsym}{#1}{#2}{#3}} 
 \def\Mpasconj(#1,#2,#3){\kpq{M\Unsymconj}{#1}{#2}{#3}} 

\def\3{ ---\nobreak\ }

\renewcommand*\set[1]{{\let|\mid \{#1\}}}	
\renewcommand*\etc{\ETC\kill@period}
\def\skip@period\fi.{\fi}

\newif\ifBIG

\def\BIG{\BIGtrue} 
\def\SMALL{\BIGfalse}

\SMALL		

\def\@BIGinsert#1{\ifBIG#1\fi}

\let\sc@le\empty

\def\SIZE#1{\edef\sc@le{\expandafter\noexpand\csname #1\endcsname}}

\def\MID{\mathbin{\hbox{\vrule height\ht0 depth\dp0 width0.2pt}}}

{\activebar
\gdef\<#1>{{\def\arg{\ifBIG\displaystyle\fi
\ifx\sc@le\empty
	\def\sc@lel{\@BIGinsert{\left}}
	\def\sc@ler{\@BIGinsert{\right}}
\else
	\let\sc@lel\sc@le
	\let\sc@ler\sc@le
\fi
\sc@lel
\langle
#1
\sc@ler
\rangle
}
\let|\mid
\setbox0\hbox{$\arg$}
\ifBIG
	\let|\MID
\fi
\arg
}
\global\BIGfalse 	
\global\let\sc@le\empty
}
}

\def\BIGavg#1>{\left\langle#1\right\rangle} 

\def\bra#1|{\mathopen{}\ifBIG\left\fi\langle
	\,#1\,
	\ifBIG\BIGfalse\right\fi\verticalbar
	\mathclose{}}		

\def\ket#1>{\mathopen{}
	\ifBIG\left\fi\verticalbar
	\,#1\,
	\ifBIG\BIGfalse\right\fi\rangle
	\mathclose{}}		

\def\DOTS.{\unskip\ \dots\,.} 

\def\CLandau{C_{s,\sbar}\on{f} &=& -2\pi\Fr{q^2,m}_s(\nbar
	q^2)_\sbar\ln\Lambda_{s,\sbar} 
		\Partial{,\vv}\bdot\Int d\vvbar\,\mU(\vv-\vvbar)
\NN
\\
	&&\quad\bdot
		\(\fr{1,m_s}\Partial{,\vv} -
	\fr{1,m_{\sbar}}\Partial{,\vvbar}\)\!f_s(\vv) f_\sbar(\vvbar)}

\renewcommand*\l@belletters[1]{\def\Jtemp{#1}\ifx\Jtemp\empty\else\eq{#1}\fi}

\renewcommand*\App[1]{\hyperref[#1]{\Appabbrev}~\seco{#1}}
\renewcommand*\Apps[1]{\hyperref[#1]{\Appsabbrev}~\seco{#1}}

\renewcommand*\Eq[1]{\hyperref[#1]{Eq.}~(\ref{#1})}
\renewcommand*\Eqs[1]{\hyperref[#1]{Eqs.}~(\ref{#1})}
\renewcommand*\Equation[1]{\hyperref[#1]{Equation}~(\ref{#1})}	

\renewcommand*\Fig[2][]{\hyperref[Fg.#2]{figure}~\seco[#1]{Fg.#2}}

\renewcommand*\Sec[1]{\hyperref[#1]{Sec.}~\seco{#1}}

\def\FIGURE{\@ifnextchar[\@FIGURE{\@FIGURE[1]}}

\def\@FIGURE[#1]#2#3{\begin{figure}
\vbox{%
\centerline{\includegraphics[width=#1\columnwidth]{#2.eps}}%
\nobreak
	    \caption{#3}%
	    \edef\@currentlabel{\thefigure}%
	    \label{Fg.#2}
}
\end{figure}}

\newcommand\footnoteref[1]{\protected@xdef\@thefnmark{\ref{#1}}\@footnotemark}

\renewcommand*\Partial[2]{\frac{\partial#1}{\partial#2}}
\renewcommand*\PartiaL[2]{\left(\frac{\partial#1}{\partial#2}\right)}

\renewcommand*\Total[2]{\frac{\dd#1}{\dd#2}}
\renewcommand*\TotaL[2]{\frac{{\rm D}#1}{{\rm D}#2}}

\renewcommand*\fr{\frac}
\renewcommand*\fR[2]{\left(\frac{#1}{#2}\right)}

\renewcommand*\[{\left[}
\renewcommand*\]{\right]}

\let\asterisk*
\renewcommand\.{\boldsymbol{\cdot}}

\renewcommand*\Ext{^{\rm ext}}
\renewcommand*\Tot{^{\rm tot}}

\renewcommand*\At[2]{\mathord{\left.#1\right|\setbox0=\hbox{$\displaystyle#1$}%
\dimen0=\ht0
\advance\dimen0 by\dp0
\dimen0 = 0.3\dimen0
\!\lower\dimen0\hbox{$\scriptstyle#2$}}}

\renewcommand*\cEk{\cE(\vk)}

\newcommand*\Df{\Delta f}

\newcommand*\Dp{\Delta\phit}

\renewcommand*\H[1]{H_{#1}}

\renewcommand*\set[1]{\{#1\}}

\newcommand*\st{\Tilde s}

\renewcommand*\Tilde{\widetilde}

\renewcommand*\U[1]{U_{#1}}

\newcommand*\A{{\rm A}}

\newcommand*\B{{\rm B}}

\newcommand*\C{{\rm C}}
\renewcommand*\cE{\pmb{\mathbb{E}}}
\renewcommand*\cEk{\cE_\vk}
\renewcommand*\Chat{\Hat{\C}}
\let\chio\chi
\renewcommand*\chi{\Delta\chio}
\renewcommand*\chibar{\Delta\Bar{\chio}}

\newcommand*\cu{\chio_\mu}
\renewcommand*\ck{\chio_{\rm v}}
\newcommand*\cv{c_{\rm v}}
\renewcommand*\cp{c_{\rm p}}

\let\Delta\rmDelta
\newcommand*\Dhat{\Hat{\rm D}}
\renewcommand*\Dn{\D n}
\renewcommand*\Dp{D_p}
\newcommand*\Dskip[1]{\mskip#1.2mu}
\newcommand*\Dva{\D\va}
\newcommand*\Dvahat{\D\Hat\va}

\newcommand*\ebar{\Bar{\epsilon}}
\newcommand*\Ehatpar{\Hat{E}_\parallel}
\newcommand*\Ehatperp{\Hat{E}_\perp}

\newcommand*\fMbar{\Bar{f}_{\rm M}}

\renewcommand*\G{{\rm G}}
\renewcommand*\Ghat{\Hat{G}}
\newcommand*\GQ{\G_\Q}
\newcommand*\GQhat{\Ghat_\Q}
\newcommand*\gradT{\unit{t}}

\renewcommand*\H{{\rm H}}
\newcommand*\He{\Mathop{He}}

\newcommand*\kparbar{\kbar_\parallel}

\renewcommand*\L{{\rm L}}
\newcommand*\lbar{\Bar{\lambda}}
\newcommand*\LE{\L_\vE}
\newcommand*\LL{\cL}
\newcommand*\LM{\L_{\rm M}}
\newcommand*\lon{^{\rm long}}
\newcommand*\LQ{\L}
\newcommand*\Lvec{{\vec{\L}}\hspace{0.1em}{\setbox0=\hbox{$L$}\rule{0pt}{1.1\ht0}}}

\renewcommand*\M{{\rm M}}
\renewcommand*\Mhat{\Hat{\M}}
\newcommand*\mbar{\Bar{m}}
\newcommand*\mbeta{\boldsymbol{\beta}}
\newcommand*\mChat{\Hat{\mC}}
\newcommand*\mdelta{\boldsymbol{\delta}}
\newcommand*\meta{\boldsymbol{\eta}}
\newcommand*\mE{\tensor{E}}
\newcommand*\mSigma{\boldsymbol{\Sigma}}
\newcommand*\mSigmahat{\Hat{\mSigma}}
\newcommand*\MZ{Mori--Zwanzig}
\newcommand*\MZf{\MZ\ formalism}

\renewcommand*\nhat{\Hat{n}}
\newcommand*\nuebar{\Bar{\nu}_e}
\newcommand*\nuibar{\Bar{\nu}_i}

\newcommand*\one{\hbox{\bf 1}}

\renewcommand*\P{{\rm P}}
\newcommand*\psix{\psi_x}
\newcommand*\psiy{\psi_y}

\newcommand*\Q{{\rm Q}}

\newcommand*\rhoe{\rho_e}
\newcommand*\rr{r_e}

\let\Ssymb\S
\let\SECTION\S
\renewcommand*\S{S}
\renewcommand*\Sbar{\Bar{\S}}
\newcommand*\Shat{\Hat{S}}

\newcommand*\Term[1]{$\rm (#1)$}
\newcommand*\To{T}
\newcommand*\trans{^{\rm trans}}

\renewcommand*\U{{\rm G}}
\newcommand*\UQ{\GQ}

\newcommand*\vc{\v{c}}
\newcommand*\ve{\v{e}}

\newcommand*\vg{\vv\.\vgrad}
\newcommand*\vJhat{\Hat{{\bf J}}}
\newcommand*\vtM{v_{{\rm t}M}}
\newcommand*\vto{\vt}
\newcommand*\vtmu{v_{{\rm t}\mu}}
\renewcommand*\vus{\vu_s}

\newcommand*\Wpar{\Omega_\parallel}
\newcommand*\Wperp{\Omega_\perp}
\newcommand*\Wcross{\Omega_{\cross}}

\renewcommand*\Total[2]{\frac{\dd#1}{\dd#2}}

\def\bra#1|{\mathopen{}\ifBIG\left\fi\langle
\Dskip{1}#1\Dskip{1}
	\ifBIG\BIGfalse\right\fi\verticalbar
	\mathclose{}}		

\def\ket#1>{\mathopen{}
	\ifBIG\left\fi\verticalbar
\Dskip{1}#1\Dskip{1}
	\ifBIG\BIGfalse\right\fi\rangle
	\mathclose{}}		

{\activebar
\gdef\<#1>{{\def\arg{\ifBIG\displaystyle\fi
\ifx\sc@le\empty
	\def\sc@lel{\@BIGinsert{\left}}
	\def\sc@ler{\@BIGinsert{\right}}
\else
	\let\sc@lel\sc@le
	\let\sc@ler\sc@le
\fi
\sc@lel
\langle
\Dskip{1}#1\Dskip{1}
\sc@ler
\rangle
}
\def|{\Dskip{1}{\mid}\Dskip{1}}%
\setbox0\hbox{$\arg$}
\ifBIG
	\def|{\Dskip{5}{\MID}\Dskip{5}}%
\fi
\arg
}
\global\BIGfalse 	
\global\let\sc@le\empty
}
}

\def\CLandau{C^{\rm L}_{s\sbar}\on{f} &= -2\pi\,(\nbar m)_s\m1 
		\Sbar_{s\sbar}\Partial{}{\vv}\bdot \Int
                \dd\vvbar\,\mU(\vv-\vvbar) 
\bdot
		\(\fr{1}{m_{s}}\Partial{}{\vv}- \fr{1}{m_\sbar}
	\Partial{}{\vvbar}\)\!f_s(\vv) f_\sbar(\vvbar)}

\renewcommand\subsubsection{%
  \@startsection{subsubsection}{3}{\z@}
    {9\p@ \@plus 3\p@ \@minus 3\p@}
    {3\p@}
    {\raggedright\normalfont\normalsize}%
}

\def\HBOX#1{\hbox to 0pt{\hss #1\hss}}%

\def\subcenter#1{_{%
\dimen0 = 0pt 
\toks0={}
\def\Hbox##1{\hbox to \dimen0{\hss ##1\hss}}%
\vtop{\small\subparse#1{}}}}

\def\subparse#1{%
\def\next{#1}%
\ifx\next\empty
	\the\toks0 
\else
	\setbox0=\hbox{\small#1}%
	\ifdim\wd0>\dimen0
		\dimen0=\wd0 
	\fi
	\toks0=\expandafter{\the\toks0\Hbox{#1}}
	\let\next\subparse
\fi
\next
}

\def\subnoparse#1{%
\def\next{#1}%
\ifx\next\empty
	\the\toks0 
\else
	\setbox0=\hbox{\small#1}%
	\toks0=\expandafter{\the\toks0\Hbox{#1}}
	\let\next\subparse
\fi
\next
}

\def\Underbrace#1#2{\underbrace{#1}\subcenter{#2}}

\def\Multiple#1#2#3{{\count0=#2
\toks0={#3}%
\loop\advance\count0 by -1
\ifnum\count0<0
\else
	\edef\temp{\toks0={#1{\the\toks0}}}%
	\temp
\repeat
\the\toks0}}

\comment
\newenvironment{Quote}
{%
\bgroup
\parindent0pt%
\leftmargini=0pt%
\smallskip
\begin{quote}
}
{\end{quote}
\smallskip
\egroup
}
\endcomment

\comment
\newenvironment{Quote}
  {\begin{quotation}} 
  {\end{quotation}}
\endcomment

\def\precite#1#2{{\def\NAT@open{(#1; }\cite{#2}}}

\def\@hangfrom@appendix#1#2#3{#1\@if@empty{#2}{\MakeTextUppercase{#3}}{#2\@if@empty{#3}{}{:\ \MakeTextUppercase{#3}}}}
\def\@appendixcntformat#1{\MakeTextUppercase{\appendixname\ \csname
    the#1\endcsname}} 

\comment
\def\@makefntext#1{\def\baselinestretch{1}
\leftskip1em
\parindent1em
\noindent
\nobreak
\hskip-\leftskip
\hb@xt@
\leftskip{\hss\@makefnmark\ }#1\par}

\def\@footnotetext{\insert\footins\bgroup\def\baselinestretch{0.75}\make@footnotetext} 

\show\footnote
\endcomment

\usepackage{xr}
\comment 
\renewcommand*\Sec[2][]{\setxdelim{#1}Sec.~#1\xdelim\seco{#1\xdelim#2}}

\renewcommand*\Eq[2][]{\setxdelim{#1}\hyperref[#1\xdelim#2]{Eq.}~(#1\xdelim\ref{#1\xdelim#2})} 
\renewcommand*\EQ[2][]{\setxdelim{#1}(#1\xdelim\ref{#1\xdelim#2})} 
\renewcommand*\Eqs[2][]{\setxdelim{#1}\hyperref[#1\xdelim#2]{Eqs.}~(#1\xdelim\ref{#1\xdelim#2})} 
\endcomment

\renewcommand*\Sec[2][]{\setxdelim{#1}\SECTION#1\xdelim\seco{#1\xdelim#2}}

\renewcommand*\Eq[2][]{\setxdelim{#1}(#1\xdelim\ref{#1\xdelim#2})} 
\renewcommand*\EQ[2][]{\setxdelim{#1}(#1\xdelim\ref{#1\xdelim#2})} 
\renewcommand*\Eqs[2][]{\setxdelim{#1}(#1\xdelim\ref{#1\xdelim#2})} 
\renewcommand*\App[2][]{{\setxdelim{#1}\hyperref[#1\xdelim#2]{\Appabbrev}~#1\xdelim\seco{#1\xdelim#2}}}
\renewcommand*\Apps[2][]{{\setxdelim{#1}\hyperref[#1\xdelim#2]{\Appsabbrev}~#1\xdelim\seco{#1\xdelim#2}}}

\newcommand*\nopdflink{\def\pdfmark[##1]##2{##1}}

\newcommand\setxdelim[1]{\def\temp{#1}%
\ifx\temp\empty
 \def\xdelim{}%
\else
 \def\xdelim{:}%
 \nopdflink
\fi
}

\let\pio\pi
\renewcommand*\pi{\mathup\pio}
\let\Pi\pio

\renewcommand\eg{e.g.}
\newcommand\dd{\mbox{d}}
\newcommand\ee{\mbox{e}}
\newcommand\ii{{\rm i}}

\def\wavenumber{wavenumber}

\def\condition(#1){\hbox{if\ }#1}

\def\NAT@exlab{} 

\def\Footnote#1{footnote~\ref{#1}%
\edef\curpage{\the\c@page}%
\ on page \pageref{#1}%
}

\def\alternate{alternative}
\def\analog{analogue}
\def\analyze{analyse}
\def\analyzing{analysing}
\def\behavior{behaviour}

\def\centre{centre} 

\def\favor{favour}

\def\labeled{labelled}
\def\Magnetize{Magnetise}
\def\magnetize{magnetise}

\def\pages#1#2{#1 (#2 pages)}
\def\range#1{\rangeo#1.}
\def\rangeo#1-#2.{#1-#2}
\def\press{in press}

\def\addperiod#1{#1}
\def\killcomma#1{<KC>}
\def\commatosemicolon{<CS>}

\long\def\pg#1#2#3{#1%
\ifx#2,%
 \ifx#3[%
  \ #3
 \else
  \ifx#3;%
   ; %
  \else
   #2 #3%
  \fi
 \fi
\else
 #2 #3%
\fi
}

\catcode`\@=12

 \title{Projection-operator methods for classical transport in \magnetize d
 plasmas.\\  I.  Linear response, the Braginskii equations, and fluctuating
 hydrodynamics}

\author{John A. Krommes\corresp{\email{\protect\url{krommes@princeton.edu}}}} 

\affiliation{Princeton Plasma Physics Laboratory,
P. O. Box 451, MS 28,
Princeton, New Jersey  08543--0451
USA}

\begin{document}

\maketitle

\begin{abstract}
An introduction to the use of projection-operator methods for the
derivation of classical fluid transport equations for weakly coupled,
\magnetize d, multispecies plasmas is given.  In the present work, linear
response (small 
perturbations from an absolute Maxwellian) is addressed.  In the
Schr\"odinger representation, projection onto the hydrodynamic subspace
leads to the conventional linearized Braginskii fluid equations, while the
orthogonal projection leads to an \alternate\
derivation of the Braginskii correction equations for the nonhydrodynamic
part of the one-particle distribution function.  Although ultimately
mathematically 
equivalent to Braginskii's calculations (at linear order), the
projection-operator approach 
provides an
appealingly intuitive way of discussing the derivation of transport equations
and interpreting the significance of the various parts of the perturbed
distribution function; it is also technically more concise.  A special case
of the Weinhold metric is used to provide a covariant representation of the
formalism; this allows a succinct demonstration of the Onsager symmetries
for classical transport.
The Heisenberg
representation is used to derive a 
generalized Langevin system whose mean recovers the linearized Braginskii
equations 
but that also includes fluctuating forces.  Transport coefficients are
simply related to the two-time correlation functions of those forces, and
physical pictures of the various transport processes are naturally couched
in terms of them.
A number of appendixes review the traditional \CE\ procedure; record
some properties of the linearized Landau collision operator; discuss the
covariant representation of the hydrodynamic projection; provide an example
of the calculation of some transport effects; describe the
decomposition of the stress tensor for \magnetize d plasma; introduce the
linear eigenmodes of the Braginskii equations; and, with the aid of several
examples, mention some caveats for the use of projection operators.
\comment
This leads to
a derivation of equations for hydrodynamic correlation functions that is more
efficient than \alternate\ approaches based on the BBGKY hierarchy.

The methodology is
extended to nonlinear order, which includes the Burnett transport coefficients,
in Part~II.
\endcomment
\end{abstract}

\clearpage

{\makeatletter

\global\let\@makempfntext\@makempfntexto
\global\let\@mpfootnotetext\@mpfootnotetexto
\global\let\footnoterule\footnoteruleo
\global\let\@makefnmark\@makefnmarko
\global\let\thempfootnote\thempfootnoteo
\global\let\thefootnote\thefootnoteo

\global\let\@fnsymbol\@fnsymbolo

\global\def\@makefntext#1{\strut $^{\the\c@footnote}$#1}

\global\setlength\footnotesep{10\p@}
}

\setcounter{tocdepth}{3} 
{
\def\boldmath{}
\def\bfit{}
\tableofcontents
}


\section{Introduction}
\label{Introduction}

The review article by Braginskii (published in Russian in 1963 and in
English translation in 1965)\hbox to 0pt{\hphantom{\Ref{Braginskii}}}
on classical transport in weakly
coupled, \magnetize d, multispecies plasmas has served as
an invaluable 
reference for multiple generations of plasma physicists.  For a
two-component plasma with small electron-to-ion mass ratio,
$\mu \doteq \me/\mi \ll 1$ ($\doteq$~denotes a definition), 
Braginskii
described a path to the derivation of the so-called \emph{correction
  equations} for the nonhydrodynamic parts of the distribution function,
from which the classical transport coefficients are ultimately derived.
The methodology, first published by
\Ref{Braginskii57},\footnote{Braginskii's original 1957 paper contains a
footnote 
  indicating that the work was performed in 1952.}
can be traced back to the pioneering work of 
\Ref{Chapman} and \Ref{Enskog} on the kinetic theory of rarified gases; for
many details, see \Ref{Chapman-Cowling}.  The traditional \CE\ procedure is
reviewed for the simple case of a one-component plasma (OCP) in \App{CE}.

Although 
the relevant mathematics was described clearly by \Ref{Braginskii57},
experience
shows that many 
students do not take the time to work through those calculations and
consequently do not always grasp the beautiful underlying structure of the
transport problem.  The techniques described in this article provide an
\alternate, heuristically appealing and technically efficient approach that
for neutral fluids is 
known to unify a number of
threads of nonequilibrium statistical mechanics.  That unification carries
over to the more complicated plasma.  Appreciation of the methods enables
one to avoid reinventing the wheel and provides one with a concise, workable
formalism on which nontrivial generalizations can be built.  An example of
such a generalization is the calculation of second-order (Burnett)
transport coefficients, addressed in Part~II of this series of
articles\cite{JAK_projection_II}. 

In the same year that the English translation of Braginskii's review
appeared, \Ref{Mori} published a seminal paper in which the transport
problem was reformulated with the aid of projection-operator methods.  
The general approach had been anticipated by
\Ref{Zwanzig61,Zwanzig61_LTP}, and the methodology is now known as the
\emph{\MZf}.  
The purpose of the present article is to describe the application of the
\MZf\ to the problem of classical transport in weakly coupled,
\magnetize d plasmas.  This leads one to an \alternate\
derivation of the Braginskii correction equations.

A great strength of Braginskii's review article is its focus on the physical
interpretation of the transport coefficients, knowledge of which is
essential for researchers on \magnetize d plasmas.  But although Braginskii's
interpretations of the mathematics are entirely correct, here too the present
projection methods are helpful in providing additional intuition.  They enable an
extension of 
Braginskii's equations for the macroscopic, mean hydrodynamic variables
(density, flow, and temperature) to generalized Langevin equations that include
fluctuating forces.  Such equations have previously been derived from the BBGKY
hierarchy\cite{Bixon69,Hinton}, but the projection-operator methods are
arguably 
more transparent 
and efficient.  Fluctuating forces appear implicitly in Braginskii's
heuristic explanations of the various transport processes, and their
effects are contained 
in his systematic mathematics.  However, explicit definition of
those forces brings additional clarity to the transport calculations.

\comment
 Although ultimately mathematically
equivalent to Braginskii's calculations, the aproach provides a
usefully intuitive way of discussing the derivation of transport equations
and interpreting the significance of the various parts of the perturbed
distribution function.
\endcomment


A projection operator~$\P$ is linear and idempotent (\ie, $\P^2 = \P$).
It extends to linear algebra and functional analysis the notion of
graphical projection onto an axis.
An example of a projection operator is the ensemble
average\footnote{Indeed, any 
linear, properly normalized averaging operation~$\A$ possesses the property
$\A^2 = \A$ and is thus a projection 
operator.  \Ref{Weinstock69,Weinstock70} employed such operators in his
approach to Vlasov turbulence; see some discussion by \Ref{JAK_PR} and
\Ref{JAK_tutorial}. }
 $\<\dots>$.
 Thus, all of
statistical closure theory\cite{JAK_PR,JAK_tutorial} can be said to involve
projection operators.  However, this is stretching the point.  The specifically
\MZ-style projection-operator techniques have a particular flavor
and do not naturally generalize to all possible methods of statistical
closure.  They have been little used in
plasma physics.  \Ref{JAK_thesis} and \Ref{JAK_cells} employed them to
discuss the 
phenomenon of long-time tails in \magnetize d, thermal-equilibrium plasmas,
earlier identified in the molecular-dynamics neutral-fluid computer
experiments of 
\Ref{Alder-Wainwright}.\footnote{Some discussions of long-time tails are
  given by \Ref[Sec.~21.5]{Balescu}, \Ref[Sec.~S11.A]{Reichl2}, and
  \Ref[Chap.~9]{Zwanzig}.} 
  They are mentioned briefly by \Ref{Diamond_I}.
However, they have not been used systematically in the context of plasma
transport theory.  In particular, there is no published account of their
application to the derivation of the Braginskii transport
equations.\footnote{Some of the material discussed here was taught by
  the author for many years in the course Irreversible Processes in
  Plasma offered at the second-year graduate level in Princeton
  University's Department of Astrophysical Sciences.  One purpose of this
  article is to make this material more accessible and to suggest that it
  is worthy of a core topic in a plasma-physics educational curriculum.}

One possible reason that the \MZf\ has not seen much use
in plasma physics is that the most general
nonequilibrium-statistical-mechanical treatments 
of the transport problem produce formulas for the transport coefficients in
terms of averages over the $N$-particle equilibrium ensemble and dynamics
evolved with a modified $N$-particle Liouville propagator, where $N$~is
the total number of particles in the system.  Such formulas
go back to \Ref{Green54} and \Ref{Kubo57}. (Subtleties with the
interpretation of the Green--Kubo formulas are discussed in \Sec{Plateau}.)
While elegant, they are difficult to 
work out in the general case.  However, in the special but very important
regime of weak coupling, those formulas simplify dramatically.  Moreover,
with the further approximation of linear response an \alternate\ approach
becomes possible wherein a kinetic 
equation for the one-particle distribution function is first derived, then
processed to produce formulas for the transport coefficients.  It is that
processing that is clarified by the projection-operator techniques
described in the present paper.  I shall show
that a suitably modified \MZf\ can be easily applied to the linearized
plasma kinetic 
equation, leads efficiently to the standard results, clarifies the
structure of the hydrodynamic system, and fosters heuristic understanding of
the first-order transport processes.

This paper is part~I of a two-part series of articles whose goal is to show
how to formulate classical, weakly coupled plasma transport theory by using
projection-operator methods. 
Part~I describes a self-contained rederivation of the Braginskii correction
equations for the special case of linear perturbations of an absolute
Maxwellian equilibrium.  This is less than what Braginskii accomplished
(his equations are nonlinear),
but it serves to familiarize one with the basic methodology in the simplest
possible context.  It leads to what plasma physicists call the \emph{Braginskii
transport coefficients}, better known in nonequilibrium statistical
mechanics as the 
\emph{\NS\ coefficients}.   I shall also derive a set of
\emph{generalized Langevin equations} that extend
Braginskii's equations to include fluctuating forces. 

In Part~II, the problem is reformulated in order to embrace
nonlinearity.  The methodology used there is a generalization of the
formulation of \Ref{Brey} to include a background magnetic field and
multiple species.
The nonlinear Braginskii equations are recovered with ease.   The
techniques also allow one to obtain the 
next-higher-order \emph{Burnett equations}.  Those
were originally obtained for plasmas described by the Landau collision
operator by \Ref{Mikhailovskii67},
\Ref{Mikhailovskii-Tsypin71,Mikhailovskii-Tsypin84},  and
\Ref{Catto-Simakov}, none of whom remarked on 
connections to previous calculations of Burnett coefficients for neutral
  gases.  In Part~II  it is shown that
 the \NS\ and Burnett transport coefficients calculated by Catto and
 Simakov are special cases of the complete set of coefficients for which
\Ref{Brey} gave general expressions
in terms of integrals of two-time correlation functions.\footnote{A proviso
  is that \Ref{Brey} did not include a magnetic field, but that is easy to
  add.} Thus, the present series of articles serves to unify a variety of  
previous research.  

I now return to an overview of the present Part~I\@.  
I illustrate the basic approach in \Sec{OCP} by deriving linearized
transport equations for the un\magnetize d one-component plasma.  Then in
\Sec{Braginskii} I extend the calculations to embrace multispecies and
\magnetize d 
plasmas, and I derive the linearized version of Braginskii's correction
equations.  

The formalism employed in \Sec{OCP} and \Sec{Braginskii} uses what is
known as the \emph{\Schr\ representation}, which means that one takes
time-independent velocity moments of the time-dependent distribution
function~$f$ (which is taken to evolve according to the Landau kinetic 
equation).  Alternatively, one can use the \emph{Heisenberg
  representation}, in which equations are written for time-dependent,
random hydrodynamic operators whose statistical \behavior\ evolves from
statistics (velocity dependence) that reside in the distribution  
function at some initial time.  The averaged equations are the same as
before.  However, the raw equations for the random variables can be cast
into the form of 
generalized Langevin equations, which include fluctuating forces and
imply a theory of hydrodynamic correlation functions.  This is done in
\Sec{GLE}. 
The body of the paper concludes with a brief discussion in \Sec{Discussion}. 

Several appendixes are included.  In \App{CE} I review the
traditional \CE\ calculation for the one-component plasma, in \App{Chat}
I record various properties of the linearized Landau collision operator,
in \App{Covariant} I describe a technically
efficient covariant
representation of the transport equations,
in \App{qe} I provide an example of the evaluation of the general formulas by
considering the electron heat flow in the limit of small collisionality,
in \App{Viscosity} I review Braginskii's tensorial decomposition of the stress
tensor, 
in \App{Eigenmodes} I consider some important special cases of the linear
eigenmodes of the Braginskii equations, 
and in \App{Caveats} I focus on some important caveats
regarding the use of projection operators.  Key notation is summarized
in \App[II]{Notation},\footnote{Sections relating to
  Part~II are prefaced by `II:' (\eg, appendix II:J).}
 which merges the definitions from both Part~I and Part~II.

This paper is intended to be useful to both graduate students just
beginning their study of classical plasma transport as well as seasoned
researchers interested in advanced techniques.  Although it rederives some
of Braginskii's principal results and quotes others of them for
completeness, it is neither a complete replacement for Braginskii's article
nor a comprehensive review of classical plasma transport theory.  For
initial study of classical transport, a reasonable plan of attack would be
to begin by skimming the first 
half of \Ref{Braginskii} (through the end of \Ssymb5, p.~262).  Additional 
perspective should then be provided from the present paper by
\Sec{Introduction}, \Sec{OCP}, 
\Sec{Braginskii} 
and \Sec{Discussion} as well as \Apps{CE}, \APP{Chat},
\APP{qe},
\APP{Viscosity}, and \APP{Eigenmodes},
with \Sec{GLE} and \Apps{Covariant} and \APP{Caveats}
containing more advanced material.  To fully appreciate the material in
\App{Caveats}, the reader may find it useful to first study the concise modern 
introduction to nonequilibrium statistical mechanics given by
\Ref{Zwanzig}, particularly Chap.~8 on projection operators.

\section{Linearized hydrodynamics for the one-component plasma}
\label{OCP}

In a one-component, weakly coupled\footnote{Weakly coupled means that the
 plasma is almost an ideal gas [\ie, that the plasma discreteness parameter
 $\ep \doteq 1/(n\lD^3)$ is very small].  Here $n$~is the density and
 $\lD$~is the Debye length.}
 plasma (for example, a discrete ion
plasma with 
smooth electron neutralizing background) in which the collision
frequency~$\nu$ is 
ordered large, it is well known that \CE\ theory singles out the number
density, momentum density, and kinetic-energy density  (or temperature, or
entropy density) as 
preferred \emph{hydrodynamic variables}.  That is, those quantities are
conserved by the Landau collision operator.  Mathematically, with
$\nbar$~being the spatially uniform mean density,
the velocity-dependent
functions~$\nbar$, $\nbar m\vv$, and $\half\nbar m v^2$ are the left null
eigenfunctions of the collision operator (in terms of a natural scalar
product to be defined shortly).  No dissipative transport is associated
with the 
null eigenfunctions.  Rather, transport is carried by the nonhydrodynamic
corrections to the local Maxwellian distribution function.  The traditional
\CE\ approach is reviewed in \App{CE}.

\subsection{Basic idea of the hydrodynamic projection}

These ideas can be codified by dividing the single-particle velocity space
into two orthogonal 
subspaces:  the \emph{hydrodynamic subspace}, spanned by the null
eigenfunctions of the collision operator, and the \emph{nonhydrodynamic},
\emph{orthogonal}, or  
\emph{vertical subspace}.\footnote{The notion of a vertical projection is
  common in the modern theory of differential geometry; see, for example,
  \Ref{Fecko}.} 
 
  The fluid evolution equations `live in
  the hydrodynamic subspace'.
The nonhydrodynamic part of the distribution function
determines the values of the transport
  coefficients. Transport coefficients enter the fluid equations
  because the dynamics of the hydrodynamic and nonhydrodynamic subspaces
  are coupled by the evolution equation for~$f$.  That coupling is a
  special case of the statistical closure problem\cite{JAK_tutorial} for
  passive equations with random coefficients.  (Such equations are said to
  possess a 
  \emph{stochastic nonlinearity}.)  Here the random variable is 
  velocity and the stochastic nonlinearity is the $\vv\.\vgrad f$ term in
  the kinetic equation.

The decomposition into hydrodynamic and nonhydrodynamic parts can be
expressed by the introduction 
of appropriate projection operators.  If $\P$~projects into the hydrodyamic
subspace and $\Q \doteq 1 - \P$ (where $1$~is the identity operator), and
if the perturbed (denoted by~$\D$) distribution function 
is written as $\Df = \Df_{\rm 
  h} + \Df_\perp$, then $\Df_{\rm h} = \P\Df$ and $\Df_\perp
= \Q\Df$.  
The basic idea is illustrated in \Fig{PQ}; the precise realization
of~$\P$ is given in the next section.

\FIGURE[1]{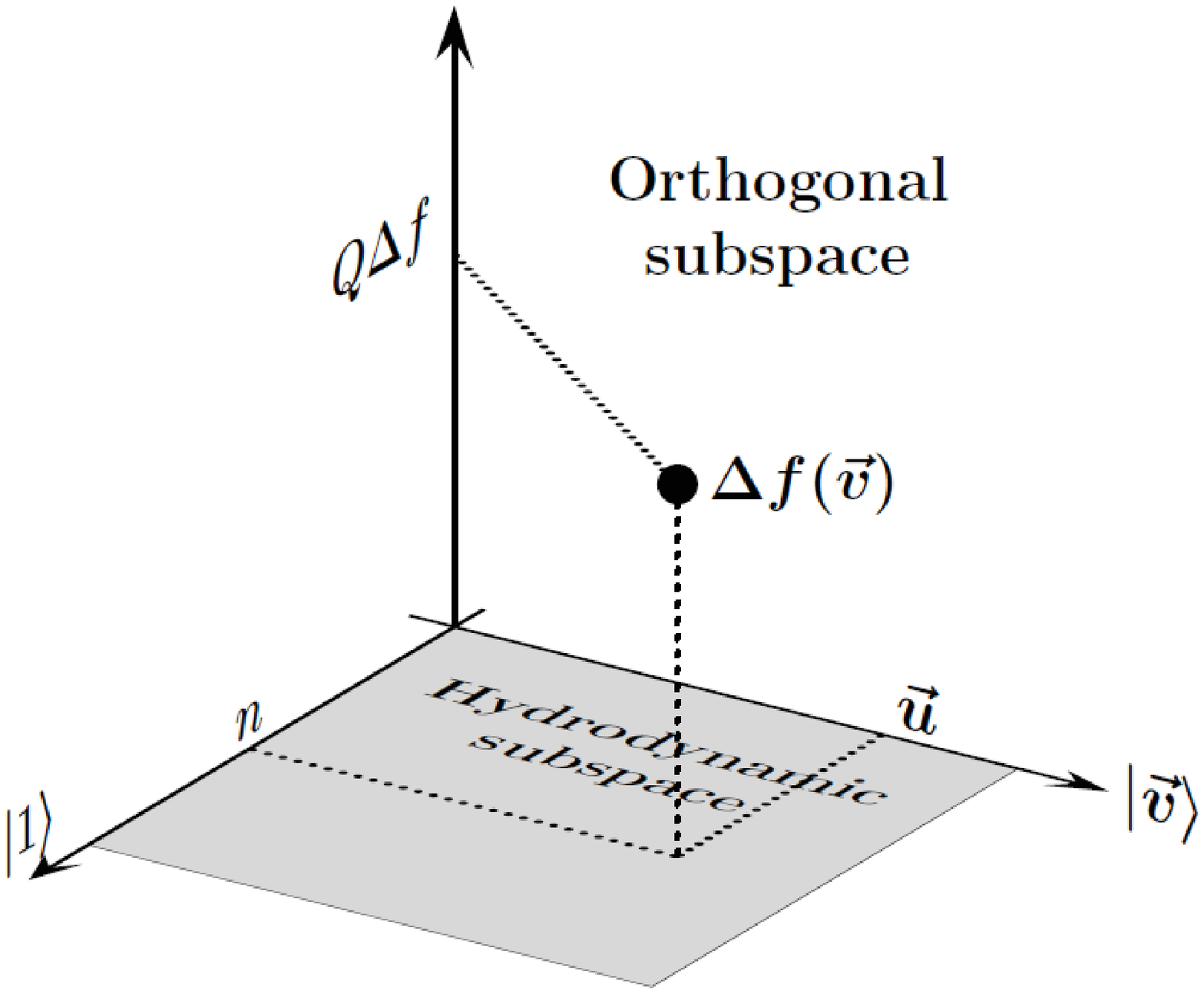}{Illustration of the $\P$ and~$\Q$ projections.  The
kinetic-energy axis of the hydrodynamic subspace is omitted for clarity.}

The formulation described in the present paper focuses on the decomposition
of the  
one-particle distribution function~$f_s(\vx,\vv,t)$, which lives in the
so-called 
\Muspace\ (the 6D phase space for one generic particle of species~$s$).  In
contrast, the instantaneous state of the entire collection of $N$~particles
can be described by a single point in the so-called \Gspace\ (of
$6N$~dimensions).  A well-known 
consequence of the reduction from \Gspace\ to \Muspace\ is that
the kinetic evolution equation 
for~$f$ is nonlinear (for example, it involves the nonlinear Landau collision
operator), which leads to various complications.
\gspace\ dynamics, on the other hand, are formally linear, being described
by the Liouville equation.  That linearity combined with the linearity of
projection operators was usefully exploited by Mori, who
was able to write 
exact equations for two-time correlation functions with apparent ease;
ultimately, the transport coefficients are expressed in terms of various
integrals over those correlation functions.
However, the simplicity is only formal.  Mori's projection operators
involve the full $N$-particle dynamics, and his equations contain
nonlinearities to all orders; in the general case, it is difficult to
extract from them 
quantitative expressions for the transport coefficients. As I
shall show in Part~II, the
formulas simplify in the limit of weak coupling, which fortunately is
appropriate for research on magnetic fusion and various other
applications.  However, that still leaves the dichotomy between the
linearity of projection operators and the nonlinearity of \muspace\ dynamics
to be dealt with.  It might seem that a projection-operator formalism is
restricted to the derivation of linearized transport equations.  That is
not correct, although the generalization is nontrivial. I
shall address nonlinear transport in Part~II\@. In the present paper, I
restrict attention 
to linear response (formally, infinitesimal perturbations of thermal
equilibrium), where the basic features of the projection-operator
methodology can be explained in the simplest possible context.

\subsection{Transport equations for the un\magnetize d, one-component, weakly
  coupled plasma} 

In this section I shall introduce the formalism by using projection methods
to derive the 
linearized fluid equations for the one-component plasma (OCP) in the limit
of weak coupling and in the
electrostatic approximation with background magnetic field $\vB =
\v0$.  (An even simpler example, the Brownian test particle, is discussed
in \Sec{Brownian}.)
It is well known and easy to show that the fully nonlinear moment
equations for the density~$n$, flow velocity~$\vu$, and temperature~$T$ of the
OCP are 
\BM
\BE
	\delt n + \div(n\vu) = 0,
\eq{n_dot}
\EE
\BE
	nm\Total{\vu}{t} = nq\vE - \vgrad p  - \div\mPi,
\eq{u_dot}
\EE
\BE
	\Case32 n\Total{T}{t} = -p\div\vu  - \div\vq - (\vgrad\vu):\mPi.
\eq{T_dot}
\EE
\EM
Here\footnote{The inclusion of the mean density~$\nbar$ in the definition
  \EQ{n_def} of 
  density is a normalization convention that makes $f\,\dd\vv$ dimensionless.
Specifically, $f$~is normalized such that $V\m1\Int
  \dd\vx\,\dd\vv\,f(\vx,\vv,t) = 1$, where $V$~is the volume of 
  the system.  Thus, $f$~differs, though inessentially, from a true
  probability density function (PDF), whose normalization would not include
  the $V\m1$~factor.  An example of an~$f$ normalized with this convention
  is the local Maxwellian given in \Eq{fLM}.
\endgraf
\hspace{5pt}
In the general case of arbitrarily coupled plasma, there are
potential-energy corrections to \Eq{T_def}.   A kinetic theory
for such a system was derived by \Ref{Forster-Martin}.  One of the
few deficiencies of Braginskii's review article is that he does not emphasize
the restriction to a nearly ideal gas.} 
\BALams
n(\vx,t) &\doteq \Int \dd\vv\,\nbar f(\vx,\vv,t),
\eq{n_def}
\\
n\vu(\vx,t) &\doteq \Int \dd\vv\,\nbar\vv f(\vx,\vv,t),
\\
\Case32 nT(\vx,t) &\doteq \Int \dd\vv\,\Half\nbar m\,w^2 f(\vx,\vv,t),
\eq{T_def}
\EALams
where $\vw \doteq \vv - \vu$ is the so-called peculiar velocity
and $p \doteq nT$ is the pressure of an ideal gas.
The total advective time derivative is
\BE
\Total{}{t} \doteq \Partial{}{t} + \vu\.\vgrad,
\EE
and the stress tensor~$\mPi$ and the heat-flow vector~$\vq$ are defined by 
\BALams
	\mPi &\doteq \Int \dd\vv\,(\nbar m\vw)\vw f - p\mone,
\\
	\vq &\doteq \Int \dd\vv\,\(\Half \nbar m w^2\)\vw f.
\eq{q_def0}
\EALams
These equations are closed
in the collisional limit by the results
\BE
\mPi \doteq 
-nm\mu\((\vgrad\vu) + (\vgrad\vu)\Tr - \Case23(\div\vu)\mone\),
\eq{Pi_def}
\EE
where T~denotes transpose and $\mu$~is the kinematic viscosity\footnote{In context, there
should be no confusion between the kinematic viscosity and the mass ratio
$\me/\mi$, which I also denote by~$\mu$.  The distinction is clearer in
multispecies plasma, where the viscosities carry species
subscripts. I also use~$\mu$ for the reduced mass $\mu_{ss'}$ as well as a
contravariant or covariant index of a hydrodynamic vector.}; and 
\BE
\vq \doteq - n\k\vgrad T,
\EE
where $\k$~is the thermal conductivity.
The linearizations of these equations around an absolute
thermal equilibrium (which possesses no gradients
or flows) of density $n_0 = \nbar$ are 
\BM
\BE
\Partial{}{t}\fR{\D n}{\no} + \div\D\vu = 0,
\eq{Dn_dot}
\EE
\BE
\no m\Partial{\D\vu}{t} = \no q\D\vE - \vgrad\D p - \div\D\mPi,
\eq{Du_dot}
\EE
\BE
\Case32\no\Partial{\D T}{t} = -p_0 \div\D\vu - \div\D\vq,
\eq{DT_dot}
\EE
\EM
where
\BALams
\D\mPi &\doteq -\no m\mu_0\((\vgrad\D\vu) + (\vgrad\D\vu)\Tr -
\Case23(\div\D\vu)\mone\),
\eq{DPi_def}
\\
\D\vq &\doteq -\no\k_0\vgrad\D T.
\eq{Dq_def}
\EALams
Specific formulas for the transport coefficients~$\mu_0$ and~$\k_0$ are  
available (\App{CE}).  The immediate goal is to rederive these linear
results from the projection-operator formalism. 

The 
starting point is the nonlinear Landau kinetic equation
\BE
\TotaL{f}{t} \doteq \Partial{f}{t} + \vv\.\vgrad f + \vE\.\vdel f = -\C\on{f},
\eq{f_dot}
\EE
where $\vdel \doteq (q/m)\del/\del\vv$ and $\C\on{f}$ is the nonlinear
Landau collision operator,\footnote{The unconventional choice of minus
  sign on the \rhs\ of \Eq{f_dot} is made so that the collision operator is
  effectively positive, consistent with the usual convention of a positive
  collision frequency.   The linearized operator is positive-semidefinite;
  see \App{Chat}.}
 written here as a general nonlinear functional (denoted by the square
 brackets) 
of~$f$.  The electric field~$\vE = -\vgrad\phi$ (I consider only the
electrostatic approximation) is obtained from Poisson's
equation
\BE
-\grad^2\phi(\vx,t) = 4\pi\r = 4\pi\sum_\sbar(\nbar q)_\sbar\Int
\dd\vvbar\,f_\sbar(\vx,\vvbar,t). 
\EE
For the OCP, one allows perturbations to only one species, so a species
label is dropped in this section; the other
species merely serve to provide an overall charge-neutral background.
A stationary and stable solution of this system is the absolute
Maxwellian,
\BE
\fM(\vv) \doteq (2\pi\,\vt^2)\m{3/2}e^{-v^2/2\vt^2},
\EE
where $\vt \doteq (T/m)\ehalf$.  Infinitesimal perturbations to that
equilibrium obey
\BE
\delt\Df + \vv\.\vgrad\Df + \D\vE\.\vdel\fM = -\Chat\Df,
\eq{Df_dot}
\EE
where $\Chat$~is the linearized Landau operator (discussed in \App{Chat}).
This is the basic dynamical equation used in this article.  (I shall add a
background magnetic field in \Sec{Braginskii}.)  As a linear dynamical 
evolution equation, it is analogous
to the \Schr\ equation of quantum mechanics.  There are, of course,
important differences:  the \Schr\ equation is time-reversible whereas the
present equation is time-irreversible due to the presence of~$\Chat$, and
the quantum-mechanical wave function is a probability amplitude (whose
squared modulus is a probability) whereas
$f$~is directly an actual probability (density).

In this and the next several sections I shall use the \Schr\
representation, in which averages of time-independent functions of velocity
are taken with the time-evolved~$f(t)$, which plays the role of a
time-dependent state vector.  It is useful to treat those
averages as projections, and it is convenient to realize those projections
by using a time-independent
scalar product.  I shall use a Dirac bra--ket notation with a hidden weight
function~$\fM$.  Thus, for any functions $A(\vv)$ and~$B(\vv)$,
\BALams
\bra A| &\doteq A(\vv),
\\
\ket A> &\doteq A(\vv)\fM(\vv),
\\
\<A | \L | B> &\doteq \Int
\dd\vv\,\dd\vvbar\,A(\vv)L(\vv,\vvbar)B(\vvbar)\fM(\vvbar),
\eq{ALB}
\EALams
where $\L$~is a linear operator with two-point kernel\footnote{For example, if $\L \doteq \del/\del\vv$, then
  $L(\vv,\vv') = \delvv\Dirac{\vv - \vv'}$.  Here 
  `kernel' is used in one of its two standard meanings; it does not refer
  to the null space of the operator.}  
$L(\vv,\vv')$.
When multispecies plasmas are discussed later, the scalar product will be
extended to include species summations.
Note that no complex conjugate appears in this definition, unlike the
analogous situation in quantum mechanics.
  This scalar
product is `natural' because (\App{Chat}) $\nbar\Chat$~is self-adjoint \wrt~it:
\BE
\<A|\nbar\Chat|B> = \<B|\nbar\Chat|A>.
\eq{A_Chat_B}
\EE
A special case of \Eq{ALB} is obtained by specializing~$\L$ to the identity
operator $\mathrm{I} \to \Dirac{\vv - \vvbar}$:
\BE
\<A|B> = \Int \dd\vv\,A(\vv)B(\vv)\fM(\vv),
\EE
just the equilibrium velocity average of the product $AB$:  $\<A|B> =
\<AB>_{\rm M}$. I shall often drop the $\rm M$~subscript.

If one writes $\Df \doteq \chi\fM$ for unknown (dimensionless)~$\chi$,
\Eq{Df_dot} becomes 
\BE
\delt\ket\chi> + \vv\.\vgrad\ket\chi> + \ket\vdel\ln\fM>\.\D\vE =
-\Chat\ket\chi>,
\eq{chi_dot}
\EE
where
$
\ket\vdel\ln\fM> = -(q/T)\ket\vv>
$.
In collisional transport theory, the conservation properties of the
collision operator are crucial.  For the Landau operator, those
are\footnote{In \Eqs{A_Chat_B} and \EQ{OCP_cons}, the~$\nbar$ is
  unnecessary, as it is a 
  constant.  It is retained 
 so that the formula looks identical to that for multispecies
  plasma (\Sec{Braginskii}), where the scalar product includes a species
  summation.} 
\BE
\bra\nbar\breve\vA|\Chat = 0,
\eq{OCP_cons}
\EE
where $\nbar$~is the
spatially constant density of the equilibrium,
\BE
\breve\vA 
\doteq 
(
1\;\;
\vP\;\;
K
)\Tr,
\EE
$\vP \doteq m\vv$, $K \doteq \half m v^2$, and the
breve accent is used to distinguish nonorthogonal functions from more
convenient orthogonal ones to be introduced shortly.\footnote{The choice of
upper 
case for the velocity-dependent functions~$\vP$ and~$K$ is made for
consistency with the convention of 
\Ref{Brey}, whose approach I shall discuss in Part~II.  Another notational
possibility would have been to use a tilde, a notation I have
often used in other articles to denote a random variable.} 
Thus, $\bra\breve\vA|$ are the five null left eigenfunctions of~$\nbar\Chat$.
(In fact, there is no need to distinguish left and right
eigenfunctions because for the Landau operator $\nbar\Chat$~is self-adjoint
\wrt\ the chosen scalar product.)
Those eigenfunctions
span a preferred five-dimensional \emph{hydrodynamic subspace} (see
\Fig{PQ}).   

From the definitions of~$n$, $\vu$, and~$T$, it follows that
\BE
\<1|f/\fM> = n/\nbar,
\EE
so
\BE
\<1|\chi> = \D n/\nbar.
\EE
Similarly, 
\BE
\<\vv|\chi> = \D\vu
\EE
(the absolute equilibrium has no flow).  Finally,
\BE
\<K|\chi> = \Case32 \DT + \Case32\Bar{T}\fR{\Dn}{\nbar}.
\EE
Thus, the perturbed temperature can be extracted by
\BE
\Case32\DT = \<K'|\chi>,
\EE
where 
\BE
K' \doteq K - \<K> = \Half m v^2 - \Case32\Bar{T}.
\eq{K'_def}
\EE
This motivates the introduction of
\BE
\vA \doteq (1\;\;\vP'\;\;K')\Tr,
\eq{vA_def}
\EE
where $\vP' = \vP - \<\vP> = \vP$; with this definition, the components
of~$\vA$ are 
orthogonal.  Then the deviations of the statistically averaged 
hydrodynamic variables from their equilibrium values are given by
\BE
\Dva \doteq \<\vA|\chi> = 
\(
\fr{\Dn}{\nbar}\;\;
\D\vp\;\;
\Case32\DT
\)\Tr,
\eq{Dva_moment}
\EE
where $\D\vp \doteq m\D\vu$.

These results lead one to define the hydrodynamic projection operator as
\BE
\P \doteq \ket\vA\Tr>\.\mM\m1\.\bra\vA|,
\EE
where the normalization matrix
\BE
\mM \doteq \<\vA\vA\Tr>
=
\begin{pmatrix}
1 & \v0\Tr & 0\\
\v0 & N_\vp\mone & \v0\\
0 & \v0\Tr & N_T
\end{pmatrix}
\eq{M_def}
\EE
ensures that $\P^2 = \P$;
one easily calculates that
\BE
N_\vp \doteq m^2\vt^2,
\quad
N_T \doteq \Case32 T^2.
\eq{N_def}
\EE
(I now drop the overlines on the equilibrium temperature
when there is no possibility of confusion.)
The projection of the state vector is thus
\BE
\P\ket\chi> = \ket\vA\Tr>\.\mM\m1\.\<\vA|\chi> 
= \ket\vA\Tr>\.\mM\m1\.\Dva,
\eq{P_chi}
\EE
and the hydrodynamic variables can be extracted by taking the scalar
product of~$\vA$ with that projection: 
\BE
\Dva(t) = \<\vA|\P|\chi(t)>.
\eq{Dva}
\EE
\comment
\BALams
\P &\doteq \ket 1>\bra 1| + \ket m\vv>(mT)\m1\bra m\vv|
+ \BIG\ket \Half m v^2 - \Case32T>\(\Case32 T^2\)\m1\BIG\bra \Half m v^2 -
\Case32T|
\eq{Pa}
\\
&= \ket 1>\bra 1| + \ket \vv>\vt\m2\.\bra\vv| + \BIG\ket\fr{mv^2/2}{T} -
\Case32>\fR{3}{2}\m1\BIG\bra\fr{mv^2/2}{T} - \Case32|,
\eq{Pb}
\EALams
where $T$~is the temperature of the equilibrium.  (From here on I eschew
the use of 0~subscripts to denote equilibrium quantities.)
Note that $\P$~is
properly normalized such that $\P^2 = \P$, which is a requirement for a
projection operator.  For arbitrary function~$G$, let us adopt the
notation $G' \doteq G - \<G>$ except for the case $G = 1$, for which I
define $1' \doteq 1$.  Define
, $K \doteq m
v^2/2$, and $K' \doteq K - 3/2$.
Also introduce the normalization matrix
\BE
\mM \doteq \<\vA\vA\Tr> = 
\begin{pmatrix}
1 & \v0\Tr & 0\\
\v0 & N_v\mone & \v0\\
0 & \v0\Tr & N_T
\end{pmatrix},
\EE
where T~denotes transpose and
\BE
N_v \doteq \vt^2,
\quad
N_T \doteq \Case32 T^2.
\EE
In a properly covariant notation, the elements of~$\mM$ are labelled
$M^\mu_\nu \doteq \<A'{}^\mu A'_\nu>$.  
Then \Eq{Pb} can be written as
\BE
\P = \ket\vA\Tr>\.\mM\m1\.\bra\vA|.
\EE
That is, the components of~$\P$ are $P_\mu^\nu = (M\m1)_\mu^\nu$.

To understand the content of~$\P$,
Then
\BALams
\BIG\<\Half m v^2 - \Case32\To | \fr{f}{\fM}> &= \BIG\<\Half m\d v^2 + \Half m
u^2 - 
\Case32\To|\fr{f}{\fM}>
\\
&= \fr{n}{\nbar}\(\Case32T + \Half m u^2 - \Case32\To\),
\EALams
and
\BE
\BIG\<K'| \chi> = \Case32\D T.
\EE
Thus, the perturbations in the hydrodynamic variables are the (???)
components of the projected state vector:
\BE
\P\ket\chi> = \BIG\ket 1>\fr{\D n}{\nbar}
+ \BIG\ket \fr{\vv'}{\vto}>\.\fr{\D\vu}{\vto}
+ \BIG\ket \fr{m v^2/2}{\To} - \Case32>\fr{\D T}{\To}.
\EE
Alternatively
\BALams
\begin{pmatrix}
\D n/\nbar
\\
\D\vu
\\
3\D T/2 
\end{pmatrix}
= \<\vA | \P | \chi>.
\eq{components}
\EALams
\endcomment

The projection formalism can be couched in a manifestly covariant fashion
that turns out to be very convenient.
Namely, if the components of~$\vA$ are \labeled\ as $A^\mu$, where $\mu =
1,\:\dots,\:5$ (or $\mu \in\set{n,\:\vp,\:T}$), then $\P$~can be written as
\BE
\P = \ket A_\mu>\bra A^\mu|,
\EE
where Einstein's convention for summation over repeated indices is adopted.  
That is,
$(\mM\m1)_{\mu\nu}$ plays the role of a metric tensor $g_{\mu\nu}$
that lowers indices according to $A_\mu = g_{\mu\nu}A^\nu$.  One has
\BE
\<A^\mu | A_{\mu'}> = \d^\mu_{\mu'},
\EE
the hydrodynamic projection of the state vector is
\BE
\P\ket\chi(t)> = \ket A_\mu>\D a^\mu(t)
\EE
[\cf~\Eq{P_chi}],
and the hydrodynamic variables themselves are
\BE
\D a^\mu(t) = \<A^\mu|\chi(t)> = \<A^\mu|P|\chi(t)>
\EE
[\cf~\Eqs{Dva_moment} and \EQ{Dva}].   This representation, which
identifies the~$\D a^\mu$'s as 
the contravariant components of a \emph{hydrodynamic vector}, is further
discussed in \App{Covariant}.  

Because $\P$~is a time-independent linear operator, one can extract the
linearized fluid moment equations by applying~$\P$ to \Eq{chi_dot}, or
equivalently by taking the time derivative of \Eq{Dva}.
Define
\BE
\LL \doteq \LQ + \LE
\quad
\hbox{and}
\quad
\LQ \doteq \L_\vv + \L_\C,
\eq{LL0}
\EE
where
\BE
\ii\L_\vv \doteq \vv\.\vgrad,
\quad
\ii\L_\C \doteq  \Chat,
\quad
\ii\LE \doteq -(q/T) \ket\vv>\.\cE,
\eq{iLs}
\EE
and $\cE$~is the linear operator that solves Poisson's equation
for~$\D\vE$ in terms of~$\chi$; in Fourier space,
\BE
\cE_\vk = \ek\bra \nbar q|,
\quad
\ek \doteq -4\pi\,\ii\vk/k^2
\eq{cEk_def}
\EE
($\ek$~is the Fourier transform of the electric field of a unit point charge).
Then the hydrodynamic projection of the linearized kinetic equation is
\BE
\delt \P\ket\chi> + \P\ii\LL\ket\chi> = 0.
\EE
From \Eq{iLs}, one sees that the $\D\vE$~term
lies entirely in the hydrodynamic subspace [which is why $\LE$~was broken
out separately in \Eq{LL0}].  A consequence is that $\P$
and~$\LE$ commute:  $[\P,\,\LE] \doteq \P\LE - \LE\P = 0$.
The $\P \ii\LQ$ term, however, is
problematical because $\P$ and~$\LQ$ do not commute:  $[\P\LQ,\,\LQ\P] 
 \neq 0$.
(For the OCP,
 $[\P,\,\Chat] = 0$ as a consequence of the conservation laws,\footnote{For
 the multispecies case discussed in 
 \Sec{Braginskii}, also $[\P,\,\Chat] \neq 0$ since $\P$~is chosen to project
 into a species-dependent subspace but the conservation laws involve species
 summation.}  but $[\P,\,\vv]  \neq 0$.) 
This is one instance
of the famous \emph{statistical closure problem}\cite{JAK_tutorial},
applied here to averages over a random velocity variable.
To deal with this, I follow Mori and insert the identity $\P + \Q = 1$ so
that 
\BE
\P \LL\ket\chi> = \P \LL(\P + \Q)\ket\chi> = \P \LL \P\ket\P\chi> + \P \LL
\Q\ket\Q\chi>.
\eq{PLchi}
\EE
(In writing the last form, I used $\P^2 = \P$ and $\Q^2 = \Q$.  I also
slightly abused the notation\footnote{Strictly speaking, the~$\P$ should
remain outside of the ket because it operates on the hidden~$\fM$ as well as
on~$\chi$.} 
 to write $\P\ket\chi> \equiv \ket\P\chi>$.)
Let us define the \emph{frequency 
  matrix}~$\mOmega$ as
\BE
\mOmega \doteq \<\vA | \LL | \vA\Tr>\.\mM\m1;
\EE
this is a normalized version of the hydrodynamic matrix element of~$\P\LL\P$.
The nomenclature is consistent with the fact that under
Fourier transformation $\vv\.\vgrad \to \ii\vk\.\vv$, which corresponds to the
streaming frequency $\Omega_\vv \doteq \vk\.\vv$.  
In general, the `frequency' is complex because 
$\cL$~involves 
the dissipative term~$\L_\C \doteq -\ii\Chat$.  However, for the OCP
$\Chat$~does not contribute to the frequency matrix since $\P\Chat = 0$ as
a consequence of the conservation laws. 
In any event, if the $\P\LL\Q$ term in \Eq{PLchi}
were to vanish, then one would have a matrix equation for
$\ket\P\chi>$ (or for the $\D\va$'s) and the closure problem would be solved.  Unfortunately, this
is not the case because $[\P,\,\vv] \neq \v0$.  (This is one way of stating
that the kinetic equation is stochastically nonlinear in the velocity
variable.) Thus, the dynamics in the hydrodynamic subspace,
\BE
\delt\ket\P\chi> + \ket\vA\Tr>\.\mM\m1\.\ii\mOmega\.\<\vA|\P\chi> +
\P\ii\L\Q\ket\Q\chi> = 0
\eq{Pchi_dot}
\EE
or, upon applying $\bra \vA|$ to \Eq{Pchi_dot},
\BE
\delt\Dva + \ii\mOmega\.\Dva + \<\vA|\ii\LQ\Q|\Q\chi> = 0,
\eq{Dva2}
\EE
are coupled to those in the orthogonal subspace.
Note that only~$\L$ ($\LL$ \sans~$\LE$) enters the last term of \Eq{Dva2};
$\Q\LE = \LE\Q = 0$ since $\LE$~has a component only in the hydrodynamic
subspace. 

In order to obtain a closed system, it is necessary to
eliminate~$\ket\Q\chi>$ in \favor\ of some function of~$\Dva$.  Upon
applying~$\Q$ to \Eq{chi_dot}, one obtains
\BE
\delt\ket\Q\chi> + \Q\ii\LQ\Q\ket\Q\chi> + \Q\ii\LQ\P\ket\P\chi> = 0.
\eq{Qchi_dot}
\EE
Since $\Q\ii\LQ\Q$ is a linear operator, \Eq{Qchi_dot} can be solved by
means of a Green's function:
\BE
\ket\Q\chi>(t) = \UQ(t;0)\ket\Q\chi(0)> - \I0t
\dd\tbar\,\UQ(t;\tbar)\Q\ii\LQ\P\ket\P\chi(\tbar)>,
\eq{Qchi_IC}
\EE
where
\BE
\UQ(t;t') \doteq H(t-t')\exp[-\Q\ii\LQ\Q(t-t')]
\eq{U}
\EE
(since~$\Q$ and~$\LQ$ are time-independent).  Here $H(\t)$~is the unit step
function.\footnote{$H(\t) = 0$ if $\t < 0$, $\half$~if $\t = 0$, or~1 if
  $\t > 0$.} 
Only the time arguments are
written 
explicitly in the previous two equations, although in reality the kernels
of two-point operators such as~$\UQ$ also
depend on two space and two velocity variables, so convolutions over those
variables are implied.  Because the background is
spatially homogeneous, one can Fourier transform in space if that is desired.
It is conventional to ignore the
initial condition $\ket\Q\chi(0)>$.  Physically, this means that the system is
prepared to lie entirely in the
hydrodynamic subspace.  (This choice is further discussed later.)  Then one
obtains 
\BE
\delt\Dva(t) + \ii\mOmega\.\Dva(t)+ \I0t
\dd\t\,\mSigma(\t)\.\Dva(t-\t) = 0,
\eq{Dva_dot}
\EE
where\footnote{Elementary discussion of the role of the `mass
  operator'~$\mSigma$ in more general contexts is given by
  \Ref{JAK_tutorial}.  More advanced details can be found in \Ref{JAK_PR}.}  
\BE
\mSigma(\t) \doteq \<\vA|\LQ\Q\UQ(\t)\Q\LQ|\vA\Tr>\.\mM\m1.
\eq{mSigma_def}
\EE

Although \Eq{Dva_dot} is now closed in terms of~$\Dva$, that system does
not yet have the form of conventional linearized 
fluid equations because it is nonlocal in time.  That nonlocality can also
be represented in frequency space.
The one-sided temporal Fourier transform of \Eq{Dva_dot} is
\BE
[-\ii\w\mone + \ii\mOmega + \mSigmahat(\w)]\.\Dvahat(\w) =
\Dva(0).
\EE
This equation is completely general (modulo the neglect of the vertical 
initial condition), but it
introduces the frequency-dependent matrix~$\mSigmahat(\w)$.
Collisional 
transport theory emerges if one can approximate $\mSigmahat(\w) \approx 
\mSigmahat(0)$ [and also take $\vgrad \to \ii\vk \to \v0$
  in~$\G_{\Q,\vk}(\t)$].  To 
see whether this 
is possible, let us look more
closely at~$\mSigma(\t)$.  
Because in \Eq{U} the~$\L(\t)$ is sandwiched between two
$\Q$~operators, the null eigenvalues of~$\Chat$ do not contribute and the
effect of~$\Chat$ is roughly to provide a damping $e^{-\l \t}$, where
$\l$~is of the order of the collision frequency~$\nu$.  (For discussion and
examples of
this assertion, which for the Landau operator is nontrivial, see
\Sec{spectrum}.)  Then for scale lengths that are long 
compared to the mean 
free path, one can approximate ${\exp[-\Q(\vv\.\vgrad + \Chat)\Q\t]} \approx
\exp(-\Q\Chat\Q\t)$.  Thus, since $\mSigma(\t)$~has a finite decay
time~$\sim \nu\m1$, 
a Markovian approximation is justifiable.  This coarse-graining leads to a
conventional set of time-local (Markovian) fluid equations:
\BE
\delt \Dva(t) + (\ii\mOmega +
\meta)\.\Dva(t) = 0,
\eq{fluid_eqns}
\EE
where
\BE
\meta \doteq \InT \dd\t\,\mSigma(\t) \approx
\<\vA|\L\Q(\Q\Chat\Q)\m1\Q\L|\vA\Tr>\.\mM\m1,
\eq{meta}
\EE
or covariantly
\BE
\eta^\mu_\nu = \<A^\mu|\L\Q(\Q\Chat \Q)\m1\Q\L|A_\nu>.
\eq{meta_cov}
\EE
This~$\meta$ is just the temporal Fourier transform~$\mSigmahat(0)$.

To see that the formalism recovers the conventional linearized fluid
equations for the OCP, one needs to work out the various matrix elements.
Let us begin with the frequency matrix:
\BE
\ii\mOmega = \ii\mOmega_\vv + \ii\mOmega_\C + \ii\mOmega_\vE,
\EE
where
\BALams
\ii\mOmega_\vv &\doteq \<\vA | \vv\.\vgrad | \vA\Tr>\.\mM\m1,
\\
\ii\mOmega_\C &\doteq \<\vA | \Chat | \vA\Tr>\.\mM\m1,
\eq{mOmega_C_def}
\\
\ii\mOmega_\vE &\doteq \<\vA | \ii\LE | \vA\Tr>\.\mM\m1.
\EALams
In these expressions, the role of the~$\mM\m1$ is to change contravariant
indices to covariant ones.  Thus,
for example,
\BE
(\ii\mOmega_\vv)^\mu_\nu = \<A^\mu|\vv\.\vgrad|A_\nu>.
\EE
(Note that this mixed tensor has the dimensions of frequency.)
For the OCP, $\mOmega_\C$~vanishes because $\bra\vA|$ is a left null
eigenfunction of~$\Chat$.  The $\mOmega_E$ term is readily shown to give
rise to the electric-force term in the momentum equation.
Finally, with $i$, $j$, and~$k$ denoting Cartesian vector components, one finds
\BE
\<\vA|v_k|\vA\Tr> = \begin{pmatrix}
0 & T\Kron{jk} & 0\\
T\Kron{ik} & \tensor0 & T^2\Kron{ik}\\
0 & T^2\Kron{jk} & 0
\end{pmatrix}.
\EE
Upon multiplying on the right by $\mM\m1$, one finds
\BE
(\ii\mOmega_\vv)^\mu_\nu  = \begin{pmatrix}
0 & m\m1\grad_j & 0\\
T\grad_i & \tensor0 & \displaystyle\Case23\grad_i\\
0 & \To m\m1\grad_j & 0
\end{pmatrix}.
\eq{mOmega}
\EE
Note that the spatial Fourier transform of this matrix is proportional to~$\vk$.

To illustrate the content of~$\mOmega$, consider the density
component  of \Eq{fluid_eqns}.  It is easy to see that there
is no $\eta$ 
contribution because $\bra 1 |(\vv\.\vgrad + \Chat)\Q = \div\bra \vv |\,\Q
+ \bra 1|\Chat\Q = 0$.  The last equality follows since $\ket\vv>$~is in
the hydrodynamic 
subspace, which is orthogonal to~$\Q$, and since $\Chat$~conserves
number. But $\Omega^n_\vp$ couples density 
to momentum, and one obtains 
\BE
\Partial{}{t}\fR{\D n}{\nbar} + \div\D\vu = 0,
\EE
which reproduces the linearized continuity equation \EQ{Dn_dot}.
As another illustration, consider the velocity component of
\Eq{fluid_eqns}.  One readily finds that 
\BE
\Partial{\D\vp}{t} = q\D\vE - \nbar\m1\vgrad\D p + \cdots = 0,
\EE
where $\D p = \To\D n + \nbar\D T$ and the \centre d dots
denote the contribution from~$\meta$, which will be discussed in the next
paragraph. The explicit terms reproduce the 
correct nondissipative (Euler) part of the linearized momentum equation
\EQ{Du_dot}.  
Similarly, the temperature projection leads to
\BE
\Case32\Partial{\D T}{t} = -\To\div\D\vu + \cdots,
\EE
the explicit terms being the Euler part of the linearized temperature
equation \EQ{DT_dot}. 

Now consider the dissipative contributions from~$\meta$.  I shall discuss
the stress tensor in detail, leaving the heat-flow vector as an
exercise for the reader.  One has (now dropping the prime on~$\vv'$ since
there is no flow in the equilibrium)
\BE
\eta^\vp_\nu = -\< m\vv|\vv\.\vgrad\Q(\Q\Chat\Q)\m1\Q\vv\.\vgrad|
A_\nu>.
\eq{v_eta_chi}
\EE
\comment
(In these expressions I wrote $\ket\chi>$ instead of $\P\ket\chi>$ because
$\meta$ already contains a~$\P$ on the right.) 
\endcomment
\comment
Now
\BE
\Q\vv\.\vgrad\P\ket\chi> = 
(1 - \P)\vv\.\vgrad\P\ket\chi> = \vv\.\vgrad\P\ket\chi> - \mOmega\ket\chi>.
\EE
\endcomment
The density component~$\eta^\vp_n$ vanishes because $A_n = 1$ and
$\Q\ket\vv> = \v0$. 
The temperature contribution vanishes by a
symmetry argument that uses the fact that $\Chat$~is rotationally invariant.
For the self-coupling~$\eta^\vp_\vp$, note that
\BE
\Q\vv\.\vgrad\ket A_\vp> = (1 - \P)\vv\.\vgrad\ket A_\vp>.
\EE
One has
\BALams
\P\vv\.\vgrad\ket A_\vp> &= \ket A_\mu>\<A^\mu|\vv\.\vgrad|A_\vp>
\\
&= \ket A_n>(\ii\mOmega_\vv)^n_\vp + \ket A_T>(\ii\mOmega_\vv)^T_\vp
\\
&= \Third T\m1\ket v^2>\vgrad.
\EALams
Thus,
\BE
\Q\vv\.\vgrad\ket A_\vp> = \fr{1}{T}\BIG\ket \vv\,\vv - \Third
v^2\mone>\.\vgrad ,
\eq{Qvv}
\EE
and then
\BE
\eta^\vp_\vp\.\D\vp = \nbar\m1\div\D\mPi,
\EE
where the linearized stress tensor is
\BE
\D\mPi = -\nbar m\tensor{m}:\vgrad\D\vu
\EE
and the fourth-rank tensor~$\tensor{m}$ is
\BE
\tensor{m} \doteq \vto\m2\BIG\<\vv\,\vv - \Third v^2\mone|(\Q\Chat\Q)\m1|
\vv\,\vv - \Third v^2\mone>.
\eq{m_tensor}
\EE
This is the same result that follows from the traditional approach
described in \App{CE}.  There the matrix element is written in terms of
$\vw \doteq \vv - \vu$, where $\vu$~is the lowest-order flow velocity.
Here we are perturbing from an equilibrium with no flow, so $\vw \to \vv$.

The quantity $\ket m\vv\vv - \third m v^2\mone>$ [see \Eq{Qvv}] is called
the \emph{subtracted momentum flux}; the subtraction arises from the~$\P$ in $\Q
= 1 - \P$.  All dissipative transport coefficients are defined in terms of
subtracted fluxes; the physical reason is that the $\P$~terms represent
the fluxes that already exist in local thermal equilibrium, and those must be
subtracted from the total flux in order to obtain the gradient-driven
corrections that are responsible for the net transport (which determines
the relaxation of a small perturbation to thermal equilibrium).
In classical \CE\ theory (\App{CE}), subtracted fluxes
arise when the first-order solvability conditions (Euler equations) are
used to 
 simplify the correction equations by eliminating time derivatives in \favor\ of
 spatial gradients.  The use of projection operators executes that task
substantially more efficiently.

The reduction of~$\tensor{m}$ to a form involving a single scalar viscosity
coefficient~$\mu$ is given in \App{CE}.  
(In the special case of the OCP, one can replace~$\Q\Chat\Q$ by~$\Chat$
because $\P\Chat = \Chat\P = 0$.)
One thus recovers the proper expression \EQ{DPi_def} for the linearized stress 
tensor~$\D\mPi$ of the un\magnetize d OCP, where $\mPi$~is defined by 
\Eq{Pi_def}. Similar considerations lead to \Eq{Dq_def} for the linearized heat
 flow.
 
\section{The Braginskii equations:  Multispecies, \magnetize d, classical,
  collisional fluid equations}
\label{Braginskii}

I now turn to the important problem of classical transport in multispecies,
\magnetize d plasma (in the limit of weak coupling).  This is somewhat more
technically 
complicated than is the OCP because of interspecies collisional coupling and
the loss of symmetry due to the magnetic field, but the basic idea of
projection into
hydrodynamic and vertical subspaces still applies.  Again, in this paper
I only consider 
linear response (\NS\ transport coefficients).  Nonlinear (Burnett)
corrections will be considered in Part~II.  

\subsection{Exact form of the moment equations, and summary of the results
  for a two-species plasma}    
\label{exact}

Before I consider perturbations from thermal equilibrium, it is useful to
have the exact form of the moment equations in mind.
The starting point will be the Landau equation for the one-particle
distribution, including the effect of an external magnetic field~$\vB$:
\BE
	\delt f_s + \vg f_s + (\vE +
	c\m1\vv\times\vB)\.\vdel_s f = -\C_s\on{f},	
\eq{kinetic}
\EE
where as usual
$
	\vE_\vk = \sum_{\sbar}(nq)_{\sbar}\Idvbar\,\vepsilon_\vk
	f_{\sbar,\vk}(\vvbar) 
$,
$\vdel_s \doteq (q/m)_s\delvv$,
and $\C$~is the Landau operator defined in \Eq{C_Landau}.
That operator conserves number density (separately for each species), total
(summed over species) momentum density, and total kinetic-energy density:  
\BALams
\sum_\sbar\nbar_\sbar\!\Int \dd\vvbar
\begin{pmatrix}
\Kron{s\sbar}\\
m_\sbar\vvbar\\
\half m_\sbar \Bar{v}^2
\end{pmatrix}\C_\sbar\on{f} = 0.
\EALams

Upon taking the number-density, momentum-density, and
kinetic-energy-density moments of 
\Eq{kinetic}, one 
is led to
\BM
\eq{moment}
\BE
	\delt n_s + \div(n_s\vu_s) = 0,
\eq{moment_n}
\EE
\BE
	(nm)_s\Total{_s\vu_s}{t} = (nq)_s(\vE + c\m1\vus\cross\vB) -\grad
	p_s 
-\div\mPi_s + 
	\vR_s,
\eq{moment_u}
\EE
\BE
	\Case32n_s\Total{_sT_s}{t} = -p_s\div\vu_s - \div\vq_s - \mPi_s :
	\mS_s + Q_s, 
\eq{moment_T}
\EE
\EM
where
\BM
\BE
	p \doteq nT,
\EE
\BE
	\Total{_s}{t} = \Partial{}{t} + \vu_s\.\vgrad,
\EE
\BE
	\mS \doteq \Half[\vgrad\vu + (\vgrad\vu)\Tr],
\eq{S_def}
\EE
\EM
and the stress tensor~$\mPi$, the interspecies momentum
transfer~$\vR$, the  
heat-flow vector~$\vq$, and the heat generation~$Q$ are
defined by 
\begingroup
\allowdisplaybreaks
\BALams
	\mPi_s &\doteq \Int \dd\vv\,(\nbar m\vw)_s\vw_s f_s - p_s\mone,
\\
	\vR_s &\doteq -\Int \dd\vv\,(\nbar m\vw)_s \C_s\on{f},
\eq{R_def}
\\
	\vq_s &\doteq \Int \dd\vv\,\Big(\Half \nbar m w^2\Big)_s\vw_s f_s,
\eq{q_def}
\\
	Q_s &\doteq -\Int \dd\vv\,\Big(\Half \nbar m w^2\Big)_s \C_s\on{f},
\eq{Q_def}
\EALams
\endgroup
with $\vw_s \doteq \vv - \vu_s$.
Braginskii uses the notation $\vR
\equiv \vR_e$. Conservation of total momentum ensures that $\sum_i\vR_i =
-\vR$. 
For a one-component plasma, $\vR$~vanishes because the
collision operator 
conserves momentum for like-species collisions; similarly,
$Q$~vanishes for the OCP by energy conservation.  In the general case,
application of the conservation laws leads to
\BE
\sum_i Q_i = - Q_e - \sum_i \vR_i\.(\vu_e - \vu_i),
\EE
which reduces to
\BE
Q_e = -Q_i + \vR\.\vu
\EE
($\vu \doteq \vu_e - \vu_i$) for a single species of ions.

For a two-species, strongly \magnetize d electron--ion plasma (charges $q_e =
-e$ and $q_i = Ze$) possessing
overall charge neutrality and with mass ratio $\mu \ll 1$ and
$(\nu/\abso{\wc})_s  
\ll 1$, Braginskii quotes the following results, 
which constitute the hydrodynamic closure (the numerical coefficients are
valid for $Z = 1$):
\BI
\item The electron and ion collision times are
\BE
\taue = \fr{3\me\ehalf\Te^{3/2}}{4\sqrt{2\pi}\ln\Lambda\, Z^2 e^4\ni},
\quad
\taui = \fr{3\mi\ehalf\Ti^{3/2}}{4\sqrt{2\pi}\ln\Lambda\, Z^4 e^4 \ni}.
\EE

\item The gyrofrequencies are $\wcs \doteq (q B/mc)_s$.  (In my convention,
unlike Braginskii's, $\wce$~is negative.)  The gyroradii are $\r_s \doteq
\vts/\abso{\wcs}$, where $\vts \doteq (T/m)_s\ehalf$.

\item The interspecies momentum transfer is denoted by $\vR_e \equiv \vR$
and $\vR_i = -\vR$.  It consists of two parts, $\vR = \vR_\vu + \vR_T$:

\BI
\item The friction force~$\vR_\vu$ is, with $\vu \doteq \vu_e - \vu_i$, 
\BE
\vR_\vu = -(mn)_e\taue\m1(\a\vu_\parallel + \vu_\perp),
\eq{vR_vu}
\EE
where $\a \doteq 0.51$.

\item The thermal force~$\vR_T$ is
\BE
\vR_T = -\b\ne\gradpar\Te + \Case32\fr{\ne}{\wce\taue}\bhat\cross\vgrad\Te,
\eq{vR_T}
\EE
where $\b \doteq 0.71$.

\EI

\item The electron heat flux is $\vq_e = \vq_{e,\vu} + \vq_{e,T}$, where
\BALams
\vq_{e,\vu} &= \b(nT)_e\vu_\parallel -
\Case32\fr{(nT)_e}{\wce\taue}\bhat\cross\vu,
\\
\vq_{e,T} &= -\ne\(\k_{\parallel e}\vgradpar\Te + \k_{\perp e}\vgradperp\Te +
\Case52\fr{\vte^2}{\wce}\bhat\cross\vgrad\Te\),
\eq{qeT}
\EALams
where $\k_{\parallel e} = 3.16 \vte^2\taue$ and $\k_{\perp e} = 4.66
\re^2/\taue$.  (The coefficient $4.66$ is derived in \App{qe}.)

\item The ion heat flux is
\BE
\vq_i = -\ne\(\k_{\parallel i}\vgradpar\Te + \k_{\perp i}\vgradperp\Te 
+ \Case52\fr{\vti^2}{\wci}\bhat\cross\vgrad\Te\),
\EE
where $\k_{\parallel i} = 3.9 \vti^2\taui$ and $\k_{\perp i} =
2\ri^2/\taui$.

\item The ion heat generation is
\BE
Q_i = Q_\D \doteq 3\fR{\me}{\mi}\taue\m1\ne(\Te - \Ti).
\EE

\item The electron heat generation is
\BE
Q_e = -\vR\.\vu - Q_\D.
\EE

\item The stress tensor and viscosity coefficients are discussed in
\App{Viscosity}. 

\EI

\subsection{The hydrodynamic projection for multispecies and  \magnetize d 
  plasma}
\label{multispecies}

Before proceeding with a hydrodynamic projection, one must decide whether to
derive one-fluid or $S$-fluid equations, where $S$~is the number of
species.  Because the null space of the 
collision operator is five-dimensional, it would be simplest to project into
a five-dimensional, one-fluid hydrodynamic subspace.  The 
resulting single-fluid equations would be identical in form to those
derived earlier for the OCP 
except for the presence of the Lorentz force term.  Unfortunately,
such a one-fluid hydrodynamics disguises the important fact that when the
mass ratio is small the physics of the electrons and the ions are
quite different.  Thus, following Braginskii, I shall derive a set of
$S$-fluid equations.  However, this leads to technical complications
because one now has $5S$~equations although the null space remains
merely five-dimensional.  A consequence is that the interpecies collisional
coupling described by~$\vR$ and~$Q$ results in some eigenvalues of the
linearized problem
being of the order of a collision frequency~$\nu$ (fast relaxation on the
kinetic timescale) instead 
of being proportional to~$k^2$ (slow relaxation constrained by conservation
laws). 

To begin defining the relevant projection operator, let us introduce the
natural scalar product, which is the 
generalization of \Eq{ALB} to include a species summation:
\BE
\<A | \L | B> \doteq \sum_{\sbar}\Int \dd\vvbar\sum_{\sbar'}\Int
\dd\vvbar'\,A_\sbar(\vvbar)L_{\sbar\sbar'}(\vvbar,\vvbar')
B_{\sbar}(\vvbar)f_{{\rm M},\sbar'}(\vvbar').
\eq{ALBs}
\EE
This is natural for several reasons:
\BI
\item[(i)]  The linearized collision operator (times~$\nbar$) is self-adjoint
\wrt\ it:
\BE
\<\psi | \nbar \Chat | \chio> = \<\chio | \nbar \Chat | \psi>
\EE
(see \App{Chat}).  

\item[(ii)] For $\vA$~defined as in \Eq{vA_def}, the
conservation properties of~$\Chat$ take the simple form\footnote{The species sum
included in the definition \EQ{ALBs} is crucial to this result.}
\BE
\bra\nbar\vA |\, \Chat = 0.
\EE

\item[(iii)]
The first-order electric field follows as
\BE
\D\vE_\vk = \<\cEk | \chik>,
\EE
where
\BE
\cE_{\vk,s\sbar}(\vv,\vvbar) \doteq (\nbar q)_\sbar\ek
\EE
(independent of~$s$, $\vv$, and~$\vvbar$).

\EI

Although projections are often discussed in terms of scalar products, and
the scalar product \EQ{ALBs} is natural from several points of view, its
use in defining an $S$-species projection operation is somewhat problematical
because of the implied species summation.  If one insists on defining
bras and kets in terms of a scalar product, that summation must be
inhibited in one way or another in order that one end up with a
species-dependent result.  
\comment
One possibility is to define
\BE
\P_s = \ket\vA\Tr>\.\mM_s\m1\.\bra \vA 1_s|,
\eq{P_s}
\EE
where $1_s \doteq \Kron{s\sbar}$ is inserted in order to extract the fluid
variable 
of species~$s$.  That is,  the~$1_s$ inhibits the implied species sum in the
scalar product.  Thus,
\BE
\P_s\ket\D\chi>
\EE
\footnote{It is important to be clear about the meaning of
  \Eq{P_s}. The Dirac notation signifies an expectation \dots}
\endcomment
A definition that generalizes naturally to the \gspace\ theory discussed
in Part~II is the following.  Subsume the species index into a generalized
field index:  $\vA_s \equiv A^\mu_s \to A^\mu$.  (When a species index
is written explicitly, then the superscript refers to just the usual field
index.) Then define
\BE
\P \doteq \ket A^\mu>M_{\mu\mu'}\m1\bra A^{\mu'} |.
\eq{Pmm'}
\EE
Also define
\BE
\D a^\mu = \<A^\mu | \chi>.
\EE
Here one is treating~$\chi$ as the collection of all~$\chi_s$'s.  Normally,
the natural scalar product would require that the~$s$ be changed to~$\sbar$
and a summation over~$\sbar$ be done.  However, $A^{\mu}$~depends on a
specific species~$s_{\mu}$, not the entire collection $A^\mu_s$ for
all~$s$.  
A consistent interpretation is that the presence of a specific species
index inhibits the species summation in the scalar product.  Effectively,
$A^\mu_{s_\mu}$~behaves inside a scalar product as the $s$-dependent
quantity $\Kron{s_\mu (s)}A^\mu_{(s)}$, where parentheses
inhibit the summation convention.  Then 
\BE
\<A^\mu | \chi> = \sum_\sbar\Int
\dd\vvbar\,\Kron{s_\mu\sbar} A^\mu_\sbar(\vvbar)\chi_\sbar(\vvbar) = \Int
\dd\vvbar\,A^\mu_{s_\mu}(\vvbar)\chi_{s_\mu}(\vvbar) \doteq \D a^\mu.
\EE
To extract all hydrodynamic field components of specific species~$s$, I
shall use the notation
\BE
\va_s = \<\vA 1_s | \chi>,
\EE
where the~$1_s$ inhibits the species summation in the scalar product.

Because $\vA$~is defined in terms of purely kinetic quantities that do not
couple species, in fact $M^{\mu\mu'}_{ss'}$ is diagonal in the species
indices.  It is also diagonal in the field indices because of the
orthogonality built into the definition of~$\vA$ [see \Eq{M_def}].  Thus,
$M^{\mu\mu'} = M^{(\mu)}\Kron{(\mu)\mu'}$ ($M^{\mu} \doteq M^{(\mu)(\mu)}$)
and
\BE
\P = \sum_\mu\ket A^\mu>M\m1_\mu\bra A^\mu|.
\eq{Pm}
\EE
In Part~II, potential-energy contributions will be added to~$\vA$,
$\mM$~will no longer be diagonal, and (a space-dependent generalization of)
\Eq{Pmm'} will be used.

\comment
The definition \EQ{P_s} can sometimes be confusing to work with in
practice.  The basic problem is that whereas in the one-component case the
scalar product is equivalent to an average over the background Maxwellian,
with $\vv$~playing the role of a random variable, for multiple species
scalar products and expectations are distinct.  While velocity is a random
variable, the species index is not; there is no discrete probability
density~$P_s$.  
\endcomment

Define the magnetic-field operator by
\BE
\ii\Mhat \doteq c\m1\vv\cross\vB\.\vdel = \wc\Partial{}{\zeta},
\EE
where 
$\zeta$~is the negative\footnote{$\z$~is defined such that it increases
with time during ion gyration.  A diagram illustrating the coordinates of a
gyrospiraling particle, originally published as figure~1 of
\Ref{JAK_ARFM}, can be found on Sheldon's whiteboard in Episode~14,
Season~6 (January 31, 2013)
of CBS's \textsl{The Big Bang Theory}.} 
 of the polar angle in velocity space.  
It is then a
straightforward 
exercise to show that the kinetic equation linearized around an absolute
Maxwellian can be written as
\BE
\Partial{\ket\chi>}{t} + \ii\LL\ket\chi> = 0,
\eq{chi_lin}
\EE
where 
\BE
\LL \doteq \L + \LE + \LM,
\quad
\L \doteq \L_\vv + \L_\C,
\eq{LL}
\EE
with
$
\LM \doteq \Mhat
$.
\comment
Although the scalar product \EQ{ALBs} simplifies the representation of the
electric field and the conservation properties of the collision operator,
it needs to be used with case for the derivation of $S$-fluid equations.  I
shall define the $S$-species hydrodynamic projector as 

  Subsequently, I shall write $\P_s \equiv \P$.
\endcomment
The projections of the kinetic equation are formally the same as
\Eqs{Pchi_dot} and \EQ{Qchi_dot}.
Working out the components of the frequency matrix is straightforward.
There are no surprises for $\ii\mOmega_\vv$; it leads to the same
(species-dependent) Euler contributions as for the~OCP\@.  It is easy to
show that $(\ii\mOmega_\M)^\vp_\vp$ gives rise to the usual Lorentz
force,
while all other magnetic contributions to the frequency matrix vanish.\footnote{$\Mhat$~conserves number and kinetic energy, but not vector 
momentum.  However, the momentum integral of~$\Mhat$ lives solely in the
hydrodynamic subspace.  That is, upon integrating by parts,
\BE
\<A^\vp_s | \ii\Mhat_s | \chi_s> = 
\<m_s\vv | \wcs\vv\cross\bhat\.\delvv|\chi_s>
= -(m\wc)_s\<\vv\cross\bhat | \chi_s>
= -\wcs\D\vp_s\cross\bhat.
\NN
\EE
} 
It is shown in \App{Chat} that
\BE
\ii\mOmega_\C\.\D\va = m_s\sum_{s'}\nu_{ss'}(\D\vu_{s} -
\D\vu_{s'})
+ 3\sum_{s'}\fR{m_s}{M_{ss'}}\nu_{ss'}(\D T_{s} -
\D T_{s'}), 
\eq{iOmegaC}
\EE
where $\nu_{ss'}$ is a generalized interspecies collision frequency defined
by \Eq{nu_ss'}
and $M_{ss'} \doteq m_s + m_{s'}$.
At this point, we have obtained
\BALams
\Partial{}{t}\fR{\D n}{\nbar}_s &= -\div\D\vu_s,
\\
(\nbar m)_s\Partial{\D\vu_s}{t} &= (\nbar q)_s(\D\vE + c\m1\D\vu_s\cross\vB)
- \vgrad\D p_s
- (\nbar m)_s\sum_{s'}\nu_{ss'}(\D\vu_s - \D\vu_{s'})
\NN\\
&
\qquad
- \<\nbar \vP' 1_s | \P\ii\L\Q|\Q\chi>,
\eq{Du_dot_B}
\\
\Case32\nbar_s\Partial{\D T_s}{t} &= -p_s\div\D\vu_s
- 3\nbar_s\sum_{s'}\fR{m_s}{M_{ss'}}\nu_{ss'}(\D T_s - \D T_{s'})
\NN\\ 
&\qquad
- \<\nbar K' 1_s |\P\ii\L\Q|\Q\chi>.
\EALams
Obviously, the forms of the linearized fluid equations are beginning to emerge,
with the parts involving $\ket\Q\chi>$ to be determined by
closure.  Note that only $\L \doteq \L_\vv + \L_\C$ enters those parts; it is
easy to show 
that the magnetic-field operator does not couple the subspaces, so
$\P\ii\Mhat\Q = \Q\ii\Mhat\P = 0$, and a similar argument holds for~$\LE$.
Also note that the frequency-matrix term involving $\D\vu_s -
\D\vu_{s'}$ in \Eq{Du_dot_B} is not, in general, the complete contribution
to the 
interspecies momentum transfer.  As is well known\cite{Braginskii}, the
effective collision 
frequency for the parallel momentum transfer differs by a numerical factor
from the $\nu_{ss'}$ defined by \Eq{nu_ss'}; for $Z = 1$, that factor
is~$\a = 0.51$.  The physics is that perturbations to an absolute Maxwellian
background do not in general merely produce a shifted Maxwellian; the $v\m3$
dependence of the electron--ion collision rate leads to a high-energy tail
that enhances the parallel current for fixed electric field.  This
manifests as a reduction in the effective $\nu_{ss',\parallel}$.  This
effect is not seen in the frequency-matrix portion of the hydrodynamic
projection, which 
only involves the perturbations of the quantities $n$, $\vu$, and~$T$ that
would 
appear in a local Maxwellian distribution.  Thus, the
physics of the high-energy tail must be contained in the last,
$\ket\Q\chi>$ term of \Eq{Du_dot_B}, as I shall now demonstrate.

\subsection{Hydrodynamic closure}
\label{hydro_closure}

The straightforward generalization of \Eq{meta_cov} to the multispecies,
\magnetize d case is
\BE
\eta^\mu_\nu \doteq \InT \dd\t\,\Sigma^\mu_\nu(\t) \approx
\<A^\mu|\L\Q[\Q(\ii\Mhat + 
  \Chat)\Q]\m1\Q\L|A_\nu>.
\eq{meta_B}
\EE
A different way of representing the content of \Eq{meta_B} is to rewrite
the solution for the orthogonal projection,
\BE
\ket\Q\chi> \approx -[\Q(\ii\Mhat + \Chat)\Q]\m1\Q\ii\L\P\ket\P\chi>.
\EE
as the equation
\BE
(\ii\Mhat + \Q\Chat)\ket\Q\chi> = -\Q\ii\L\P\ket\P\chi>.
\eq{Qchi}
\EE
Here the result $\Q\ii\Mhat\Q = (1 - \P)\ii\Mhat\Q = \ii\Mhat\Q$ was used.
Unlike in the OCP, one cannot reduce $\Q\Chat\Q \to
\Chat$ because the present hydrodynamic projection inhibits the species
summation, so $\P\Chat \neq 0$. Following Braginskii, I now restrict the
calculation to a single species of ions.  Then one finds
\BAams
\Q\ii\L\P\ket\P\chi> &= \Half\BIG\ket\fr{\vv\,\vv -
  (v^2/3)\mone}{\vt^2}>:\mW\on{\D\vu} 
+ \BIG\ket\(\Half\fr{v^2}{\vt^2} - \Case52\)\vv>\.\vgrad\fR{\D T}{T}
\NN\\
&\quad
+ \Q\Chat\P\ket\P\chi>,
\eq{QLP_B}
\EAams
where the traceless tensor used by Braginskii is
\BE
\mW\on{\vu} \doteq \vgrad\vu + (\vgrad\vu)\Tr - \Case23(\div\vu)\mone.
\EE

To simplify the last term of \Eq{QLP_B}, use $\Q\Chat\P = \Chat\P -
\P\Chat\P$, where the last term is evaluated in \Sec{PCPchi}
and is given by 
\Eq{PCP}.  The quantity $\Chat\P\ket\chi>$ is evaluated in \Sec{CPchi} for
small mass ratio.  For the electrons, the $\D T$~part of
$\Q\Chat\P\ket\chi>$ vanishes to lowest order in the mass ratio; for the
ions, $\Q\Chat\P\ket\chi>$ vanishes altogether to lowest order.  Thus, with
the results of \App{Chat}, all pieces of \Eq{Qchi} for $\Q\ket\chi>$ are
known.  To rearrange that equation into Braginskii's form, use $\Q\Chat =
\Chat - \P\Chat$ and place the $\P\Chat$ terms on the \rhs.  For the
electrons, one finds [\cf\ Braginskii's equation~(4.12)]
\BAams
-(\ii\Mhat + \Chat_e)\ket\Q\chi>
= \Half\BIG\ket\fr{\vv\,\vv -
  (v^2/3)\mone}{\vte^2}>&:\vW\on{\D\vue} 
+ \BIG\ket\(\Half\fr{v^2}{\vte^2} - \Case52\)\vv>\.\vgrad\fR{\D T}{T}_e
\NN\\
\qquad\qquad+ \fr{1}{\vte^2\taue}\BIG\ket
\bigg[3\sqrt{\fr{\pi}{2}}\fR{\vte}{v}^3 - 
1\bigg]\vv>&\.\D\vu
- \ket\vv>\fr{1}{\vte^2}\<1_e\vv | \Chat^{\rm Lor}|\Q\chi>,
\eq{Q_chi}
\EAams
where the last term is the Lorentz approximation to $\P\Chat\Q\ket\chi>$;
only the momentum projection 
appears because the Lorentz collision operator conserves kinetic energy.
The coefficient of that term is just the lowest-order approximation to the
electron momentum transfer:
\BE
- \ket\vv>\fr{1}{\vte^2}\<1_e\vv | \Chat^{\rm Lor}|\Q\chi>
= \fr{1}{\nbar T}\ket\vv>\.\D\vR,
\eq{DR}
\EE
where [see \Eq{R_def}]
\BE
\D\vR \doteq -\<(\nbar m)_e\vv 1_e | \Chat^{\rm Lor} | \Q\chi>.
\eq{DR_def}
\EE

For the ions, one finds that the $\Q\Chat\P$ term in \Eq{QLP_B} is
negligible for small mass ratio, so [\cf\ Braginskii's equation~(4.15)]
\BE
-(\ii\Mhat + \Chat_{ii})\ket\Q\chi>
= \Half\BIG\ket\fr{\vv\,\vv -
  (v^2/3)\mone}{\vti^2}>:\vW\on{\D\vui} 
+ \BIG\ket\(\Half\fr{v^2}{\vti^2} - \Case52\)\vv>\.\vgrad\fR{\D T}{T}_i;
\EE
this is identical in form to the correction equation for the OCP
[see \Eq{Chat_chi_2}]. 

Thus, the projection-operator methodology has reproduced Braginskii's
correction equations --- as, of course, it must since physics content is
invariant to mathematical representation.  Although we have not obtained any
new results,
it is hoped that the use of
projection operators clarifies the underlying structure of the transport
equations, the key import of the null eigenspace, and the distinction
between a perturbed local Maxwellian distribution and the true perturbed
distribution that includes a high-energy, non-Maxwellian tail driven by the
various thermodynamic forces.

From this point
forward, the route to the final values of the transport coefficients, namely
the evaluation of the matrix elements \EQ{meta_B},
follows
that of Braginskii and other authors.  In general, numerical or approximate
analytical work is
required; there is no need to repeat such analysis here.  But as an
illustration of the content of the correction equations and with the goal
of providing further insight into the various orthogonal projections, I
show in \App{qe} how to work out the perpendicular electron heat
flow~$\vq_{\perp,e}$ in the limit
of small $\nue/\abso{\wce}$.

\subsection{Onsager symmetries}

Onsager's symmetry theorem\cite{Onsager31,Onsager31b,Casimir45,JAK_Onsager}
is one of the deepest results in classical
statistical physics.  It is a statement about the relaxation of an
arbitrarily-coupled  
\hbox{$N$-body} system
slightly perturbed from a Gibbsian thermal equilibrium; in the present
covariant notation, it reads
\BE
\etahat^{\mu\nu}(\vB) = \etahat^{\nu\mu}(-\vB).
\eq{Onsager}
\EE
Here $\Hat{\meta} \doteq \mE\.\meta$, where $\mE$~is the \emph{parity matrix}
such that under a time-reversal transformation $\vA \to \mE\.\vA$.  (In a
diagonal representation, $E\up{i}_{(i)} = \pm1$ depending on whether the
$i$th variable is even or odd under time reversal.  For my choice of~$\vA$,
$\mE = \Mathop{diag}[1,\,-1, 1]$ for each species.)
  Fundamentally, this symmetry
is a consequence of the time-reversibility of the microscopic dynamics.
It is critical to observe that the theorem
applies to the fully contravariant (or fully covariant) transport
tensor, not the mixed tensor~$\eta^\mu_\nu$ that appears naturally in the
hydrodynamic 
equations.  Failure to recognize this fact has led to confusion in the
literature; a thorough discussion is given by \Ref[\SECTION III]{JAK_Onsager}.

As a consistency check, I shall sketch a proof that the present representation,
involving weakly coupled 
dynamics represented by the Landau collision operator, possesses
Onsager symmetry (as already discussed by Braginskii from a more
traditional point of view).  The fully contravariant version of
\Eq{meta_B} is
\BE
\eta^{\mu\nu} = \<A^\mu|\L\Q[\Q(\ii\Mhat + \Chat)\Q]\m1\Q\L|A^\nu>.
\EE
First suppose that instead of~$A^\mu$ and~$A^\nu$ one had generic
functions~$\psi_s$ and~$\chio_s$, where as usual the~$s$ is to be summed
over in the 
standard scalar product.  (Such functions would arise in a one-fluid
hydrodynamics.)  
Then one could proceed by rearranging the scalar
product with the aid of adjoint operators as follows:
\BE
F\on{\psi,\chio;\vB} \doteq \<\psi|\L\Q[\Q(\ii\Mhat + \Chat)\Q]\m1\Q\L|\chio>
= \<\chio|\L\adj\Q\adj[\Q(\ii\Mhat + \Chat)\Q]\adj{}\m1\Q\adj\L\adj|\psi>.
\EE
Now \wrt\ the standard scalar product (which does not include complex
conjugation), the operators $\Q$, $\nbar\Chat$, $\L_\vv \doteq -\ii\vv\.\vgrad$,
and $\nbar\L_\C 
\doteq 
-\ii\nbar\Chat$ are self-adjoint,
whereas
$\Mhat$~is anti-self-adjoint.\footnote{$\L_\vv$ and~$\Mhat$ are diagonal in
  the species index, so $\nbar\L_\vv$ and~$\nbar\Mhat$ possess the same
  adjoint properties as do~$\L_\vv$ and~$\Mhat$.}
  Therefore,
\BE
F\on{\psi,\chio;\vB} = \<\chio|\L\Q[\Q(-\ii\Mhat + \Chat)\Q]\m1\Q\L|\psi> =
F\on{\chio,\psi;-\vB}. 
\eq{F_pc}
\EE
This is a restricted form of Onsager's symmetry.  However, notice that
if~$\psi$ and~$\chio$ were the hydrodynamic vector~$\vA$, then the
properties $\bra\nbar\vA|\Chat = \v0$ (conservation) and $\Chat\ket\vA> = \v0$
(null eigenvectors) remove the~$\Chat$ from~$\L$ and lead to the representation
\BE
F\on{\vA,\vA\Tr;\vB} = \grad_i\<\vA|v_i\Q[\Q(\ii\Mhat + \Chat)\Q]\m1\Q
v_j|\vA\Tr>\grad_j. 
\EE
Given the way $\vA$ was constructed (its components are orthogonal),
symmetry in velocity space implies that there is no cross coupling between
elements with opposite parity.  Thus, Onsager's symmetry \EQ{Onsager}
applies for the unhatted form of the transport matrix in this case.


In the more interesting case in which one projects onto a particular
species, the transport matrix is constructed from  
the specific~$A^\mu$ and~$A^\nu$, whose species
indices~$s_\mu$ and~$s_\nu$ are not to be summed.  Because the outer
summations are
inhibited, one cannot use the self-adjoint property of the~$\Chat$ that
appears in the $\L$~operators, nor can one remove~$\Chat$ from~$\L$ by
means of the species-summed conservation property; this implies the
existence of nontrivial cross terms in the transport matrix. I shall
illustrate for the important special case of two species.  One has
\BALams
\eta^{\vp_e\Te} &= -\<\vP_e'|\Chat_{ei}\Q(\Dhat\m1)_{ie}\Q|K_e'\vv>\.\vgrad,
\eq{eta_pe-Te}
\\
\eta^{\Te\vp_e} &= -\div\<K_e'\vv|\Q(\Dhat\m1)_{ei}\Q\Chat_{ie}|\vP_e'>,
\eq{eta_Te-pe}
\EALams
where $\Dhat \doteq[\Q(-\ii\Mhat +
\Chat)\Q]\m1$.  The inverse of the matrix
\begingroup
\def\Ahat{\Hat{\A}}
\def\Bhat{\Hat{\B}}
\BE
\mM \doteq 
\begin{pmatrix}
\Ahat & \Bhat\\
\Chat & \Dhat
\end{pmatrix},
\EE
where $\Ahat$, $\Bhat$, $\Chat$ and~$\Dhat$ are noncommuting operators,
is\footnote{When the operators commute, \Eq{noncommuting_M} correctly
  reduces to the familiar result
$$
\begin{pmatrix}
A & B\\
C & D
\end{pmatrix}\m1
=
\fr{1}{\Delta}
\begin{pmatrix}
D & -B\\
-C & A
\end{pmatrix},
$$
where $\D \doteq AD - BC$.}
\BE
\mM\m1 = 
\begin{pmatrix}
(\Ahat - \Bhat\Dhat\m1\Chat)\m1
&
-(\Dhat\Bhat\m1\Ahat - \Chat)\m1
\\
-(\Ahat\Chat\m1\Dhat - \Bhat)\m1
&
(\Dhat - \Chat\Ahat\m1\Bhat)\m1
\end{pmatrix}.
\eq{noncommuting_M}
\EE
\endgroup
Because of the complicated form of \Eq{noncommuting_M}, it is not yet
obvious that \Eqs{eta_pe-Te} and \EQ{eta_Te-pe} are equal to within a
sign.  To demonstrate that, use the momentum conservation property
\BE
\bra\nbar_e\vP_e'|\Chat_{ei} + \bra\nbar_i\vP_i'|\Chat_{ie} = 0
\EE
and the result $\eta^{\Te\vP_e} + \eta^{\Te\vP_i} = 0$, which follows from
$\Chat\ket\vP'> = 0$,
to find
\BALams
\eta^{\vp_e\Te}/T &=
\nbar_e\m1\<\nbar_i\vP'_i|\Chat_{ie}\Q(\Dhat\m1)_{ee}|\g_e\vv>\.\vgrad, 
\eq{eta_pT}
\\
\eta^{\Te\vp_e}/T &= \div\<\g_e\vv|(\Dhat\m1)_{ee}\Q\Chat_{ei}|\vP_i'>,
\eq{eta_Tp}
\EALams
where $\g(\vv) \doteq \half m v^2/T - \case52$ arises from the calculation
of $\Q K'\vv$.  Since both expressions now involve the common matrix
element\footnote{For small mass ratio, the second term of this element is
  $\OrdeR{(\me/\mi)\ehalf} \ll 1$.}
$(\Dhat\m1)_{ee} = [\Chat'_{ee} -
\Chat'_{ei}(\Chat'_{ii})\m1\Chat'_{ie}]\m1$, where $\Chat' \doteq \ii\Mhat +
\Chat$ 
[\cf\ \Eq{noncommuting_M}],
they can be easily compared.  Upon referring to the form 
\EQ{delv_chi} of the linearized Landau operator, one sees that an
integration by parts of $\bra\nbar_i\vP_i'|\Chat_{ie}$ in the expression
\EQ{eta_pT} introduces a minus sign and that the \Eqs{eta_pT} and
\EQ{eta_Tp} are otherwise equal with $\vB \to -\vB$.  Thus, we have
recovered \Eq{Onsager}.
It is interesting to contemplate that the microscopic
time-reversibility used in Onsager's original (and more general) derivation
shows up in the above proof as the constraint of macroscopic momentum
conservation.   

Braginskii remarked upon the Onsager symmetry between the electron
temperature-gradient contribution to the friction force and the flow-driven 
contribution to the electron heat flux.  (Those effects are absent for the
ions to lowest order in the mass ratio.)  He failed to mention that the
stress tensor~$\mPi$ also affords an example of the symmetry.  As discussed
for the case of the OCP, $\mPi$~can be written for infinitesimal
perturbations as a fourth-order tensor~$\tensor{m}$ applied to
$\vgrad\D\vu$. The ultimate effect in the momentum equation is $-\div\mPi =
-\div\tensor{m}:(\vgrad\D\vu)$, 
which in Fourier space can be written as $(\vk\.\tensor{m}\.\vk)\.\D\vu \to
\eta^i_j \D u^j = \eta^{ij}\D u_j$; here the lowering of the index just
involves an index-independent normalization factor.  As discussed
in \App{Viscosity}, $\tensor{m}$~is constructed from symmetrized tensor
products of the matrices $\mB \doteq
\bhat\,\bhat$, $\mdelta^\perp \doteq 
\mone - 
\bhat\,\bhat$, and $\mbeta \doteq \bhat\cross$.  The contributions
to~$\meta$ that do not involve~$\mbeta$ are easily seen to be symmetric and
invariant under a change of sign of~$\vB$.
The remaining terms (\ie, the gyroviscous stresses),
involve either $\set{\mdelta^\perp\mbeta}$ or 
$\set{\mB\,\mbeta}$, where the symmetrization is denoted by the braces. Thus,
the gyroviscous contributions to~$\meta$ involve
$\vk\.\set{\mdelta^\perp\mbeta}\.\vk$ or $\vk\.\set{\mB\,\mbeta}\.\vk$.  These
tensors are antisymmetric because of the factor of~$\mbeta$, but since the
gyroviscous terms are proportional to one power of the signed
gyrofrequency, symmetry 
is restored under the replacement $\vB \to -\vB$.  Therefore, all contributions
to~$\eta^{ij}$ obey the Onsager symmetry \EQ{Onsager}.

\section{Generalized Langevin equation for the hydrodynamics of \magnetize d
plasmas}
\label{GLE}

In the previous sections I used the \Schr\ representation, in which the
state vector $\ket\chi(t)>$ changes with time while the hydrodynamic
operators~$\bra\vA|$ are time-independent.  
   An \alternate\
representation uses the Heisenberg picture, in which averages are taken with
the initial state $\ket\chi(0)>$ while the operators become
time-dependent.  In this section I shall use the Heisenberg representation
to develop a generalized Langevin equation for the random hydrodynamic
operators $\bra\vA(t)|$.  The mean of that equation [its contraction with
  $\ket\chi(0)>$] reduces to the usual fluid equations, but the random Langevin
equation also contains
fluctuating forces, analogous to the Langevin theory for classical Brownian
motion.  It will be seen that the transport coefficients are intimately 
related to the two-time correlations of those forces.

\subsection{Heisenberg versus \Schr\ representations}

While the Heisenberg representation is familiar from quantum mechanics,
there are some technical differences in the present application that need to be
appreciated; therefore, I digress for a brief review.  In quantum
mechanics, the \Schr\ equation
\BE
\ii\hbar\,\delt\psi = \H\psi
\EE
can be written as
\BE
\delt\psi = -\ii\LL\psi,
\eq{psi_dot}
\EE
where $\LL \doteq \H/\hbar$.  For time-independent~$\H$, the solution is
given by
\BE
\psi(t) = \U(t)\psi(0),
\EE
where $\U(t) \doteq \ee^{-\ii\LL t}$.  Because $\H$~is self-adjoint \wrt\ the
usual complex-valued scalar product, $\U$~is a unitary operator:  $\U\U\adj =
1$.  

In statistical mechanics, the $N$-particle PDF~$P_N(\Gamma,t)$, where
$\Gamma$~is 
the set of all phase-space coordinates, obeys the Liouville equation
\BE
\delt P_N(\Gamma,t) = -\ii\LL P_N,
\EE
where $\LL$~is the Liouville operator.  Thus, the state evolves as
\BE
P_N(\Gamma,t) = \ee^{-\ii\LL t}P_N(\Gamma,0).
\EE
Since $\LL$~is anti-self-adjoint
\wrt\ a real-valued scalar product, time dependence can be transferred to
operators (functions of~$\Gamma$ that are to be averaged) according to
\BALams[op_transfer]
\<A(\Gamma)> &\doteq \Int \dd\Gamma\,A(\Gamma)P_N(\Gamma,t)
\\
&= \Int \dd\Gamma\,A(\Gamma)\ee^{-\ii\LL t}P_N(\Gamma,0)
\\
&= \Int \dd\Gamma\,[e^{\ii\LL t}A(\Gamma)]P_N(\Gamma,0)
\\
&= \<A(t;\Gamma)>_0,
\eq{L_anti}
\EALams
where $A(t;\Gamma) \doteq \ee^{\ii\LL t}A(\Gamma)$ and the average is now
\wrt\ the  
initial PDF\@.  Thus, if the states are evolved with~$\G(t) \doteq
\ee^{-\ii\LL t}$, the trajectories evolve with~$\G(-t)$.
This well-known
result is a consequence of the fact that the microscopic dynamics are
time-reversible.   

In the present situation governed by the linearized Landau kinetic
equation, the state $\chi(\vv,t)$ again evolves according 
to an equation of the form \EQ{psi_dot}, where $\LL\doteq \L_\vv + \L_\C +
\LE + \LM$~is given by \Eq{LL}. But $\LL$ has no 
definite symmetry.  The operators $\L_\vv \doteq \vk\.\vv$ and~$\L_\C
\doteq -\ii\Chat$ are self-adjoint 
\wrt\ the natural (real-valued) scalar product, $\LM$~is
readily shown to be 
anti-self-adjoint, and the electric-field operator $\LE\propto
\ket\vv>\bra 
1|$ has no symmetry.  Thus, the best one can do is to transfer the time
dependence from the state to the operators according to
\BE
\<\vA| \chi(t)> = \<\vA(t)| \chi(0)>,
\EE
where
\BE
\vA(t) \doteq \G\adj(t)\vA(0)
\EE
with
\BE
\G(t) \doteq \ee^{-\ii\LL t},
\quad
\G\adj(t) \doteq \ee^{-\ii\LL\adj t}.
\EE
As a consistency check, note that the magnetic-field operator is a special
case of the Liouville operator and possesses the same (anti)symmetry as is
demonstrated by \Eqs{op_transfer}.

\subsection{Derivation of the generalized Langevin equation}

To derive the generalized Langevin equation, consider the time evolution of
the hydrodynamic variables:
\BE
\delt\bra\vA(t)| = -\ii\bra\LL\adj\vA(t)|= -\ii\bra\LL\adj\G\adj\vA(0)| =
-\ii\bra\G\adj\LL\adj\vA(0)| ,
\eq{bra_A_dot}
\EE
the last result following since $\LL$~commutes with~$\G$ (the latter being
constructed from powers of~$\LL$).
As in previous manipulations, this result will be manipulated by a judicious
insertion of the identity $\P + \Q = 1$.   If that were done directly in the
last form, virtually all of the symbols in the resulting expressions would
be adorned with~daggers. 
That could be avoided by working with the adjoint of \Eq{bra_A_dot}.
Alternatively, one can write formally
\BE
-\ii\bra\G(t)\adj\LL\adj\vA(0)| = -\ii\bra\vA(0)|\LL\G(t),
\EE
anticipating that this bra will ultimately be combined with the
Heisenberg state $\ket\chi(0)>$.  Proceeding similarly to the manipulations
in the \Schr-picture projection, I rewrite this as
\BE
-\ii\bra\vA(0)|\,\LL\G = -\ii\bra\vA(0)|\,\LL(\P + \Q)\G.
\eq{P+Q}
\EE
The $\P$~part of this becomes
\BE
-\ii\<\vA(0)|\LL|\vA(0)>\.\mM\m1\.\bra\vA(0)|\G
= -\ii\mOmega\.\bra\vA(t)|.
\EE
For the $\Q$~part, Mori, Zwanzig, and others have shown that it is useful
to express the final~$\G$ in \Eq{P+Q} in terms of the modified
propagator~$\GQ$ defined by \Eq{U}.  To do so, consider the Fourier
transform of $\G(\t)$,  
\BE
\Ghat(\w) = \InT \dd\t\,\ee^{\ii\w\t}\ee^{-\ii\LL\t} = [-\ii(\w - \LL + \ii\e)]\m1
\EE
and use the identity, valid for arbitrary noncommuting operators~$\A$
and~$\B$ (assuming that $\A\m1$~is defined), 
\BE
(\A + \B)\m1 = \A\m1 - \A\m1\B(\A+\B)\m1
\eq{ABm1}
\EE
with
\BE
\A \doteq -\ii(\w - \Q\LL\Q + \ii\e),
\quad
\B \doteq \ii(\LL - \Q\LL\Q).
\eq{AB_choices}
\EE
Thus,\footnote{\label{identities}%
The inverse Fourier transform of \Eq{G_GQ} leads to
\BE
G(\t) = \GQ(\t) - \I0\t \dd\taubar\,\GQ(\taubar)\ii(\LL - \Q\LL\Q)G(\t -
\taubar).
\NN
\EE
This, or \Eq{ABm1}, is part of a family of similar identities.  For
example, one also has
\BE
(\A + \B)\m1 = \A\m1 - (\A+\B)\m1\B\A\m1.
\NN
\EE
And instead of using $\Q\LL\Q$ in the choices \EQ{AB_choices}, one
could choose $\Q\LL$ instead.  That leads to the identity
\BE
\ee^{-i\LL\t} = \ee^{-\ii\Q\LL\t} - \I0\t
\dd\taubar\,\ee^{-\ii\LL\taubar}\ii\P\LL \ee^{-\ii\Q\LL(\t - \taubar)}.
\NN
\EE
\Ref{Fox_PR} calls such identities \emph{disentanglement theorems} and
cites \Ref{Feynman_operators}.
In the uses made of the modified propagator in practice, the final~$\Q$ in
$\Q\LL\Q$ is never necessary.  [\Equation{Qchi_dot} could have been written
without the final~$\Q$ before the second ket.]  However, I prefer to work
with the symmetrical construction $\Q\LL\Q$.}
\BE
\Ghat(\w) = \GQhat(\w) - \GQhat(\w)[\ii(\LL - \Q\LL\Q)]\Ghat(\w).
\eq{G_GQ}
\EE
Now
\BE
\LL - \Q\LL\Q = \LL -[(1 - \P)\LL(1-\P)]
= \P\LL + \LL\P - \P\LL\P.
\eq{LQLQ}
\EE
One requires $\Q\G$ for use in \Eq{P+Q}.  Because $\Q\GQ = \GQ\Q$ and $\Q\P =
0$, the first and 
last terms of \Eq{LQLQ} do not contribute to \Eq{G_GQ}.  Therefore, upon
noting that $\LE$ and~$\LM$ do not contribute to~$\Q\LL$, one finds
\BE
\Q\G = \Q\GQ - \Q\GQ\ii\L\P\G.
\eq{QGQGQ}
\EE
Upon inserting the explicit form of~$\P$ into \Eq{QGQGQ}, one can rewrite
the last term of \Eq{P+Q} as
\BE
-\ii\bra\vA(0)|\L\Q\GQ(t) -
\<\vA(0)|\L\Q\GQ(t)\L|\vA\Tr(0)>\.\mM\m1*\bra\vA(t)|, 
\eq{split}
\EE
where $*$~denotes time convolution.
The first term of \Eq{split} can be written as a random force
$\bra\vf(t)|$, where 
\BE
\ket\vf\Tr(t)> \doteq -\ii\GQ\adj(t)\Q\LL\adj\ket\vA\Tr(0)> =
\Q\GQ\adj(t)\ket\dot\vA\Tr(0)>. 
\eq{vf_def}
\EE
The last term of \Eq{split} can be written as
\BE
- \<\vA(0)|\L\Q\GQ(t)\L|\vA\Tr(0)>\.\mM\m1 *\bra\vA(t)| = -\mSigma*\bra\vA(t)|,
\EE
where, upon recalling \Eq{vf_def},
\BE
\mSigma(t) \doteq \-\<\vf(t)\vf\Tr(0)>\.\mM\m1.
\EE

In summary, we have found the exact generalized Langevin
equation\footnote{More commonly, this is written without the explicit bra
  notation.  Given my definition of a bra, the content is identical;
  however, use of the bra emphasizes that the hydrodynamic operators are
  covectors, not vectors.} 
\BE
\delt\bra\vA(t)| + \ii\mOmega\.\bra\vA(t)| + \I0t
\dd\taubar\,\mSigma(\taubar)\.\bra\vA(t - \taubar)| = \bra\vf(t)|.
\eq{genL}
\EE
To demonstrate compatibility with the previous results, one may apply
$\ket\chi(0)>$ to \Eq{genL}, thus performing the statistical average. One
needs 
\BE
\<\vf(t)|\chi(0)> = -\ii\<\vA(0)|\L\GQ(t)\Q|\chi(0)>.
\EE
This generates the same contribution to the $\P$~equation from
the initial condition that one would have found by retaining the first term
of \Eq{Qchi_IC}.  When the system 
is prepared in the hydrodynamic subspace, one has $\ket\Q\chi(0)> = 0$ and
the contribution from the random force vanishes.  The resulting equation,
\BE
(\delt + \ii\mOmega + \mSigma*)\<\vA(t)|\chi(0)> = 0,
\EE
is identical to that previously derived from the \Schr\ representation [see
\Eq{Dva_dot}].

\subsection{Fluctuating hydrodynamics and transport coefficients}

While \Eq{genL} has the form of a generalized Langevin equation, it must
not be assumed that it is always justifiable to treat $\bra \vf(t)|$ as
being white noise, as is often done in simple models (for example,
see the discussion of the Brownian test particle in \Sec{Plateau}).  The
random noise involves the modified propagator~$\GQ$, which encapsulates
complicated details of the dynamics.
Generalized Langevin equations can be derived for projections into
essentially any subspace whatsoever, and the properties of~$\bra\vf|$ depend
on the dimensionality of the subspace and the choice of variables~$\vA$ that is
made.  (For some important caveats 
relating to the choice of projection operators, see \App{Caveats}.)
The issue is particularly clear when one follows \Ref{Mori} and projects the
Liouville equation.  Then \Eq{genL} merely describes an exact rearrangement
of the $N$-particle dynamics, with both (some) linear and nonlinear physics
being buried 
in~$\vf$.  Note that the precise way in which physics content is
apportioned between~$\mOmega$, $\mSigma$, and~$\vf$ depends on the choice of
the projection operator.  In particular, for arbitary~$\P$ there is a
first-order part of~$\vf$ that lives partly in the hydrodynamic subspace and
whose mean does not vanish.  However,
\Ref[p.~156]{Zwanzig} shows that provided that one chooses~$\P$ as I have
done (using the standard scalar product), that mean vanishes to first
order.  Furthermore, the specific choice of the hydrodynamic
variables~$\vA$ that I have used to build~$\P$ ensures that the
long-wavelength limit of~$\mSigma$ is well behaved for the evolution of the
conserved quantities.
Thus, the exact generalized Langevin equation \EQ{genL} is useful
for the treatment of first-order perturbations from thermal equilibrium, to
which this paper is restricted.  (In Part~II, I show how to generalize the
procedure to include second-order effects.)

Indeed, for the standard hydrodynamic projection,
several classical results for neutral fluids, as well as their extensions
to \magnetize d plasmas, readily follow from the previous results.   The
topic of hydrodynamic fluctuations and their relation to 
transport coefficients has a long history that I shall not attempt to fully
review here.  In brief:  \Ref{Landau57,Landau_fluids} argued that the transport
coefficients of 
a classical fluid are intimately related to the two-time correlation
functions of certain fluctuating forces; for example, the thermal
conductivity is related to the autocorrelation of a random heat flow.
\Ref{Kadanoff-Martin} stressed the importance of 
the double (ordered) limit $\lim_{\w\to 0}\lim_{k \to 0}$ in extracting
transport coefficients from certain response formulas.  (See the discussion
of the plateau phenomenon in \Sec{Plateau}.)  The 
Landau--Lifshitz formulas were derived more systematically from kinetic
theory by \Ref{Bixon69}, whose work was slightly generalized by
\Ref{Hinton}.  A review article that provides useful background is by
\Ref{Fox_PR}. 

\subsubsection{Transport coefficients as current--current correlations}

To tie those discussions of hydrodynamic fluctuations to the present
formalism, compare \Eq{mSigma_def} with 
formula \Eq{vf_def}, which defines the fluctuating force.  One readily sees
that  
\BE
\Sigma^\mu_\nu(\t) = \<f^\mu(\t)f_\nu(0)>.
\eq{Sigma_tau}
\EE
To obtain the Markovian transport matrix~$\eta^\mu_\nu$, \Eq{meta_B}, one
takes $k,\w \to 0$ in~$\GQ$ (the order of the limits is immaterial).  To
illustrate, first consider the 
un\magnetize d OCP\@.  Then only $\L_\vv$ contributes to
formulas \EQ{meta_B} and \EQ{vf_def}.  With $\vgrad \to \ii\vk$, one finds 
\BE
\eta^\mu_\nu \to \vk\,\vk:\<A^\mu|\vv\Q(\Q\Chat\Q)\m1\Q\vv|A_\nu>.
\EE
If one writes $f^\mu = \div\vJhat^\mu$ for generalized (subtracted)
currents~$\vJhat^\mu$, 
then one has\footnote{For the OCP, $\Q\Chat\Q = \Chat$.}
\BE
\eta^\mu_\nu = \vk\,\vk:\<\vJhat^\mu|\Chat\m1|\vJhat_\nu>.
\EE
By the symmetry of the un\magnetize d system, the expectation must be
proportional to the unit tensor,  
\BE
\<\vJhat^\mu|\Chat\m1|\vJhat_\nu> = D^\mu_\nu\mone,
\EE
so one can obtain the generalized transport coefficients~$D^\mu_\nu$ by
\BE
D^\mu_\nu = k\m2\eta^\mu_\nu.
\EE
As an example, the thermal conductivity follows as
\BE
\k = \<\Hat{J}^T|\Chat\m1|\Hat{J}_T>,
\EE
where
\BE
\Hat{J}^T \doteq \(\Half \fr{v^2}{\vt^2} - \Case52\)v_z.
\quad
\EE
Note that~$\case52nT$ is the ideal-gas value of the enthalpy.  
The role of the enthalpy subtraction and the thermodynamic interpretation
of~$\Hat{J}^T$ as a heat current is discussed 
by \Ref[p.~441]{Kadanoff-Martin}. 

\subsubsection{Fluctuating forces and collision-driven fluxes}

In the multispecies case, $\Chat$~also contributes to the~$\L$
in~$f^\mu$.  That gives rise to a fluctuating friction force~$\d\vR$
and a fluctuating thermal force~$\d\vq$.
\comment
(There is also a small 
fluctuating heat generation~$\d Q$, ignored by Braginskii, that I shall
discuss further below.)  
\endcomment
The 
autocorrelation of~$\d\vR$ leads to the nonhydrodynamic part of the
friction force.  The cross correlation of~$\d\vR$ and~$\d\vq$ leads to
the temperature-gradient-driven part of the 
momentum transfer and, by Onsager symmetry, to the flow-driven part of the
heat flow.  

The existence of all of these effects was well known to Braginskii, who
interpreted the systematic \CE\ mathematics with simple physical pictures.
Those arguments are entirely 
correct, and I have nothing to add to the physics.  However, since
Braginskii does not
explicitly mention fluctuating forces in the sense of the present
formalism, it is useful to understand the connection between the various
approaches.  As an example, consider the temperature-gradient
contribution to the electron momentum transfer.  This arises from
\BE
\eta^\vp_T
= \InT 
\dd\t\,\<f^\vp(\t)f_T(0)> = \InT \dd\t\,\<f^\vp(0)|\GQ(\t)|f_T(0)>
\EE
when $f^\vp$~is evaluated with~$\L_\C$ and $f_T$~is
evaluated with~$\L_\vv$.

The streaming contribution to the fluctuating heat flow is
\BALams
\ket f_T(0)> &= -\ii\Q\L_\vv \ket A_T(0)> 
\\
&=  -\Q\vv\.\vgrad \(\Case32
T\)\m1\BIG\ket\Half\fr{v^2}{\vt^2} - \Case32>
\\
&= -\(\Case32 T\)\m1\BIG\ket\vv\(\Half\fr{v^2}{\vt^2} - \Case52\)>\.\vgrad.
\EALams
This describes the fact that a microscopic velocity stream carries with it
the ideal-gas value of the enthalpy, which must be subtracted from the
kinetic-energy flux to give the gradient-driven heat flow.

Next, one has
\BE
\InT \dd\t\,\lim_{k\to 0}\GQ(\t)\ket f_T(0)> = [\Q(\ii\Mhat + \Chat)\Q]\m1
\ket f_T(0)>. 
\eq{f_T(tau)}
\EE
For the parallel physics,
this states that the characteristic autocorrelation time of the fluctuations
is the collision time, and it introduces the collisional mean free
path~$\lmfp$ as the characteristic characteristic length.  In Braginskii's
discussion, the 
macroscopic temperature profile is expanded in the small ratio $\lmfp /
L_\parallel$, $L_\parallel$~being a macroscopic parallel scale length.
  For perpendicular motions, the
characteristic timescale is the gyroperiod, the characteristic extent of
the interactions is the gyroradius $\r \doteq \vt/\abs{\wc}$,
and in the limit of $\nu/\abs{\wc}
\ll 1$ the net autocorrelation
time is the gyroperiod reduced by the small ratio\footnote{Gyration is
  nondissipative.  The ratio $\nu/\abs{\wc}$ is the fractional amount of
  dissipation per cycle.}
 $\nu/\abs{\wc}$, namely
$\tac = 
(\nu/\abs{\wc})\abs{\wc}\m1$.  

The microscopic velocity stream mentioned above suffers
the fluctuating friction force
\BE
\ket f_\vp(0)> = -\Q\Chat\ket A_\vp>.
\eq{f_p}
\EE
For the electrons, one has
\BE
\Chat\ket A_\vp> = \Chat_{ei}\ket T\m1\vv>.
\EE
The distinction here between the contravariant component $A^\vp = m\vv$ and
covariant component $A_\vp = A^\vp/N_\vp = \vv/T$ [see \Eqs{M_def} and
\EQ{N_def}] is important:  the contravariant 
component contains a mass, whereas the covariant one does not.  For the
latter, this means that the conventional orderings in the mass ratio may be
used, so $\Chat_{ei} \approx \Chat^{\rm Lor}$ and
\BE
\Chat_{ei}\ket T\m1\vv> \approx 2T\m1\nu(v)\ket\vv>.
\EE
Sans the temperature factor, this can be interpreted as the
velocity-dependent friction on a microscopic 
velocity stream, which is one of the principal ingredients in Braginskii's
heuristic pictures (\cf\ Braginskii's discussion of his figure~1).  Note that
upon applying the~$\Q$ that is required in 
\Eq{f_p}, one obtains a ket that is orthogonal to~$\bra\vv|$:
\BE
\fr{1}{\vte^2\taue}\BIG\ket \bigg[3\sqrt{\fr{\pi}{2}}\fR{\vte}{v}^3 -
1\bigg]\vv>.
\eq{f_p(0)}
\EE
This is, in fact,
exactly the ket that multiplies~$m\D\vu$ in the second line of \Eq{Q_chi};
it describes the nonhydrodynamic part of the flow-driven tail on the
perturbed distribution function.

The net frictional effect on the microscopic heat flow is given by the
cross correlation between the 
fluctuating friction force \EQ{f_p(0)} and the fluctuating heat flow
\EQ{f_T(tau)}.  It 
is easy to see that that correlation gives rise to the same matrix element
calculated by Braginskii for the off-diagonal contribution to the heat flow.

\section{Discussion}
\label{Discussion}

The purpose of this paper has been to describe the application of
projection-operator methods to classical plasma transport for the special
case of linear response and the Braginskii (or \NS) transport coefficients.

In general, there are two routes to the derivation of irreversible
transport coefficients:  (i)~first derive an irreversible kinetic equation from
the reversible Liouville equation, then project into the hydrodynamic
subspace; (ii)~project the Liouville equation, then perform a \gspace\
ensemble average in order to obtain the irreversible decay of correlation
functions whose time integrals are the transport coefficients.
In the present paper, route~(i) was followed:  projection was done on the
(linearized) irreversible 
kinetic equation.  [In Part~II, I shall instead follow route~(ii).]
It is useful to compare method~(i) with the traditional
\CE\ approach, which is reviewed in
\App{CE} for the special case of the one-component plasma.
Obviously, both that method and the projection-operator approach
capture the same physics and make compatible predictions when their regimes
of validity overlap.  The traditional approach allows for a background
zeroth-order flow, so it contains nonlinear advective derivatives.  Those
are 
absent in the 
linear-response formalism (developed \via\ either projection operators or
in any other way)
when perturbations are made around an absolute Maxwellian distribution.
However, both methods predict the same
hydrodynamic fluxes to first order in the gradients.  

At the level of linear response, the principal difference in the formalisms is
the way in which the solvability constraints are satisfied.  In the
projection-operator method, the frequency operator $\P\LL\P$ leads to the
Euler part of 
the hydrodynamic equations, and
use of the orthogonal projector~$\Q$ in the correction terms replaces
the traditional \CE\ substitution of the partial time derivatives in the Euler
equations by spatial gradients [see \Eq{combos}].  The projection-operator
method provides an 
optimally concise representation of that algebra, which leads to the
subtracted fluxes.

It must be emphasized, however, that the methods are equivalent only when
the proper hydrodynamic projector is used.  For classical transport, the
natural projection operator is clear; it is built from the null
eigenvectors of the linearized collision operator.  However, one can
project into any 
subspace whatsoever.  Since the physics is invariant to the mathematical
representation, the same results must ensue in the long-wavelength,
low-frequency limit regardless of the choice of~$\P$.  However, one must be
extremely cautious because if the projection is chosen inaptly the
Markovian approximation will not 
be satisfied.  This issue is explored in \Sec{nonlocality}--\Sec{Plateau}.
However, a simple example 
given in \Sec{Brownian} shows that provided that one projects at least into
all of the null subspace of the collision operator, a higher-dimensional
Markovian projection can also be used if one desires information additional
to that contained in the natural transport equations.

In conclusion, the projection-operator approach to the derivation of
linearized fluid 
equations is intuitive and technically efficient.  It embeds the
classical plasma derivations of transport equations into more general
and modern formulations 
of statistical dynamics.  
Although projection is a linear operation, the methodology is
useful even for nonlinear response, as \Ref{Brey} have shown.  That topic
is addressed in Part~II, where it is shown how to obtain nonlinear fluid
equations and the next-order Burnett corrections to the classical transport
coefficients.  
It should also be clear that the formalism is not
restricted to classical transport; one can contemplate applications to
neoclassical theory and to \GK s\cite{JAK_ARFM}, for example.
Projection-operator  
methods should be in the toolbox of every serious plasma theorist.

\begin{acknowledgments}
This paper is dedicated to Prof.\ Allan Kaufman, one of the pioneers of
classical plasma transport theory, whose concise and beautiful technical
approaches to the calculation of various linear and 
nonlinear plasma processes provided some of the inspirations for this work.
I am grateful to G.~Hammett for useful suggestions on the manuscript.
This work was supported by the U. S. Department of Energy Contract
DE-AC02-09CH11466.
\end{acknowledgments}

\appendix

\section{The traditional \CE\ calculation for the one-component plasma}
\label{CE}

\bgroup
\let\chi\chio
\def\t#1{t_{#1}}
\def\x#1{\vx_{#1}}
\def\f#1{f_{#1}}

It is instructive to compare the traditional \CE\ approach to hydrodynamic
equations and transport coefficients with the projection-operator formalism
described in the main text.  In the present appendix, I present my own
version of the traditional calculation for the simplest case of the
un\magnetize d one-component plasma (OCP).  The procedure was described by
\Ref{Robinson-Bernstein}.
In essence, my discussion is little more than a transcription of
their outline to the notation of the present paper, but I have also
attempted to 
include some additional pedagogical content. 

I assume that the plasma consists
of discrete ions with a smooth neutralizing electron background.  Then the
relevant collision operator is the ion--ion Landau operator $\C_{ii}$; I
shall subsequently drop the subscripts.  That velocity-space operator
conserves the densities of number, 
momentum, and kinetic energy without the necessity for summation over species.
The governing kinetic equation can be written as
\BE
\TotaL{f}{t} = -\C\on{f},
\eq{Df/Dt}
\EE
where ${\rm D}/{\rm D}t$ denotes the Vlasov operator defined in \Eq{f_dot}
and the 
brackets denote functional dependence.

I shall use the method of multiple time and space
 scales\cite[and references therein]{multiple_scales} and consider time 
variations slow \wrt\ the collision time and spatial variations much longer
than the collision mean free
path~$\lmfp$, which is taken to be much
smaller than the box size or characteristic gradient scale length~$L$.  I
thus use an ordering 
parameter~$\e \doteq \lmfp/L\ll 1$ ($\e$~is called the \emph{Knudsen number}~Kn)
and assume that
\BE
\nu\m1\delt = \Order{\e},
\quad
\lmfp\vgrad = \Order{\e}.
\eq{e-ordering}
\EE
I also assume that the electric field is small enough that the entire \lhs\
of \Eq{Df/Dt} is small.  The method then proceeds by asymptotically expanding
\Eq{Df/Dt} order by order in~$\e$, using the multiple-scale definitions
$t_n \doteq \e^n t$ and 
$\vx_n \doteq \e^n \vx$.
Three physically distinct time and space scales are relevant:
\begin{quote}
\begin{enumerate}
\parskip=\baselineskip

\item \textbf{\boldmath $\Order{\e^0}$:  kinetic scales $\t0$ and~$x_0$} ---
  irreversible phenomena related to  90\degrees\ collisions: $\w/\nu =
  \Order{1}$, $k\lmfp = 
  \Order{1}$.  On the kinetic timescale, the distribution function relaxes
  to a local Maxwellian distribution.  I shall assume that that process has
  gone to completion [\ie, that $\del_{t_0}$ and~$\vgrad_0$ vanish or,
  equivalently, that \Eq{e-ordering} holds]. 

\item \textbf{\boldmath $\Order{\e^1}$:  transit scales $\t1$ and~$x_1$}
  --- reversible phenomena related to 
  particles free-streaming across the box. The transit timescale is $t_{\rm 
  transit} = L/\vt = (L/\lmfp)(\lmfp/\vt) = \e\m1\nu\m1$, one order longer than
  the kinetic timescale.\footnote{With the definition $\t1 \doteq \e t$, it
takes a time of $\Order{\e\m1}$ to achieve an order-unity change in~$\t1$;
thus, the 
transit time scale is one order longer than the kinetic timescale.}  

\item \textbf{\boldmath $\Order{\e^2}$:  transport scales $\t2$ and~$x_2$}
  --- irreversible phenonema 
  related to classical diffusion and dissipation.  If $\mu$~is a spatial
  transport coefficient with the classical random-walk scaling $\mu \sim
  \lmfp^2\nu$, then the characteristic diffusion rate $\mu\grad^2 \sim
  \mu/L^2$ satisfies $(\mu/L^2)/\nu = (\lmfp/L)^2 = \e^2$.

\end{enumerate}

\end{quote}

\vspace{\baselineskip}

\noindent
In the
multiple-scale formalism, it is assumed that time and space variations are
extended to independent variations on the multiple scales:  $f(\vx,t) =
f(\x0,\x1,\x2,\dots,\t0,\t1,\t2,\dots)$.  Thus,
\BE
\Partial{}{t} = \Partial{}{\t0} + \e\Partial{}{\t1} + \e^2\Partial{}{\t2} +
\cdots,
\EE
and similarly for $\del/\del\vx$.  The distribution function is also
expanded according to
$
f = \sum_{n=0}^\infty \f{n}\e^n
$.
Because it is assumed that relaxation on the kinetic scale has already gone to
completion, one drops ${\rm D}/{\rm D}\t0$.  Then, through
$\Order{\e^2}$, one obtains
\BALams
0 &= -\C\on{\f0},
\eq{0}
\\
\TotaL{\f0}{\t1} &= -\Chat\f1,
\eq{1}
\\
\TotaL{\f0}{\t2} + \TotaL{\f1}{\t1} &= -(\Chat\f2 + \C\on{\f1,\f1}),
\eq{2}
\EALams
where $\Chat$~is the linearized collision operator (see \App{Chat}) and the
notation $\C\on{\f1,\f1}$ reflects the fact that the nonlinear Landau
operator is actually a bilinear 
form.\footnote{More generally, the \BL\ operator should be used; that is a
  more complicated nonlinear functional.  The \BL\ operator is discussed in
  \App[II]{BL}.}

\subsection{Kinetic timescale}

The unique solution to \Eq{0} is the local Maxwellian distribution
\BE
\f0(\vx,\vv,t) = \flM(\vv \mid \x1,\t1,\x2,\t2,\dots), 
\EE
where
\BE
\flM(\vv\mid\vx,t) = \fR{n'(\vx,t)}{\nbar}\[[2\pi\,\vt'{}^2(\vx,t)]\m{3/2}
\exp\(-\fr{[\vv - \vu'(\vx,t)]^2}{2\vt'{}^2(\vx,t)}\)\]
\eq{fLM}
\EE
with $\vt'{}^2(\vx,t) \doteq T'(\vx,t)/m$.  Here the primed parameters specify
the portions of the density, flow velocity, and temperature that are
carried by the local 
Maxwellian.  Below I shall argue that they can be identified with the same
quantities that are carried 
by the full distribution~$f$, so I shall subsequently drop the primes.
It is physically most instructive to isolate the density factor from the
local Maxwellian.  Thus, define~$\Fo$ as the factor in large brackets in
\Eq{fLM}, so
\BE
\flM(\vv \mid \vx,t) = \fR{n(\vx,t)}{\nbar}\Fo(\vv \mid \vx,t).
\EE
Note that for the local Maxwellian the viscous stress~$\mPi$ and the heat 
flow~$\vq$ vanish.

\subsection{Transit time scale}

The viscous stress and heat flow are determined by the first-order
correction to the local Maxwellian.  Upon writing 
\BE
\f1 = \fR{n}{\nbar}\chi \Fo
\eq{f1}
\EE
and
using~$\Fo$ as the weight function in the natural scalar product defined
by \Eq{ALB}, one can write the first-order correction equation \EQ{1} as 
\BE
\fR{n}{\nbar}
\Chat\ket\chi> = -\TotaL{}{\t1}\(\fr{n}{\nbar}\ket 1>\).
\eq{src_a}
\EE
\comment
&= -\BIG\ket\TotaL{(n/\nbar)}{\t1}> - \fR{n}{\nbar}\BIG\ket\TotaL{\ln
  \Fo}{\t1}>. 
\eq{Chat_chi}
\EALams
\endcomment
If $\Chat$~were positive-definite, a unique solution to this equation would be
guaranteed.  In fact, however, $\Chat$~is only positive-semidefinite since it
has a five-dimensional null space associated with the conservation laws.
Thus, either the solution does not exist or, if a certain solvability
condition is satisfied, the solution exists but is not unique; this is the
\emph{Fredholm alternative}.
Solvability is ensured by asserting that the \rhs\ of \Eq{src_a} is
orthogonal to the 
left null eigenvectors $\bra A^\mu|$, where the~$A^\mu$ are defined in
\Eq{vA_def}.  Because $\Fo$~is a Gaussian function of $\vw \doteq \vv
- \vu$, it is technically more convenient to use~$\dA^\mu$ instead
of~$A^\mu$, where $\dA^\mu(\vv) \doteq A^\mu(\vw)$.  Thus, the
solvability constraints are 
\BE
n\<\dA^\mu|\Chat|\chi> = 0 = -\BIG\<\dA^\mu | \TotaL{}{\t1}|
n>,
\EE
where the ${\rm D}/{\rm D}\t1$ acts on both the explicit~$n$ as well as the
space and 
velocity dependence of~$\Fo$ (which is hidden in the ket notation).

Those constraints determine the first-order evolution of the hydrodynamic
variables~$a'{}^\mu$.  It is easy to see that the required averages are
nothing but the hydrodynamic moments of the kinetic equation evaluated with
first-order derivatives and with $\mPi$ and~$\vq$ set to~$\v0$.
  These
nondissipative constraints are called the \emph{Euler equations}.  Thus,
for example, the first of the five Euler equations is the continuity
equation
\BE
\Partial{n}{\t1} + \vgrad_1\.(n\vu) = 0.
\EE

When the first-order Euler equations are satisfied, a solution to
\Eq{src_a} is guaranteed.  That solution is not unique, however, because an
arbitrary linear superposition of the null eigenvectors can be added.  Thus,
\BE
\ket\chi> = \sum_{\mu=1}^5 \a^\mu \ket A_\mu> + \ket\chi_\perp>,
\EE
where $\<A^\mu|\chi_\perp> = 0$.  To the extent that the~$\a$'s are
nonzero, they specify the amounts of the hydrodynamic variables carried by
the first-order distribution.  However,
there are no further constraints on
the~$\a$'s at this order, nor will any emerge at higher order.  Thus, one is
free to choose the~$\a^\mu$ to vanish, and this freedom will persist
through all orders.  This means that one can arrange things such that all
of the hydrodynamic variables are 
carried by the local Maxwellian; in other words, one may set $n' = n$,
$\vu' = \vu$, and $T' = 
T$.  While this choice is not necessary, it is by far the most convenient.

With the constraints satisfied, one must now solve \Eq{src_a}.
The \rhs\ of that equation can be simplified by carrying out the required
partial time, space, and velocity derivatives, then using the first-order
Euler equations to replace the terms in $\del/\del_{\t1}$ by terms
in~$\vgrad_1$. The algebra is straightforward.  However, it is instructive
to sketch it because there is an important lesson to be learned about the
relation 
of this approach to the projection-operator method.  \Equation{src_a} can be
written as
\BE
\Chat\ket\chi> = -\BIG\ket\TotaL{\ln n}{\t1}> - \BIG\ket \TotaL{\ln\Fo}{\t1}>.
\eq{Chat_chi}
\EE
One has
\BE
\TotaL{\ln n}{\t1} = (\del_{\t1} + \vv\.\vgrad_1)\ln n.
\EE
Also, since
\BE
\ln\Fo = -\fr{w^2}{2\vt^2} - \Case32\ln T + \const,
\EE
one has (temporarily dropping the 1 subscripts)
\BE
\Partial{\ln\Fo}{t} = \fr{1}{\vt^2}\vw\.\Partial{\vu}{t} +
\(\Half\fr{w^2}{\vt^2} - \Case32\)\Partial{\ln T}{t}
\EE
and similarly for $\vv\.\vgrad\ln\Fo$.  Upon using the Euler equations to
replace the partial time derivatives, one finds
\BALams
\delt\ln n &= -\div\vu + \vw\.\vgrad\ln n,
\\
\delt\ln\Fo &= \vt\m2\vw\.[-\vu\.\vgrad\vu + (q/m)\vE -
  (nm)\m1(T\vgrad n + n\vgrad T)]
\NN\\
&\quad + \(\Half\fr{w^2}{\vt^2} - \Case32\)\Case23(\vu\.\vgrad\ln T -
\div\vu).
\EALams
In \Eq{Chat_chi}, some terms cancel and others combine as follows.  One finds
\begingroup
\allowdisplaybreaks
\BALams
\Chat\ket\chi> &= -(\underbrace{-1}_{\delt\ln n} + \underbrace{1}_{\delt\ln
  T})\div\vu\ket 1>
- (\underbrace{1}_{\L_\vv\ln n} -
\underbrace{1}_{\delt\vu})\ket\vw>\.\vgrad\ln n
\NN\\
&\qquad- (
\underbrace{-1}_{\L_\vE\ln\Fo} + \underbrace{1}_{\delt\vu}
)\fr{q}{T}\ket\vw>\.\vE
\NN\\
&\qquad- \fr{1}{\vt^2}\ket\underbrace{\vw\,\vw}_{\L_\vv\vu} -
\underbrace{\Third w^2\mone}_{\delt\ln T}>:\vgrad\vu
- \bigg|\bigg(\underbrace{\Half\fr{w^2}{\vt^2} -
  \Case32}_{\L_\vv\ln\Fo} -
\underbrace{1}_{\delt\vu}\bigg)\vw\bigg\rangle\.\vgrad\ln T 
\eq{combos}
\\
&=
- \fr{1}{\vt^2}\BIG\ket \vw\,\vw -
\Third w^2\mone>:\mS_1
-\fr{1}{T}\BIG\ket\(\Half\fr{w^2}{\vt^2} -
\Case52\)\vw>\.\vgrad_1 T,
\eq{Chat_chi_2}
\EALams
\endgroup
where $\mS$~is the rate-of-strain tensor,
\BE
\mS \doteq \Half[(\vgrad\vu) + (\vgrad\vu)\Tr],
\EE
and the underbraces indicate the origins of the various terms.  Note that
in each of 
the pairwise combinations in \Eq{combos} the second term stems from a
partial time derivative (\ie, from the enforcement of an Euler equation).
This is the present algebra's way of ensuring that the \rhs\ of \Eq{src_a}
is orthogonal 
to the null eigenspace.  In the projection-operator method, the same result
is obtained by working with the orthogonal projector~$\Q$.  In that
approach, there is no 
explicit elimination of partial time derivatives; that is effectively done
by the $\P$~term in $\Q = 1 - \P$ in the constructions $\Q\ii\L\vA$.  The
physical reason for this subtraction is given in the paragraph following
\Eq{m_tensor}. 

Because $\Chat$~is linear and the \rhs\ is linear in the
gradients, the solution to \Eq{Chat_chi_2} can be determined by linear
superposition to be $\chi = \chi_\vu + \chi_T$, where
\BE
\chi_\vu = \mA(\vw):\mS_1,
\quad
\chi_T = \vB(\vw)\.\vgrad_1 T,
\EE
where
\BE
\ket\mA> = -\Chat\m1\fr{1}{\vt^2}\BIG\ket\vw\,\vw - \Third w^2\mone>,
\quad
\ket\vB> = -\Chat\m1\fr{1}{T}\BIG\ket\(\Half\fr{w^2}{\vt^2} - \Case52\)\vw>.
\eq{ketAB}
\EE
To evaluate these expressions, 
numerical work or approximate analytical methods [such as variational
techniques\cite{Robinson-Bernstein} or truncations of
expansions in orthogonal polynomials\cite{Braginskii}] are required. 

\subsection{The hydrodynamic regime}

One can now proceed to $\Order{\e^2}$, where the effects of
dissipation become apparent.  Upon rearranging \Eq{2}, one must solve
\BE
\fR{n}{\nbar}\Chat\ket\chi_2> =  - \ket\Fo\m1\C\on{\f1,\f1})> -
\TotaL{}{\t2}\BIG\ket\fR{n}{\nbar}1> -
\TotaL{}{\t1}\BIG\ket\fR{n}{\nbar}\chi_1>.  
\EE
The solvability conditions are
\BAams
\BIG\<\nbar\,\dA^\mu |\Chat | \chi_2> = 0 = -\fR{\nbar}{n}
&\bigg(\<\nbar\,\dA^\mu | 
\Fo\m1\C\on{\f1,\f1}> 
+ \BIG \<\dA^\mu | \TotaL{}{\t2}|n>
\NN\\
&\quad + \BIG\<\dA^\mu |
\TotaL{}{\t1}|n\chi_1>\bigg).
\EAams
The first term on the \rhs\ vanishes because of the conservation properties
of~$\C$.\footnote{The $\Fo\m1$ cancels against the hidden~$\Fo$ in the ket.}
The second term involves the same algebra that was 
done at first order and produces (\sans\ minus sign) the Euler moments in
the~$\x2$ and~$\t2$ variables.  Finally, consider
\BAams
\BIG\<\dA^\mu | \TotaL{}{\t1}|n\chi_1>
= &\BIG\< \dA^\mu | \del_{\t1} | n\chi_1>
+ \BIG\< \dA^\mu | \vv\.\vgrad_1 | n\chi_1>
\NN\\
&\qquad
+ \BIG\< \dA^\mu | \vE\.\vdel | n\chi_1>.
\EAams
The first term on the \rhs\ vanishes because $\ket\chi_1>$ has been
constructed to be orthogonal to the null eigenfunctions.\footnote{One has
$\<\dA^\mu | \del_{\t1} | n\chi_1> = \del_{\t1}\<\dA^\mu | n\chi_1> -
\<(\del_{\t1}\dA^\mu) | n\chi_1>$.  The first average vanishes directly by
orthogonality.  One has $\del_{\t1}\dA^\mu = \vlp0,\; {-}m\del_{\t1}\vu,\;
  {-}m\vw\.\del_{\t1}\vu - (3/2)\del_{\t1}T\vrp\Tr$. The velocity dependence
  of the last result involves only~1 and~$\vw$, both of which are
  elements of the null 
  space. Therefore, the second average vanishes as well.}
  The last term
also vanishes by orthogonality upon integration by parts.  The middle term
can be written as $\vgrad_1\.\vGamma_1^\mu$, where 
\BE
\vGamma_1^\mu \doteq n\<\dA^\mu\vv | \chi_1> = 
n\<\dA^\mu\vw | \chi_1> =
(
0\;\;
\mPi\;\;
\vq
)_1\Tr.
\EE

Finally, add $\e$~times the first-order solvability constraints (the
first-order Euler equations) and $\e^2$~times the second-order solvability
constraints, and use $\e\,\del_{\t1} + \e^2\del_{\t2} \approx \delt$ (and
similarly for~$\del_\vx$).  One thus reproduces the moment equations correct
through second order and with explicit expressions for the fluxes:
\BE
\mPi = -nm\tensor{m}:\mS,
\quad
\vq = -n\boldsymbol{\kappa}\.\vgrad T,
\EE
where
\BALams
\tensor{m} &\doteq \fr{1}{\vt^2}\BIG\<\vw\,\vw - \Third w^2\mone |
\Chat\m1 | 
\vw\,\vw - \Third w^2\mone>, 
\\
\boldsymbol{\kappa} &\doteq \fr{1}{\vt^2}\BIG \<\(\Half\fr{w^2}{\vt^2}
-\Case52\)\vw | 
\Chat\m1 | 
\(\Half\fr{w^2}{\vt^2} -\Case52\)\vw>. 
\EALams

These tensors can be simplified by using symmetry considerations.
Now $\tensor{m}$ is symmetric and traceless in both the first and the last
pair of its indices, and it depends on no preferred direction.  The most
general form of such a tensor is
\BE
m_{ijkl} = a\,\Kron{ij}\Kron{kl} + \mu(\Kron{il}\Kron{jk} +
\Kron{ik}\Kron{jl}),
\eq{m_OCP}
\EE
where $a$ and~$\mu$ are constants.
Asserting the traceless condition leads to $a = -2\mu/3$.  An
expression for the
scalar~$\mu$ (kinematic viscosity) follows, for example, from $m_{1221}$:
\BE
\mu = \vto\m2\<w_x w_y|\Chat\m1|w_x w_y>.
\EE
Note that $\mu \sim \vt^2/\nu$, which
is the correct random-walk scaling for an un\magnetize d transport
coefficient.  Contraction of $m_{ijkl}$ with~$\mS$ leads to
\BE
\mPi \doteq 
-nm\mu\mW,
\eq{Pi_def2}
\EE
where
\BE
\mW \doteq (\vgrad\vu) + (\vgrad\vu)\Tr - \Case23(\div\vu)\mone.
\EE
Similar considerations lead to
$
\boldsymbol{\kappa} = \k\mone
$,
where the thermal conductivity is
\BE
\k = \fr{1}{\vto^2}\BIG\<\(\Half w^2 - \Case52\)w_z | \Chat\m1 |\(\Half w^2 -
\Case52\)w_z>.
\EE

This completes the review of the traditional \CE\ theory of the OCP.

\egroup

\section{The linearized Landau operator}
\label{Chat}

The Landau collision operator\footnote{A clear introduction to the Landau
  operator is given by \Ref[Chap.~3]{Helander-Sigmar}.  
A pedagogical compendium of useful
 properties of that operator is by \Ref{Hazeltine_Coulomb}.}
 is
\BAams
	\CLandau,
\eq{C_Landau}
\EAams
where
\BE
\mU(\vv) \doteq v\m1(\mone - \vhat\,\vhat)
\EE
and
\BE
\Sbar_{s\sbar} \doteq (\nbar q^2)_s
(\nbar q^2)_\sbar\ln\Lambda_{s\sbar}
\EE
(obviously symmetric in~$s$ and~$\sbar$).\footnote{The overline on~$\Sbar$
  denotes evaluation with the mean density~$\nbar$.  An~$\S$ \sans\
  overline denotes evaluation with the full density~$n$.}
The proper definition of the Coulomb logarithm $\ln\Lambda$ is discussed by
\Ref{JAK_Lambda}, who cites original references.
  Useful properties of~$\mU$,
which is proportional to a projector into the direction perpendicular to
its argument, are 
\BE
\mU(\vv) = \Partial{^2v}{\vv\,\del\vv},
\quad
\Partial{}{\vv}\.\mU = -\fr{2\vv}{v^3}.
\eq{U_props}
\EE
$\C^{\rm L}$~is a bilinear operator
on~$f$, so it can be written as $\C\on{f,f}$ (henceforth dropping
the L~superscript for brevity). If one writes $f =
(1+\chi)\fM$, 
then the operator linearized around a Maxwellian involves
\BAams
&\(\fr{1}{m}\Partial{}{\vv} - \fr{1}{\mbar}\Partial{}{\vvbar}\)
[(\chi + \chibar)\fM\fMbar]
\NN\\
&\qquad = \(\fr{1}{m}\Partial{\chi}{\vv} -
\fr{\vv}{T}(\chi + \chibar)\)\fM\fMbar
 - [(\vv,s) \Leftrightarrow (\vvbar,\sbar)].
\EAams
The antisymmetrization introduces the relative velocity $\vv - \vvbar$,
which is annihilated by~$\mU(\vv - \vvbar)$.  Thus, the linearized Landau
operator is
\BAams
\Chat\ket\chi> = -2\pi\,(\nbar m)_s\m1\Partial{}{\vv}&\.\sum_\sbar
\Sbar_{s\sbar}\Int \dd\vvbar\,\mU(\vv - \vvbar)
\NN\\&
\.\(\BIG\ket\fr{1}{m_s}\Partial{\chi_s}{\vv}>\fMbar
 - 
[(\vv,s) \Leftrightarrow (\vvbar,\sbar)]\).
\eq{delv_chi}
\EAams

In practice, mass-ratio expansions of~$\Chat$ are often useful. For
electron--ion collisions, the integration velocity~$\vvbar$ is limited by
the ion Maxwellian to be $\Order{\vti}$; thus, for typical electron
velocities one has $\mU(\vv - \vvbar) \approx \mU(\vv)$ and 
\BE
\Chat_{ei}\ket\chi> \approx   -2\pi\,(\nbar
m)_e\m1\Sbar_{ei}
\Partial{}{\vv}\.\mU(\vv_e)\.\(\fr{1}{\me}\BIG\ket\Partial{\chi_e}{\vv_e}>
\hbox to 0pt{\phantom{\hss$\displaystyle\fr{1}{\mi}$\hss}}_e
 - {\BIG\ket 1\phantom{\hbox
   to0pt{\hss$\displaystyle\Partial{\chi_i}{\vv_i}$\hss}}>}_e
   \fr{1}{\mi}\BIG\<1   
   |\Partial{\chi_i}{\vv_i}>_i\). 
\eq{Chat_ei}
\EE
Terms of $\Order{\me/\mi}$ have been neglected in the electron term.  The
explicit $\Order{\mi\m1}$ ion term is not necessarily negligible because one
does not yet know the size of the~$\chi_i$ on which the operator will act.
(Indeed, in some later projection operations one will need to insert~$\chi$'s
that are 
explicitly proportional to mass, so the mass dependence will cancel out
in those cases.) 
When the ion term is in fact negligible, 
one obtains the Lorentz operator as usually defined:
\BALams
\Chat_{ei} \approx \Chat^{\rm Lor} \doteq  \nu\fR{\vte^3}{v^3}
\Lvec^2,
\EALams
where $\Lvec^2$~is the square of the angular momentum
operator.\footnote{Explicitly, in a spherical-polar $(v,\theta,\phi)$
coordinate system one has $\Lvec^2 
  = -[(\sin\th)\m1\del_\th\sin\th\,\del_\th +
    (\sin^2\th)\m1\del_\phi^2]$.  See, for example,
  \Ref[p.~79]{Gottfried}.} 
 The eigenfunctions of $\Lvec^2$ are the spherical harmonics:
\BE
\Lvec^2 Y_l^m(\theta,\phi) = l(l+1)Y_l^m,
\EE
with
$Y_l^m(\theta,\phi) \doteq P_l^m(\cos\theta)e^{\ii m\phi}$,
$P_l^m(x)$~being the 
associated Legendre functions of the first kind.  The collision
frequency~$\nu$ is related to Braginskii's collision rate $\taue\m1$,
defined in \Eq{taue} below, by
$
\nu = (3\sqrt{2\pi}/4)\taue\m1
$.

From the definition \EQ{R_def} and with the aid of integration by parts,
\Eq{Chat_ei} 
generates the electron momentum transfer
$
\vR \approx \sum_i\vR_{ei}
$,
where
\BE
\vR_{ei} \doteq -2\pi\,\Sbar_{ei}
\(\BIG\<\mU\.\fr{1}{\me}|\Partial{\chi_e}{\vv_e}>_{{\rm M}e} - \<\mU>_{{\rm
M}e}\.\fr{1}{\mi} 
\BIG\<\Partial{\chi_i}{\vv_i}>_{{\rm M}i}\).
\eq{R_approx0}
\EE
A standard reference calculation assumes (illegitimately) that the
distribution function is a local (shifted) Maxwellian: $\flM \doteq
(n/\nbar)(2\pi\,\vt^2)\m{3/2}\exp[-\abs{\vv - \vu}^2/2\vt^2] \approx \fM(1 +
\vv\.\vu/\vt^2)$ for $\abs{u}/\vt \ll 1$.  (This is incorrect because of
the formation of 
high-energy tails on~$f$, as discussed and calculated later.)  If $\chi_s =
\vv\.\vu_s/\vts^2$ is inserted into \Eq{R_approx0}, the explicit mass
dependences cancel (a possibility that was noted above) and one finds with
the aid of 
\BE
\<\mU>_{{\rm M}e} = \fr{8\pi}{3}(2\pi)\m{3/2}\vte\m1\mone
\eq{vUe}
\EE
that
\BE
\vR_{ei} \approx -(mn)_e\tau_{ei}\m1(\vue - \vui),
\eq{R0}
\EE
where
\BE
\fr{1}{\tau_{ei}} \doteq \Case43\sqrt{2\pi}\,\fr{(q^2)_e(n
  q^2)_i\ln\Lambda_{ee}}{\me T\vte}
\eq{taue}
\EE
[cf.~the definition \Eq{nu_ss'} below of the generalized collision
rate~$\nu_{ss'}$, which holds for arbitrary mass ratio].  For the case
of a single species of ions, 
Braginskii writes $\tau_{ei} \equiv \taue$.

For ion--electron collisions, one has
\BALams
\Chat_{ie}\ket\chi> &\approx -2\pi\,(\nbar m)_i\m1\Sbar_{ie}\Partial{}{\vv_i}\.
\Int \dd\vv_e\(\mU(\vv_e) - \vv_i\.\Partial{}{\vv_e}\mU(\vv_e)\)
\NN\\
&\qquad\.\(\fr{1}{\mi}\BIG\ket\Partial{\chi_i}{\vv_i}>\fMbar(\vv_e)
- \fr{1}{\me}\BIG\ket\Partial{\chi_e}{\vv_e}>\fMbar(\vv_i)\)
\\
&\approx -2\pi\,(\nbar m)_i\m1\Sbar_{ie}\Partial{}{\vv_i}\.
\bigg(
\underbrace{\<\mU>_{{\rm
M}e}\.\fr{1}{\mi}\BIG\ket\Partial{\chi_i}{\vv_i}>}_{\rm (a)} 
+
\underbrace{\ket\vv_i>\.\fr{1}{\me}\BIG\<\Partial{\mU}{\vv_e}\.|\Partial{\chi_e}{\vv_e}>}_{\rm (b)}
\NN\\
&\qquad - \underbrace{\fr{1}{\me}\BIG\<\mU\.|\Partial{\chi_e}{\vv_e}>\ket
1>_i}_{\rm (c)}
\bigg).
\eq{Cie}
\EALams
To understand the content of \Eq{Cie}, note that without linearization
the ion--electron operator is approximately
\BE
\C_{ie}\on{f} \approx -\fr{(mn)_e}{(mn)_i} \taue\m1\Partial{}{\vv_i}\.
\((\vv_i - \vu_i)f_i + \fr{\Te}{\mi}\Partial{f_i}{\vv_i}\)
- \fr{1}{(mn)_i} \vR\.\Partial{f_i}{\vv_i},
\EE
where it was assumed that the electron distribution is a local
Maxwellian.\footnote{If the approximation \EQ{R0} is used for the momentum
transfer, the operator assumes the appealing \FP\ 
form
\BE
\C_{ie}\on{f} \approx -\fr{(mn)_e}{(mn)_i} \taue\m1\Partial{}{\vv_i}\.
\((\vv_i - \vu_e)f_i + \fr{\Te}{\mi}\Partial{f_i}{\vv_i}\),
\NN
\EE
appropriate for a test ion moving through a sea of of electrons with
mean flow~$\vu_e$.}
The linearization of this operator around absolute Maxwellians with equal
electron and ion temperatures is
\BAams
\D \C_{ie}\on{f} = &-\fr{(mn)_e}{(mn)_i}\taue\m1\Partial{}{\vv_i}\.
\bigg[\underbrace{\vv_i\D f_i}_{\rm (a')} 
+ \underbrace{\fr{T}{\mi}\Partial{\D f_i}{\vv_i}}_{\rm (b')} 
- \underbrace{\D\vu_i f_i}_{\rm (c')} 
- \underbrace{\vv_i \fR{\D\Te}{T} f_i}_{\rm (d')}\bigg]
\NN\\
&- \underbrace{\fr{1}{(mn)_i}\Partial{}{\vv_i}\.(\D\vR\, f_i)}_{\rm (e')}.
\eq{Cie'}
\EAams
For consistency, \Eq{Cie} should reduce to this when $\chi_e$~is taken to be
the perturbation of a local Maxwellian.  That this is so is demonstrated in
footnote~\ref{fn_Cie}.
\addtocounter{footnote}{1}
\footnotetext{\protect\label{fn_Cie}%
Here I sketch how \Eq{Cie} reduces to \Eq{Cie'}.   Term~(a) in \Eq{Cie} can
be rewritten by pulling the velocity derivative out of the ket according to
\BE
\fr{1}{\mi}\BIG\ket\Partial{\chi_i}{\vv_i}> =
\fr{1}{\mi}\Partial{}{\vv_i}\ket\chi_i> + \fr{\vv_i}{\Ti}\ket\chi_i>
\to \fr{1}{T}\(\vv_i\ket\chi_i> + \fr{T}{\mi}\Partial{}{\vv_i}\ket\chi_i>\)
\NN
\EE
for $\Ti = \Te = T$,
which when \Eq{vUe} is used
reproduces terms~\Term{a'} and~\Term{b'}. To evaluate term~(b), note that
for a local 
Maxwellian one has
\begingroup
\allowdisplaybreaks
\BAams
\D\ln f_e = \chi_e &= \D\[\ln\fR{\ne}{\nbar} - \fr{(\vv - \vue)^2}{2\vte^2} -
\Case32\ln(2\pi \vte^2)\]
\NN
\\
&= \fr{\D\ne}{\nbar} + \fr{\vv\.\D\vue}{\vte^2} +
\(\Half\fr{v^2}{\vte^2} - \Case32\)\fr{\D\Te}{\Te}.
\NN
\EAams
\endgroup
Then term~(b) involves
\BE
\BIG\<\Partial{\mU}{\vv_e}\.|\Partial{\chi_e}{\vv_e}>
= \BIG\<\Partial{\mU}{\vv_e}\.\(\fr{\D\vue}{\vte^2} +
\fr{\vv_e}{\vte^2}\fr{\D\Te}{\Te}\)>.
\NN
\EE
Upon integration by parts, the $\D\vue$~term vanishes while the
$\D\Te$~term reduces to $\vte\m2\<\mU>(\D\Te/\Te)$ and leads to
term~\Term{d'}. 
Finally, the momentum transfer is
$
\vR = -\Int \dd\vv_e\,(\nbar m\vv)_e \C_{ei}\on{f}
$,
where
\BE
\C_{ei}\on{f} \approx
-\nu\vte^3\Partial{}{\vv_e}\.\mU(\vv_e- \vui)\.\Partial{f_e}{\vv_e}.
\NN
\EE
Then, after integration by parts,
\BE
\D\vR = -(\nbar m)_e\nu\vte^3\Int \dd\vv_e\,\[\mU\.\Partial{f_e}{\vv_e}
- \vui\.\PartiaL{\mU}{\vv_e}\.\Partial{f_e}{\vv_e}\].
\NN
\EE
The first term on the \rhs\ is recognized as being proportional to term~(c).  It
can thus be replaced by a term of $\Order{\D\vR}$ [term~\Term{e'}] and a
term of 
$\Order{\D\vui}$ [term~\Term{c'}].  Thus, one has accounted for all of the terms
in \Eq{Cie'}.  The reader can check that all of the cofficients work out
correctly.}


\subsection{Properties of the linearized Landau operator}

The linearized Landau operator inherits the conservation laws of the full
operator:
\BE
\bra
\nbar\,\vA
 |\, \Chat = 0,
\eq{CL_conservation}
\EE
where the scalar product includes species summation [see \Eq{ALBs}] and where
\BE
\vA \doteq 
(1\;\;
\vP'\;\;
K'
)\Tr.
\EE
This also follows directly from \Eq{delv_chi} upon integration by parts.
This means that $\Chat$ has a five-dimensional null space, with $\bra\nbar\vA|$
defining the left null eigenvectors.\footnote{A proof that these are the
  only null eigenvectors  is (essentially) given by
  \Ref[\SECTION 7.2]{Montgomery1}.} 
  For further discussion of the
spectrum of~$\Chat$, see \Sec{spectrum}.

It can easily be shown from \Eq{delv_chi} that $\nbar\Chat$ is
self-adjoint:\footnote{The need for the density factor can be seen from the
  elementary estimate for the collision frequency $\nu_{s\sbar}$ of a test
  particle of species~$s$ colliding with field particles of species~$\sbar$:
  $\nu_{s\sbar} \sim \s_{s\sbar}\abso{v_s - v_{\sbar}}n_\sbar$, where
  $\s$~is the 
  scattering cross section.  This formula is not symmetric in the density;
  symmetry is restored by multiplying by~$n_s$.}
\BE
\<\psi|\nbar\,\Chat|\chio> = \<\chio|\nbar\,\Chat|\psi>.
\EE
Thus, upon taking the adjoint of \Eq{CL_conservation}, one finds that the 
right null eigenvectors are $\ket\vA>$:
\BE
\Chat\ket\vA> = 0.
\EE

\subsection{Calculation of \texorpdfstring{$\P\Chat\P\ket\chi>$}{P Chat P |chi>}}
\label{PCPchi}

To calculate $\P\Chat\P\ket\chi>$, the dissipative part of the frequency
matrix [see \Eq{mOmega_C_def}], one first evaluates $\Chat\P\ket\chi>$, then  
applies~$\P$ to that.  Into the representation \EQ{delv_chi}, one must
replace~$\chi$ by its projected value according to
\BE
\chi_s \to 1\fr{\D n_s}{\nbar_s} + \vv\fr{1}{\vts^2}\D\vu_s
+ K'_s\fr{1}{T^2}\D T_s,
\EE
so
\BE
\fr{1}{m_s}\Partial{\chi_s}{\vv} = \fr{1}{T}\D\vu_s + \vv\fr{1}{T^2}\D T_s.
\EE
For like-species collisions, this vanishes under the antisymmetrization;
this is a manifestation of the conservation laws and the self-adjointness
of~$\nbar\Chat$. 
All of the remaining integrals can be performed for arbitrary mass ratio in
terms of the error function and its derivative,
but I shall 
not do so here; simplifications for small mass ratio are given in the next
section. 
However, the general result for $\P\Chat\P\ket\chi>$ is relatively simple.
After integration by parts of $\P\Chat = \P\delvv\.\vJhat$, one must apply 
$
\ket\vv>mT\m1\bra\mone| + \ket K'>mN_T\m1\bra\vv|
$
to~$\vJhat\P\ket\chi>$.  
It is clear from the formula \EQ{delv_chi} that one requires the integrals
\BE
\Int \dd\vv\,d\vvbar\,\fM(\vv)\fM(\vvbar)
\begin{pmatrix}
\mU(\vv - \vvbar)\\
\vv\.\mU(\vv - \vvbar)\\
\vv\.\mU(\vv - \vvbar)\.\vv
\end{pmatrix}.
\EE
These are best done by transforming to the relative and \centre-of-mass
coordinates
\BE
\vw \doteq \vv - \vvbar,
\quad
\vW \doteq (m\vv + \mbar\,\vvbar)/M,
\EE
where $M \doteq m_s + m_{\sbar}$,
so
\BE
\vv = \vW + \fR{\mbar}{M}\vw,
\quad
\vvbar = \vW -\fR{m}{M}\vw,
\EE
and
\BE
\Int \dd\vv\,\dd\vvbar\,\fM\fMbar \ldots  = \Int
\dd\vW\,\dd\vw\,\Phi_M(\vW)\Phi_\mu(\vw)\dots, 
\EE
where $\Phi_{\muhat}$~is a Maxwellian with variance defined by $\s^2 =
T/\muhat$ for $\muhat = \mu$ or~$M$ with $\mu$
being the reduced mass, defined by 
\BE
\mu_{s\sbar}\m1 \doteq m_s\m1 + m_\sbar\m1.
\EE
It is then easy to show that 
\BE
\Int \dd\vv\,\dd\vvbar\,\fM(\vv)\fM(\vvbar)
\begin{pmatrix}
\mU(\vv - \vvbar)\\
\vv\.\mU(\vv - \vvbar)\\
\vv\.\mU(\vv - \vvbar)\.\vv
\end{pmatrix}
 = 2\fR{2}{\pi}\ehalf\vtmu\m1
\begin{pmatrix}
(1/3)\mone\\
\v0\\
\vtM^2
\end{pmatrix}.
\EE
The final result is
\BE
\P\Chat\P\ket\chi> = \ket\vv>N_\vv\m1\.\sum_{s'}\nu_{ss'}(\D\vu_{s'} -
\D\vu_s)
+ \ket K'>N_T\m1\bigg[3\sum_{s'}\fR{m_s}{M_{ss'}}\nu_{ss'}(\D T_{s'} - \D
  T_s)\bigg],
\eq{PCP}
\EE
where the generalized collision rate is defined by
\comment
\footnote{The mean
density~$\nbar$ appears here because one is considering small perturbations
around absolute thermal equilibrium.  In fully nonlinear theory, the complete
density~$n$ should appear.}
\endcomment
\BE
\nu_{ss'} \doteq \Case43\sqrt{2\upi}\,\fr{q_s^2(n q^2)_{s'}}{m_s T
  \vtu}\ln\Lambda_{ss'}.
\eq{nu_ss'}
\EE

\subsection{Calculation of \texorpdfstring{$\Chat \P\ket\chi>$}{Chat P
|chi>}} 
\label{CPchi}

Ultimately, one requires $\Q\Chat\P\ket\chi> = (1 - \P)\Chat\P\ket\chi>$,
so one needs the action of~$\Chat$ on the hydrodynamic subspace.  
Major simplifications ensue for small mass ratio, which was assumed by
Braginskii.  For  electron--ion collisions, one has to lowest order
$\mU(\vv - \vvbar) \approx \mU(\vv)$, 
which projects into the direction perpendicular to~$\vv$.  That removes 
the $\D T_e$~term of $\P\ket\chi>$, and the integral over~$\vvbar$ removes the
$\D T_i$ term, which is odd in~$\vvbar$.  The second property of
\Eq{U_props} can be used to simplify the divergence, and the result can be
written in terms of the small-$\me$ limit of the collision rate
$\nu_{ei}$.  Thus, for small mass ratio one finds
\BE
\Chat_{ei}\P\ket\chi> \approx
3\sqrt{\fr{\pi}{2}}\fr{1}{\vte^2\tau_{ei}}\BIG 
\ket\fR{\vte}{v}^3\vv>\.\D\vu,
\EE
where
$
\D\vu \doteq \D\vu_e - \D\vu_i
$,
so the total flow-driven contribution to $\Q\Chat\P\ket\chi>$ is
\BE
\sum_i\fr{1}{\vte^2\tau_{ei}}\BIG\ket \[3\sqrt{\fr{\pi}{2}}\fR{\vte}{v}^3 -
1\]\vv>\.\D\vu.
\EE
We shall see in \Sec{hydro_closure} that this term behaves as a source that
generates a 
contribution to $\ket\Q\chi>$ [see \Eq{Q_chi}].  The physics of this result
is that under 
perturbation the electron distribution is not merely a shifted Maxwellian;
a high-energy non-Maxwellian tail develops because of the inverse velocity
dependence of 
the electron--ion collision frequency.  Thus, in the un\magnetize d plasma the
approximation \EQ{R_approx0} is not correct; the true shape of the
perturbed distribution determines, for example, the values of~$\a$ in
\Eq{vR_vu} and~$\b$ in \Eq{vR_T}. 

For ion--electron collisions, on the other hand, it is a straightforward
calculation using \Eq{Cie} to show that to lowest order in the mass ratio
$\Chat_{ie}\ket\chi>$ lies entirely in the hydrodynamic subspace (\ie, that
$\Q\Chat\P\ket\chi> \approx 0$).  

\subsection{The spectrum of the linearized Landau operator and its relation
to the Markovian approximation}
\label{spectrum}

\Ref{Lewis} has
shown that $\Chat_e$~has a continuous spectrum except for the five discrete
null eigenvalues.  While many calculations involving~$\Chat$ can be done
without explicit reference to its spectrum, the spectral
representation is the most direct way to argue for the validity of the
Markovian approximation that is used in obtaining the 
standard form of the transport equations [for example, see \Eq{meta}].
What one needs to determine is whether a construction of the form
\BE
\ket K(v,\t)> \doteq \ee^{-\Q\Chat\Q\t}\ket \Shat>,
\eq{eQCS}
\EE
where $\P\ket \Shat> = 0$, decays on the collisional timescale. 
This is easy to argue in the affirmative when the spectrum of~$\Chat$ is
discrete; however, a continuous spectrum
introduces some subtleties.  Therefore, I shall provide some discussion.

Note that since $\ket \Shat> = \Q\ket \Shat>$ by assumption, one has
\BE
\ee^{-\Q\Chat\Q\t}\ket \Shat> = \ee^{-\Q\Chat\t}\ket \Shat>.
\EE
For simplicity, assume that $\P\Chat = 0$ (\eg, the case of
self-collisions).  Then $\Q\Chat = (1 - \P)\Chat = \Chat$.  
The simplest relevant example is the 1D \FP\ operator for a test
particle of mass~$M$ in a bath of temperature~$T$:
\BE
\Chat_i f \doteq -\Partial{}{v}\(\nu v + \Dv\Partial{}{v}\)f,
\eq{Chat_i}
\EE
where the constant coefficients are related by the Einstein relation $\Dv =
(T/M)\nu$.  With velocities being 
normalized to $v_T \doteq (T/M)\ehalf$, \Eq{Chat_i} can be written as
\BE
\Chat_i\ket\chi> = -\nu\Partial{}{v}\BIG\ket\Partial{\chi}{v}>,
\eq{Chat_i_ket}
\EE
where the implicit weight function is a Maxwellian with unit variance.
From this representation, it can easily be seen that the operator is
self-adjoint \wrt\ the standard scalar product.  It is a 1D model of the 
ion--electron collision operator~$\Chat_{ie}$ with $M = \mi$, $T = \Te$,
and $\vue = \v0$;
it has a 1D null eigenspace~$\ket 1>$ associated with density conservation.
One can readily verify that the eigenfunctions are the (probabilistic)
Hermite polynomials:\footnote{The first few one-dimensional probabilistic
Hermite 
  polynomials in a standard normalization such that
$\INT \dd v\,\He_n(v)\He_{n'}(v)(2\pi)\m{1/2}\ee^{-v^2/2} \equiv
\<\He_n(v)\mid\He_{n'}(v)> = n!\,\Kron{nn'}$
are  
$\He_0(v) = 1$, $\He_1(v) = v$, $\He_2(v) = v^2 - 1$.}
\BE
\Chat_i\ket \He_n(v)> = n\nu\ket \He_n(v)>,
\EE
where $n$~is a nonnegative integer.  Thus, $\Chat_i$~has a discrete
spectrum, a 
property shared with the linearized Boltzmann operator (which has a 5D null
space). 
To determine the \behavior\ of \Eq{eQCS}, insert the completeness relation
(resolution of the identity)
\BE
\Dirac{v-\vbar} = \sum_{n=0}^\infty
\He_n(v)\fr{1}{n!}\He_n(\vbar)
\fR{\ee^{-v^2/4}}{(2\pi)^{1/4}}
\fR{\ee^{-\vbar^2/4}}{(2\pi)^{1/4}}
\EE
into \Eq{eQCS}:
\BALams
\ket K(v,\t)> &= \ee^{-\Q\Chat_i\t}\sum_{n=0}^\infty \ket
\He_n(v)>\fr{1}{n!}\<\He_n 
\mid \Shat>
\\
&= \sum_{n=1}^\infty \ee^{-n\nu\t}\fr{1}{n!}\Shat_n\ket\He_n(v)>.
\EALams
[The $n = 0$ term is excluded because $\He_0(v) = 1$ and I have assumed
that $\<1 | \Shat> = 0$.]
This clearly decays on the collisional timescale, so there is no
difficulty with justifying the Markovian approximation.

Now consider $\Chat = \Chat_{ee}$.  \Ref{Lewis} showed that the solution of
$\delt\ket f> = -\Chat_{ee}\ket f>$ has the continuous spectral
representation (mostly using Lewis's notation)
\BE
\ket f>(\vc,t) = \sum_{l,m}\InT
\dd\r(\l_l)\,\Psi_{lm}(\vc,\l_l)\Ft_{lm}(\l_l)\ee^{-\l_l\t},
\eq{f(t)}
\EE
where $\t$~is a dimensionless time (normalized to an electron--electron
collision time), 
$\r$~is the spectral measure\footnote{The proportionality constant~$a_l$ is
fixed by the chosen normalization of the 1D eigen\-functions~$\psi_l$.}
 $\dd\r(\l_l) \doteq a_l\l_l\m{1/2}\dd\l_l$,
\BALams
\Ft_{lm}(\l_l) &\doteq N_l\m1\InT \dd\vc\,\Psi_{lm}\conj(\vc,\l_l)f(\vc,0),
\\
\Psi_{lm}(\vc,\l_l) &\doteq c\m1 \ee^{-c^2/2}\psi_l(c,\l)Y_l^m(\theta,\p),
\eq{Ftilde}
\EALams
$c \doteq v/(\sqrt2 \vt)$, the $Y_l^m$~are the spherical harmonics
normalized such that $\Int \dd\Omega\, Y_l^m (Y_{l'}^{m'})\conj = N_l
\Kron{ll'}\Kron{mm'}$, and the $\psi_l(c,\l_l)$ are the eigenfunctions that
solve a particular linear, integro-differential, self-adjoint equation
[Lewis's equation~(20)] deduced from the collision operator linearized around a
Maxwellian. 
Thus, functions $u(c)$ that are square-integrable on $(0,\infty)$ have the
spectral (generalized Fourier) representation
\BE
u(c) = \InT \dd\r(\l_l)\,\psi_l(c,\l_l)\Tilde{u}(\l_l),
\quad
\Tilde{u}(\l_l) = \InT \dd c\,\psi_l\conj(c,\l_l)u(c).
\EE
Given that the measure~$\r(\l)$ is continuous at $\l = 0$, the point $\l =
0$ can be excluded from the integration in \Eq{f(t)}.  Thus, as Lewis
states, the 
spectral representation can be shown to be complete for all perturbations
conserving the densities of number, momentum, and kinetic energy.
Furthermore, since $\l = 0$ is absent, one finds from \EQ{f(t)} that
perturbations 
decay on the collisional timescale;\footnote{In general, one could
contemplate nonphysical initial conditions such that the decay was more
complicated.  However, the specific $\Q\L\P$ terms arising in the
projection-operator formalism involve benign, low-order moments of
velocities scaled to~$\vt$.  Their generalized Fourier transform
\EQ{Ftilde} thus involve~$\l$'s that are $\Order{1}$ in dimensionless
units, leading to time dependence that is $\Order{1}$ in~$\t$ (the
collisional time scale).} thus, the Markovian approximation is
justified (provided that the area under the curve is finite). 

 Note that the collisional decay described by \Eq{f(t)} need not be
 exponential (unlike the case of a 
discrete spectrum) because it involves a continuous superposition of
 exponentials.  A 1D example is obtained by considering the operator
$\Chat \to -\del_v^2$, which is the velocity-space diffusion part of the
operator \EQ{Chat_i} written in dimensionless variables with velocities
normalized to~$v_T$ and times normalized to~$\nu\m1$.  (Note that this
 operator is not self-adjoint \wrt\ the standard scalar product.)
Unlike the operator~\EQ{Chat_i}, the diffusion operator has a continuous
 spectrum with plane-wave eigenfunctions $\exp(\ii\Lambda v)$ and
eigenvalues\footnote{Notice that Lewis's measure $\dd\r \propto 
\l\m{1/2}\dd\l$ is proportional to~$\dd\Lambda$, so the decay in \Eq{f(t)}
is $\exp(-\Lambda^2\t)$, just as for the diffusion operator.  The physics
is different, however, because for~$\Chat_{ee}$ polarization drag is
captured in the solution for~$\psi_l(\l)$, which is not a simple plane
wave [\cf\ the first unnumbered equation after Lewis's equation~(32)].}  $\l = 
\Lambda^2$,
$\Lambda$~being the continuous Fourier variable conjugate to~$v$.  As an
example, consider the specific initial condition $f(v,0) = 
v [(2\pi)\m{1/2}\ee^{-v^2/2}]$.  It is straightforward to solve the
diffusion equation by Fourier transformation to find
\BE
f(v,\t) = (1+2\t)\m{3/2}v \fr{1}{\sqrt{2\pi}}\exp\[-\Half\fR{v^2}{1+2\t}\].
\eq{f2}
\EE
This function decays on the collisional timescale, although
not exponentially, and $\InT \dd\t\,f(v,\t)$ is finite.\footnote{
\comment
The same exercise done for the initial condition $\ket
f>(v,0) = \ket v>$ leads to $f(v,t) = v(1+2\t)\m1\exp[-\half v^2/(1+2\t)]$,
whose time integral is logarithmically divergent at long times.  This
divergence arises because the chosen initial condition corresponds to the
current associated with the density projection of the equation $\del_\t f +
\ii k v f - \Dv\del_v^2 f = 0$, which is the \FPE\ corresponding to the
collisionless Langevin system $\dot\xt = \Tilde{v}$, $\dot{\Tilde{v}} =
\Tilde{a}(t)$, $\Tilde{a}(t)$~being \centre d Gaussian white noise with
correlation 
function $\<\d a(t)\d a(t')> = 2\Dv\Dirac{t-t'}$.  It is well
known that the spatial variance of the solution is $\<\d x^2>(t) = 2\Dv
t^3/3$, a superdiffusive result that corresponds to an infinite spatial
diffusion coefficient.
\endgraf
\indent
\endcomment
Regarding long-time tails that arise from superpositions of exponentials, a
closely related phenomenon is the $\t\m{d/2}$~tail on the velocity correlation
function that arises in classical kinetic theory from the spatial \wavenumber\
superposition of slowly decaying hydrodynamic
modes\cite{JAK_cells,Balescu,Reichl2,Zwanzig}.  A nonintegrable
$\t\m1$~tail arises for $d 
= 2$, giving rise to vexing issues relating to nonlocality in 2D
hydrodynamics.} 

\comment
\FIGURE[0.7]{Cfcn}{The function \EQ{f2} for $v = 1$.  The plot demonstrates
that a continuous superposition of exponentials need not be exponential.}
\endcomment

\section{Covariant representation of the hydrodynamic projection}
\label{Covariant}

Here I give justification and further discussion of the covariant
representation of the hydrodynamic projection.  

In the Dirac bra--ket notation,
kets $\ket\cdot>$ are conventionally interpreted as vectors,
while bras $\bra\cdot|$ are interpreted as covectors.\footnote{For
  discussion of the distinction between vectors and covectors and of other
  related concepts, see a modern textbook on differential
  geometry such as \Ref{Fecko}.  The presentation by \Ref{MTW} is
  particularly pictorial and pedagogical.}
  In a
finite-dimensional vector space spanned by the basis vectors~$\ve_i$,
vectors~$\vv$ are represented as $\vv = v^i\ve_i$, with the~$v^i$
being called the contravariant components.  Similarly,
covectors~$\vw$ are represented as $\vw = w_i\ve^i$, where
the~$\ve^i$ are the dual basis vectors and the~$w_i$ are called the
covariant components.  In Dirac notation, one writes $\ket\vv> =
v^i\ket\ve_i> \equiv v^i\ket i>$, the underlying basis vectors
being understood in the last notation.  Similarly, $\bra\vw| =
w_i\bra\ve^i| \equiv w_i\bra i|$.

It is useful to introduce an adjoint operation~$\dagger$ that changes
vectors into covectors (kets into bras) and \versa:
\BE
\ket\vv>\adj = \bra\vv|,
\quad
\hbox{or}
\quad
(v^i\ket i>)\adj = v_i\bra i|.
\EE
Because in the application to hydrodynamics the natural scalar product is
real-valued, no complex conjugate is taken in the execution of the adjoint
operation. 

The components~$A^\mu$ of~$\vA$ ($\mu = 1,\dots,5$) are, in 
fact, the first few multidimensional 
Hermite polynomials.
  Multiplied by~$\fM$, the complete set of those polynomials spans the
velocity space.  Therefore, $\bra A^\mu|$ plays the role of a dual basis
vector~$\ve^\mu$, consistent with the interpretation of a bra as a
covector.   

Define $M^{\mu\nu} = \<A^\mu A^\nu>$.  The inverse of this matrix is
naturally written with lower indices:  $(\mM\m1)_{\mu\nu}$.  Interpret this
inverse as a metric tensor $g_{\mu\nu}$ and lower indices according to
$A_\mu = g_{\mu\nu}A^\nu$.  The ket~$\ket A_\mu>$ is consistently interpreted
as a basis vector~$\ve_\mu$.   

Given this notation, one can define the (dimensionless and
self-adjoint) hydrodynamic projector 
\BE
\P = \ket A_\mu>\bra A^\mu| \equiv \ket\mu>\bra\mu|.
\EE
The hydrodynamic projection of the state vector $\ket\chi>$ is then
\BE
\P\ket\chi> = \ket A_\mu>\< A^\mu|\chi> = \ket A_\mu>\D a^\mu,
\EE
which defines the hydrodynamic variables~$\D a^\mu$ as the contravariant
components of 
a \emph{hydrodynamic vector}.

The choice $g_{\mu\nu} = (\mM\m1)_{\mu\nu}$ is a special case of the
\emph{Weinhold metric}\cite{Weinhold}.  The use of that metric in the
context of a 
covariant representation of Onsager symmetries has been discussed by
\Ref{JAK_Onsager}.

\section{An example of the calculation of some transport coefficients:  Classical electron heat flow} 
\label{qe}

Calculation of the classical electron heat flow in the limit of
small\footnote{The classical transport 
coefficients in the small-collisionality limit were considered by
\Ref{Rosenbluth58} and \Ref{Kaufman_viscosity}.}
$\e \doteq \nu_e/\abso{\wce}$ provides a good example of
the use of the various formulas and gives insights that are not available
from purely numerical calculations.
I repeat for convenience Braginskii's result quoted in
\Sec{exact}:\footnote{As a reminder, Braginskii's 
  gyrofrequencies are unsigned, whereas mine are signed.}
\BE
\vq_{\perp,e} \approx -4.66\ne\k_{\perp,e}\vgradperp \Te
+ \Case52\ne\fR{c\Te}{eB}\bhat\cross\vgrad \Te
-\Case32\fr{(nT)_e}{\wce\taue}\bhat\cross\vgrad\vu,
\eq{q_e}
\EE
where
$
\k_{\perp,e} \doteq \re^2/\taue
$
has the usual random-walk scaling.  ($\rhoe \doteq \vte/\abs{\wce}$ is the
electron gyroradius.)
I shall show that all of  the numerical coefficients in this expression can
be calculated analytically by approximately solving \Eq{Q_chi} for small~$\e$.

Linear superposition shows that the thermodynamic forces $\D\vW \doteq
\vW\on{\D\vu}$, 
$\vgrad\D T/T$, and~$\D\vu$ drive independent contributions to
$\ket\Q\chi>$.  The $\D\vR$~term [the last term of \Eq{Q_chi}; see \Eq{DR}]
also behaves as a thermodynamic force, but 
its magnitude is determined as part of the solution.  For the calculation
of~$\vq$, defined by \Eq{q_def}, vector symmetry precludes a contribution
from~$\D\vW$.  To calculate the $\vgradperp\D T$-driven heat flux,
let $\ket\psi>$~be the part of $\ket\Q\chi>$ driven by $\vgradperp\D T/T$.
Define $\gradT \doteq \vgrad \D T/T / (\abso{\grad\D T}/T)$ (this unit
vector in the direction of the gradient is
conventionally taken to lie in the $-\xhat$ direction), $\psibar \doteq
\psi/(\abso{\grad\D T}/T)$, and $\vRbar \doteq \vR/(\abso{\grad\D
  T}/T)$.  Thus, with velocities normalized to~$\vte$, one must solve
\BE
-(\ii\Mhat + \Chat)\ket\psibar> = \vte\BIG\ket\(\Half v^2 -
\Case52\)\vvperp>\.\gradT 
+ \fr{1}{\nbar T}\ket\vvperp>\.\D\vRbar.
\eq{psibar}
\EE
The most general solution is
\BE
\psibar = a(v)\vvperp\.\gradT + b(v)\vvperp\cross\gradT\.\bhat,
\EE
where $a(v)$ and~$b(v)$ are unknown functions to be determined.
Upon applying $\ii\Mhat \doteq \wce\,\del/\del\z$ to $\vvperp = \vperp(\sin\z,\,
-\cos\z)\Tr$, one finds $\ii\Mhat\vvperp = \wce\vv\cross\bhat$ ($\Mhat$~rotates
perpendicular velocity vectors by angle~$\z$).  Upon rearranging
\Eq{psibar} in 
anticipation of iteration in small~$\e$, one finds
\BE
\ket a \vvperp>\cross\gradT\.\bhat + \ket b\vvperp>\.\gradT =
\rr\BIG\ket\(\Half v^2 - \Case52\)\vvperp>\.\gradT 
+ \Underbrace{\fr{1}{\nbar\wce
    T}\ket\vvperp>\.\D\vRbar\on{\psibar}}{{$\Order{\e}$}} 
+ \Underbrace{\fr{1}{\wce}\Chat\ket\psibar>}{{$\Order{\e}$}},
\eq{ab}
\EE
where $\rr \doteq \vte/\wce$ (a negative quantity).  Expand~$a$ and~$b$ in
powers of~$\e$
(\eg, $a = \sum_{n=0}^\infty a_n \e^n$).  One readily deduces that
\BE
a_0 = 0,
\quad
b_0 = \(\Half v^2 - \Case52\)\rr,
\eq{b0}
\EE
At $\Order{\e}$, one must satisfy
\BE
\ket a_1 \vvperp>\cross\gradT\.\bhat + \ket b_1 \vvperp>\.\gradT =
\fr{1}{\nbar\wce T}\ket \vvperp>\.\D\vRbar\on{b_0 
  \vvperp\cross\gradT\.\bhat} + \fr{1}{\wce}\Chat\ket b_0
\vvperp>\cross\gradT\.\bhat.
\eq{ab1}
\EE
From formula \EQ{DR_def}, one finds that the momentum transfer is given by
$\D\vRbar = \mR\cross\gradT\.\bhat$, where
\BE
\mR \doteq -(\nbar m\vt)_e\<\vvperp | \Chat^{\rm Lor}|b_0\vvperp>
\EE
and $\Chat^{\rm Lor} \doteq \nu v\m3 \Lvec^2$.
This integral can be evaluated by representing the 3D velocity in a spherical
coordinate system $(v,\theta,\phi)$ and recalling that the $(l=1,\,m=1)$
spherical harmonic is proportional to $\sin\theta \,\ee^{\ii\phi}$; thus,
$\Lvec^2\vvperp = l(l+1)\vvperp = 2\vvperp$.  The resulting integral is
proportional to $\mone_\perp$ by isotropy.  
With the result that the 3D Maxwellian average of~$v^n$ is\footnote{For
  even moments, formula \EQ{<v^n>} reduces to $\<v^{2n}> = (2n+1)!!\,$.}
\BE
\<v^n> = \fr{4\pi}{(2\pi)^{3/2}}2^{(n+1)/2}\Gamma\fR{n+3}{2},
\eq{<v^n>}
\EE
the matrix element can be calculated.  The final result is
\BE
\D\vR = \Case32\fr{\nbar_e}{\wce\taue}\bhat\cross\vgrad\D T,
\EE
which agrees with Braginskii's result for the perpendicular thermal force
[see \Eq{vR_T}].  

Due to the rotational symmetry of~$\Chat$, the last term of \Eq{ab1} is
proportional to $\vvperp\cross\gradT\.\bhat$.  One therefore concludes that
$b_1 = 0$.  Contributions to~$a_1$ arise from all of the~$\D\vRbar$ and the
$\Chat = \Chat_{ei}^{\rm Lor} + \Chat_{ee}$ terms.  I shall omit the
algebra relating to~$\Chat$.

One can now undo the normalizations and proceed to calculate the heat flux
\BE
\D\vq = \<\nbar K'(v)\vv | \psi>
\EE
[$K'(v)$~is defined by \Eq{K'_def}]
from
\BE
\psi = -a\vv\.\vgrad\fR{\D T}{T} + b\vv\.\bhat\cross\vgrad\fR{\D T}{T}.
\EE
The diamagnetic flux (in the direction orthogonal to the gradient)
is\footnote{Matrix elements of the 
  form $\<G(v)\vv\,\vv>$ are by symmetry equal to $A\mone$, where $A =
  \<G(v)v^2>/3$.}  
\BALams
\D\vq_\bigast &= -(\nbar T\vt)_e\<K'(v)\vv | b_0(v)\vv>\.\bhat\cross\vgrad\fR{\D
  T}{T}
 \\
&= -\Case52\nbar_e\fR{\vte^2}{\wce}\bhat\cross\vgrad\D T =
\Case52\nbar_e\fR{c\Te}{eB}\bhat\cross\vgrad\D T.
\EALams
Here I used the result \EQ{b0} together with several instances of
the formula \EQ{<v^n>}; the answer  agrees with \Eqs{qeT} and \Eq{q_e}.
The flux in the direction of the gradient has the form
\BE
\D\vq_\perp = -\nbar\k_e\vgradperp\D T,
\EE
where $\k_e = A\k_{\perp,e}$ with
\BE
A \doteq \Third\BIG\<\(\Half v^2 - \Case32\)v^2 a_1(v)>.
\EE
Given the solution for~$a_1(v)$, 
it is
straightforward to work out the required matrix elements and find that
\BE
A = \underbrace{\Case32}_{\D\vR} + \underbrace{\Case74}_{\Chat_{ei}^{\rm Lor}} + 
\underbrace{\sqrt2}_{\Chat_{ee}} \approx 4.66,
\eq{4.66}
\EE
which reproduces the numerical coefficient in \Eq{q_e}.  A message from
\Eq{4.66} is that all relevant thermodynamic forces, including
self-collisions, contribute to the transport coefficient. 

A similar calculation leads to the perpendicular heat flow driven
by~$\D\vu$.  A 
difference is that there is no zeroth-order term; the $\D\vu$ source term
is already $\Order{\e}$.  This implies that contributions from the momentum
transfer and the explicit collisional correction
[the \analog s of the last two terms in \Eq{ab}] are $\Order{\e^2}$ and can
be neglected.  Thus, with $\D\vu = -\abso{\D u}\xhat$, one finds
\BE
b_1 = -\fr{1}{\vte^2\wce\taue}\[3\sqrt{\fr{\pi}{2}}\fR{\vte}{v}^3 - 1\].
\EE
The matrix element with~$K'$ is readily calculated, and one recovers the
last term of \Eq{q_e}.

A consequence of the fact that the $\D\vu_\perp$-driven contributions to
$\Q\vpsi$ are $\Order{\e}$ is that the perpendicular 
friction force is given dominantly by its value projected into the
hydrodynamic subspace; see
\Eq{Du_dot_B}.  That is, the numerical coefficient in $\vR_\perp = -(\nbar
m)_e\taue\m1\D\vu$ is~1.  The physical explanation is that the rapid
gyromotion rapidly restores the local Maxwellian in a time that is short
compared to the time to form a high-energy tail.

\bgroup

\def\V{\tensor{V}}
\def\K#1#2{\Kron{#1#2}}
\def\mD{\boldsymbol{\delta}^\perp}
\def\D#1#2{\d^\perp_{#1#2}}
\def\mBETA{\mbeta}
\def\Be#1#2{\beta_{#1#2}}
\def\B#1#2{B_{#1#2}}

\section{Decomposition of the stress tensor}
\label{Viscosity}

In the case of the one-component, weakly coupled plasma, the fourth-rank
viscosity 
tensor~$\tensor{m}$ was shown in \App{CE} to depend on a single scalar
coefficient~$\mu$, the kinematic viscosity,
 as a consequence of symmetry;
from \Eq{m_OCP}, one has 
\BE
m_{ijkl} = \mu\(\Kron{il}\Kron{jk} + \Kron{ik}\Kron{jl} -
\Case23\Kron{ij}\Kron{kl}\).
\eq{m_OCP_ijkl}
\EE
A background magnetic field~$\vB$ breaks the symmetry and the
representation of~$\tensor{m}$ becomes more complicated.  The most general
form of~$\tensor{m}$ that is compatible with rotational symmetry in the
plane perpendicular to~$\vB$ can be argued to depend on the three tensors
\BALams
\mB &\doteq \bhat\,\bhat = b_i b_j = 
\begin{pmatrix}
0 & 0 & 0\\
0 & 0 & 0\\
0 & 0 & 1
\end{pmatrix},
\\
\mD &\doteq \mone - \mB \equiv \D ij =
\begin{pmatrix}
1 & 0 & 0\\
0 & 1 & 0\\
0 & 0 & 0
\end{pmatrix},
\\
\mbeta &\doteq \bhat\cross = -\e_{ij3} = 
\begin{pmatrix}
0 & -1 & 0\\
1 & 0 & 0\\
0 & 0 & 0
\end{pmatrix},
\EALams
where the matrix forms are valid in a coordinate system in which $\vB$~is
locally in the $z$~direction.

The tensor~$\tensor{m}$ is symmetric and traceless in both of its first and
last pairs of indices.  Introduce a
symmetrizing operation $\set{\dots}$ that creates appropriately symmetric
and traceless 
tensors out of its argument.  Then the most general representation
of~$\tensor{m}$ is
$
\tensor{m} = \sum_{p=0}^4 n \mu_p \V_p
$,
where, using Cartesian tensor notation and Braginskii's conventions for
coefficients and signs,
\BALams
	\V_0 &\doteq 3\set{\mB\,\mB}
	 = 3\(\B ij - \Third\K ij\)\(\B kj -
		\Third\K kl\),
\\
	\V_1 &\doteq \set{\mD\,\mD} 
	= \D ik \D jl + \D il \D jk - \D ij \D kl,
\\
	\V_2 &\doteq \set{\mD\,\mB} 
	= \D ik \B jl + \D il \B jk + \B ik \D jl + \B il \D jk,
\\
	\V_3 &\doteq -\Half\set{\mD\,\mBETA}
	= -\Half(\D ik \Be jl + \D il \Be jk + \Be ik \D jl + \Be il
		\D jk),
\eq{V_3}
\\
	\V_4 &\doteq -\set{\mB\,\mBETA}
	= -(\B ik \Be jl + \B il \Be jk + \Be ik \B jl + \Be il \B
		jk).
\eq{V_4}
\EALams
The multiplicative factors of~3 and~$\half$ are for later convenience.
Note that the construction $\set{\mBETA\,\mBETA}$ is not independent
because, for example, $\Be ij \Be kl = \e_{ij3}\e_{kl3} = \D ik \D jl - \D
il \D jk$.

With these definitions, the $\V$'s obey the following properties:
$\V_p\.\V_{p'} = 0$ for $p\neq p'$,
\BALams
	\V_0 : \V_0 &= 2\V_0,
\\
	\V_1 : \V_1 &= 2\V_1,
\\
	\V_2 : \V_2 &= 2\V_2,
\\
	\V_3 : \V_3 &= -2\V_1,
\\
	\V_4 : \V_4 &= -2\V_2,
\EALams
and
\BE
	\V_0 + \V_1 + \V_2 = \set{\mone\,\mone} = \Kron{il}\Kron{jk} + \Kron{ik}\Kron{jl} -
\Case23\Kron{ij}\Kron{kl}, 
\EE
which is the kinematic part of $\V(\vB = \v0)$ [see \Eq{m_OCP_ijkl}].

To calculate~$\mPi$, one needs to know the action of~$\V_i$ on~$\vgrad\vu$.
Because 
the~$\V$'s are symmetric in their last two indices, this is equivalent to
calculating $\V_i : \mS$, where $\mS$~is defined by \Eq{S_def}.  Note
that~$\mD$ and~$\mB$ are symmetric, whereas 
$\mBETA$~is antisymmetric.  Then
\BALams
	\V_0 : \mS &= 3\(\mB - \Third\mone\)\(\mB - \Third\mone\) : \mS,
\\
	\V_1 : \mS &= 2\mS_\perp - \mD\Trace\mS_\perp,
\\
	\V_2 : \mS &= 2(\mD\.\mS\.\mB + \mB \.\mS \.\mD),
\\
	\V_3 : \mS &= \mD\.\mS\.\mBETA - \mBETA\.\mS\.\mD,
\\
	\V_4 : \mS &= 2(\mB\.\mS\.\mBETA - \mBETA\.\mS\.\mB).
\EALams
These are to be compared with Braginskii's equations~(4.42).  Instead of~$\mS$,
he uses $\mW \doteq 2(\mS -\third\div\vu\,\mone)$, or
$
	\mS = \half\mW + \third\div\vu\,\mone
$.
The last term, proportional to the identity operator, does not contribute
to~$\V_0$ 
(because~$\mS$~is dotted with a traceless quantity), $\V_2$
(because~$\mD$ and~$\mB$ are orthogonal), or $\V_3$ and~$\V_4$ (because of
cancellations due to the antisymmetry).  It contributes to~$\V_1$ a term
\BE
	\(\Casefr23 \div\vu - \vgradperp\.\vu_\perp\)\mD 
		= \(\gradpar \upar - \Third\div\vu\) \mD
		= \Half W_{zz}\mD .
\EE
Thus, one finds
\BALams
	\V_0 : \mS &= \Casefr32 \(\mB - \Third\mone\)\(\mB - \Third\mone\) :
	\mW,
\eq{V:S0}
\\
	\V_1 : \mS &= \mD\.\mW\.\mD + \Half (\bhat\.\mW\.\bhat)\mD,
\eq{V:S1}
\\
	\V_2 : \mS &= \mD\.\mW\.\mB + \mB\.\mW\.\mD,
\eq{V:S2}
\\
	\V_3 : \mS &= \Half(\mD\.\mW\.\mBETA - \mBETA\.\mW\.\mD),
\eq{V:S3}
\\
	\V_4 : \mS &= \mB\.\mW\.\mBETA - \mBETA\.\mW\.\mB.
\eq{V:S4}
\EALams
These are equivalent to Braginskii's equations~(4.42).  Note that each
of the constructions $\V_i:\mS$ is symmetric, consistent with the overall
symmetry of~$\mPi$.

The magnetic-field scalings of the $\mu_p$'s are
\def\emleft{-8em}
\BALams
 \mu_0 &\sim B^0
&&\hbox{\hspace{\emleft}(parallel transport)},
\\
\mu_1,\,\mu_2 &\sim B\m1
&&\hbox{\hspace{\emleft}(nondissipative gyroviscosities)},
\\
\mu_3,\,\mu_4 &\sim B\m2
&&\hbox{\hspace{\emleft}(cross-field transport)}.
\EALams
It can be shown that $\mu_1(\wc) = \mu_2(2\wc)$ and $\mu_3(\wc) =
\mu_4(2\wc)$.  This is a consequence of (i)~the fact that the spherical
harmonics are eigenfunctions of both~$\Chat$ and~$\ii\Mhat$, and (ii)~the
aptly chosen tensorial decompositions of the~$\V_p$'s.

The nondissipative gyroviscosities~$\mu_1$ and~$\mu_2$ emerge in
collisionless Vlasov or \GK\ 
theory as well. In the collisionless limit, a \CE-truncated fluid
description is inappropriate and \GK s\cite[and references
therein]{JAK_ARFM} provides a 
much superior approach. For further discussion, see \Ref{Belova01}.
\egroup

\section{Linear eigenmodes of the Braginskii equations}
\label{Eigenmodes}

After Fourier analysis in space, the linearized
Braginskii equations can be written as
\BE
\delt \va_\vk(t) = \mK_\vk\.\va_\vk,
\EE
where $\mK$~is a $5S\times5S$ square matrix.
This leads to the $5S$-dimensional eigenvalue problem
\BE
\det(\mK - \l\mone) = 0,
\EE
where I have dropped the $\vk$~labels for simplicity.  In this appendix I
discuss some aspects of the eigenmodes for first the un\magnetize d
one-component plasma 
(five eigenmodes; \Sec{eigen_OCP}), then a two-species plasma (ten
eigenmodes) 
in the two limits $\vB = \v0$ (\Sec{eigen_B0}) and $\e \doteq
\nu/\abso{\wc} \ll 1$ 
(\Sec{eigen_B}).  (Only special cases are considered for the latter.)  This
knowledge is of intrinsic conceptual interest and 
is also useful for numerical work with the Braginskii
equations.  While one might be most interested in low-frequency phenomena,
the Braginskii equations will not totally oblige; they contain high-frequency
eigenvalues as well: Langmuir oscillations for $\vB = \v0$; hybrid
oscillations for $\vB \neq 0$.  Those may limit the time step unless
special care is used in the formulation of the numerical algorithm.

For simple cases, the eigenvalue calculations can be done by hand.
However, for the most complicated situations the algebra becomes tedious.  The
determinant of a ten-dimensional matrix all of whose entries are unique
contains $10! \approx 3.6\cdot 10^6$ terms, each of which may be a
complicated product.  The matrices for the linearized Braginskii equations
are fortunately not full; nevertheless, the fully expanded determinant for
the $\vB
\neq \v0$ case contains more than 2500 terms.  Those are of various orders
in the small parameters~$\mu$, $\e$, and~$k^2$, and one is
interested only in the dominant balances.  Since diverse orderings are
possible, machine-aided
manipulations are useful in sorting out the details.\footnote{I used
\MATH\ to guide 
and check the algebra.  The basic operation is
\texttt{Det[\,]}, which returns the symbolic determinant of an
$n$-dimensional square 
matrix; the result is an $n$th-order polynomial in the eigenvalue~$\l
\doteq 
-\ii\w$.  \texttt{CoefficientList[\,]} extracts the coefficients of~$\l$.
Each coefficient in that list can be replaced by its dominant approximation
\wrt\ a chosen variable by the user-defined module
\texttt{reducelist[list\_, var\_]}.  That module expresses each term in
\texttt{list} 
as a polynomial in \texttt{var}, then searches the coefficient list of that
polynomial and returns the lowest-order term that is nonzero.  
Consecutive uses of that module for the
various small parameters finally lead to relatively simple expressions for
the characteristic polynomial, from which the dominant balances can be easily
recognized by the use of Kruskal diagrams.} 

\subsection{Eigenmodes of the un\magnetize d one-component plasma}
\eq{eigen_OCP}

The eigenmodes of the hydrodynamic equations of a neutral fluid are well
known; however, they are nontrivially modified in the presence
of the long-ranged 
Coulomb force.  I shall illustrate that for the un\magnetize d one-component
plasma.  These results are well known\cite{Balescu}.  However, 
lessons learned here generalize to the more complicated
multispecies and \magnetize d problems, which I shall discuss in later
sections. 

The linearized continuity equation is
\BE
\Partial{}{t}\fR{\Dn}{n} = -\ii\vk\.\D\vu,
\EE
the linearized momentum equation is
\BE
\Partial{\D\vu}{t} = \fr{q}{m}\D\vE - (nm)\m1\ii\vk\D p
- \mu k^2\D\vu - \(\Third\mu + \z\)\vk\,\vk\.\D\vu
\EE
(here I allow for a bulk viscosity~$\z$)
with $\D\vE = -\ii\vk\D\phi$,
and the linearized temperature equation is
\BE
\Partial{}{t}\fR{\D T}{T} = -\Case23\ii\vk\.\D\vu - \Case23 \k k^2\fR{\D T}{T}.
\EE
Decompose~$\D\vu$ into longitudinal and transverse components (\wrt~$\vk$):
\BE
\D\vu = \D\vu\lon + \D\vu\trans,
\EE
where
\BE
\D\vu\lon \doteq \khat\,\khat\.\D\vu,
\quad
\D\vu\trans \doteq (\mone - \khat\,\khat)\.\D\vu =
\khat\cross(\D\vu\cross\khat).
\EE
These components decouple according to
\BE
\l\D\vu\trans = -\mu k^2\D\vu\trans,
\EE
which gives rise to two \emph{shear modes}, each with $\l = -\mu k^2$,
and
\BALams
\l\fR{\Dn}{n} &= -\ii k\D u\lon,
\eq{l1_n}
\\
\l\D u\lon &= -\ii\wp^2\fr{1}{k}\fR{\Dn}{n} - \ii\vt^2k \(\fr{\Dn}{n} +
\fr{\DT}{T}\) 
- \(\Case43\mu + \z\)k^2\D u\lon,
\eq{l1_u}
\\
\l\fR{\DT}{T} &= -\Case23\ii k\D u\lon - \Case23\k k^2\fR{\DT}{T}.
\eq{l1_T}
\EALams
Upon defining $\cu \doteq 4\mu/3 + \z$ and $\ck \doteq \k/\cv$, where $\cv
\doteq 3/2$ is the specific heat at constant volume for a 
three-dimensional ideal
gas,
\Eqs{l1_n}--\EQ{l1_T} can be combined to obtain the longitudinal dispersion
relation 
\BE
\l^3 + k^2(\cu+\ck)\l^2 + \(\wp^2 + \Case53 k^2\vt^2 + (k^2\cu)(k^2\ck)\)\l
+ (\wp^2 + k^2\vt^2)(k^2\ck) = 0.
\eq{l3}
\EE
Although a cubic equation has an explicit analytical solution, that
is opaque in general.  Fortunately,
of most interest is the hydrodynamic limit $k\lmfp \to 0$.  Upon noting
that each of~$\mu$ and~$\k$ has the classical random-walk scaling $\vt^2/\nu$,
one can make \Eq{l3} dimensionless by normalizing~$\l$ to~$\nu$, dividing
by~$\nu^3$, and introducing\footnote{This is the square of the \CE\
  expansion parameter used in \App{CE}.}
 $\d \doteq (k\vt/\nu)^2 = (k\lmfp)^2$.  One may
treat~$k^2$ as 
$\Order{\d}$.  For the roots of polynomials with small coefficients, an
efficient and pictorial way of \analyzing\ the dominant balances is to use a
\emph{Kruskal diagram}\cite{asymptotology} in which the terms in the
polynomial populate a 2D lattice whose abscissa measures the powers of~$\l$
and whose ordinate measures the powers of~$\d$.  Dominant balances are
found by bringing up lines from below until they rest on populated points.
The Kruskal diagram for \Eq{l3} is shown in \Fig{Kruskal}.  The balance
between the terms in~$\l^1$ and~$\l^0$ (I shall call that the 1--0 balance)
signifies a \emph{thermal-diffusion mode} with
\BE
\l \approx -k^2\ck = -k^2\k/\cv.
\eq{l_cv}
\EE
  The 3--1 balance leads to
\BE
\l^2 = \wp^2 + \Case53k^2\vt^2 + \Order{\e^2}.
\eq{53}
\EE
These are obviously \emph{plasma oscillations} with real mode frequency $\Omegak
\approx \pm\wp$, but with a thermal correction that
is incorrect in the weakly coupled limit; the proper coefficient
(which follows from collisionless kinetic theory)
is~3 rather than~$5/3$.  
The error arises because in the limit of weak coupling these modes do
not satisfy the Markovian requirement $\abs{\Omegak} \ll \nu$, so one
should not  
be taking the $\w = 0$ limit of the projection formalism.\footnote{This
  point is well known in related contexts.  For example, high-frequency 
  conductivity has been treated thoroughly by \Ref{Dawson-Oberman} and
  \Ref{Dawson_rad}.}
As is well known, the prediction \EQ{53} is easily rationalized on physical
grounds: 
$5/3$~is the ratio of specific heats $\cp/\cv = (d+2)/d$ of a
$d$-dimensional ideal gas for $d = 3$
whereas 
the correct coefficient of~3 corresponds to $d = 1$.  The Braginskii
equations incorrectly assume that strong collisions have isotropized the
wave motion.\footnote{This observation is not made in the
  otherwise excellent  massive
  tome on statistical mechanics by
\Ref{Balescu}, who obtains in his \SECTION 12.7 a result that
  reduces to the incorrect \Eq{53} for weakly coupled plasma.}
  Furthermore, although the Braginskii equations predict
(an incorrect formula for) dissipative collisional damping [not written in
    \Eq{53}], collisionless Landau damping 
is absent.\footnote{\Ref{Hammett-Perkins} discuss a useful method for
  incorporating collisionless effects into fluid equations.}

\FIGURE[0.5]{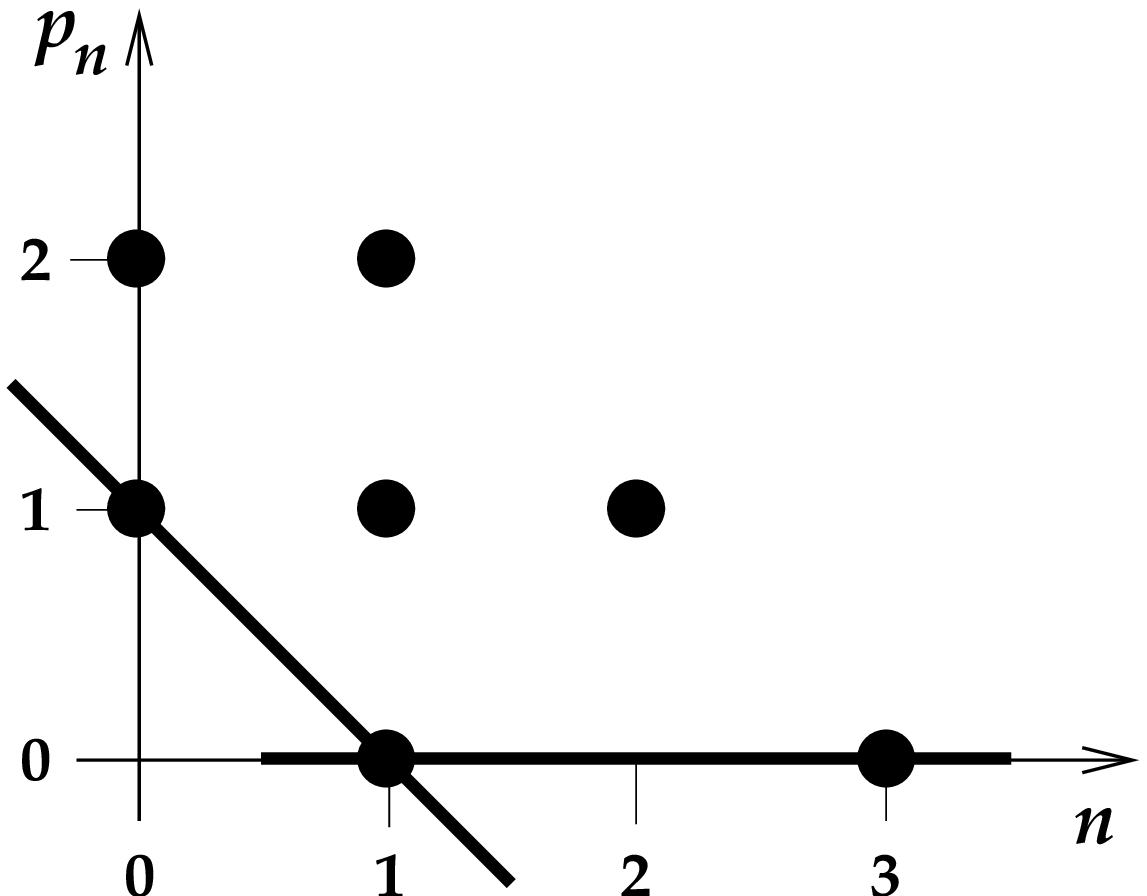}
{Kruskal diagram for the longitudinal modes of an un\magnetize d
  one-component plasma [\protect\Eq{l3}], showing the balance between the
  terms 
  in~$\l^0$ and~$\l^1$ (thermal-diffusion mode), and between~$\l^1$
  and~$\l^3$ (plasma oscillations).}

\FIGURE[0.5]{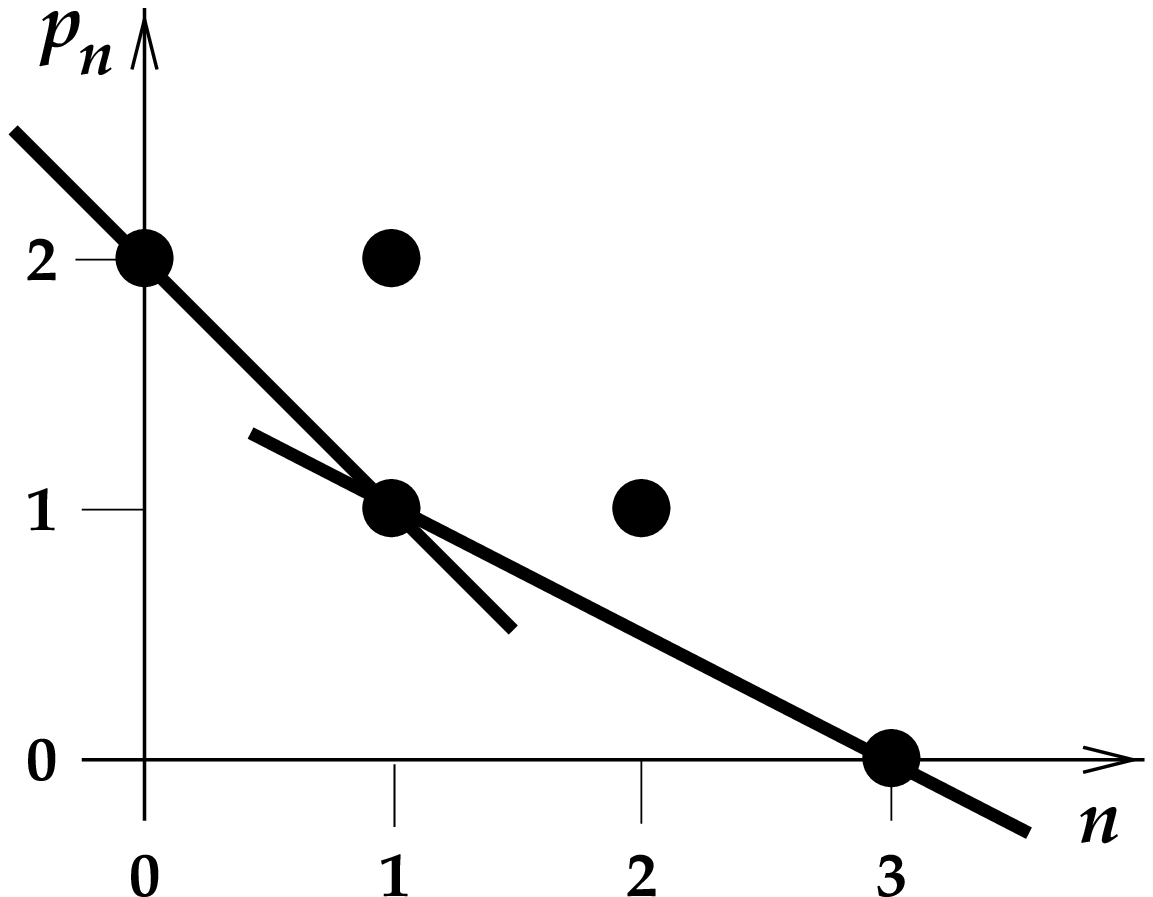}
{Kruskal diagram for the neutral gas, showing that the eigenmodes are a
  thermal-diffusion mode (1--0 balance) and two sound waves (3--1 balance).}

Of course, the plasma oscillations are a consequence of the long-ranged
nature of the Coulomb force.  It is instructive to consider the limit of a
neutral gas by letting $\wp^2 \to 0$.  Then the dominant terms for~$\l^0$
and~$\l^1$ move up to~$\infty$;
the corresponding Kruskal diagram is shown in \Fig{Kruskal_neutral}.  
The 1--0 balance now leads to
\BE
\l = -\Case35 k^2\ck = -k^2\k/\cp,
\eq{l_cp}
\EE
where $\cp \doteq 5/2$ is the specific heat at constant pressure for a
three-dimensional ideal gas. This is again a thermal-diffusion mode, but with
a thermodynamics 
that differs from that of the OCP\@.\footnote{The presences of~$\cv$ in the
plasma formula \EQ{l_cv} and~$\cp$ in the neutral-fluid result \EQ{l_cp} are
easy to understand on physical grounds.  In the neutral fluid, the dominant
balance in the thermal-diffusion mode is $\D n/n \approx -\D T/T$ (\ie, $\D
p \approx 0$); the diffusion of heat occurs at constant pressure.  In the
plasma, the dominant balance is instead between the electrical force and the
pressure force.  Because of the long-ranged nature of the Coulomb interaction,
only a small amount of density fluctuations is required in order to provide
a substantial electric field at long wavelengths; as a consequence, in the
plasma thermal-diffusion mode $\D n /
n \ll \D T/T$.  The density in a volume element of volume~$V$ containing
$N$~particles  is $n = N/V$.  Since no particles are exchanged in a
thermal-diffusion process, one has $\D n/n = -\D V/V$.  Since to lowest
order $\Dn/n =0$, the process occurs at constant volume.  The dominant
balances can be seen in a way that is more physical than the dispersion
relation \EQ{l3} by eliminating~$\D n$ and~$\D T$ in \Eq{l1_u}:
\BE
\l\D u = \Underbrace{\fr{-\ii\wp^2}{k}\fR{-\ii k\D u}{\l}}{{(a)}} - \ii\vt^2 
k\bigg(\Underbrace{\fr{-\ii k\D u}{\l} }{{(b)}}
- \Underbrace{\fr{(2/3)\ii k \D u}{\l + \ck k^2}}{{(c)}}\bigg) - \(\Case43\mu +
\z\) k^2\D u. 
\NN
\EE
The neutral-fluid balance is between terms~(b) and~(c), while the plasma
balance 
is between terms~(a) and~(c).  [In the latter, the $\Order{k^4}$ correction
to $\l = -\ck k^2$ is required in order to balance the leading-order $k\m2$
dependence of each term.]}
  The 3--1 balance
leads to two \emph{sound waves}:
\BE
\l = \pm k\cs,
\EE
where
\BE
\cs^2 \doteq \Case53\vt^2 
\EE
is the ideal-gas limit of the well-known result
\BE
\cs^2 = \fR{\cp}{\cv}\fr{1}{m}\PartiaL{p}{n}_T = \fr{1}{m}\PartiaL{p}{n}_s,
\EE
where $s$~denotes entropy density.  That long-range forces lead to profound
modifications in linear response theory is well known; good discussions are
given by \Ref{Martin_charged,Martin_book}.

\subsection{Eigenmodes of a two-species \magnetize d plasma for
  \texorpdfstring{$\vB =   \v0$}{B = 0}}
\label{eigen_B0}

Next I address the generalization of the previous results to a two-species,
un\magnetize d plasma.  The linearized continuity equations are unchanged in
form.  To the linearized momentum equations must be added the perturbed
momentum transfer:
\BE
\D\vR_e = -(nm)_e\taue\m1\a(\D\vue - \D\vui) - \b\ne\ii\vk\D\Te,
\quad
\D\vR_i = -\D\vR_e,
\EE
where $\a \doteq 0.51$ and $\b \doteq 0.71$.
Finally, to the linearized temperature equations must be added a temperature
equilibration term~$\D Q$:
\BE
\D Q_e = -3\fR{\me}{\mi}\taue\m1(\D\Te - \D\Ti),
\quad
\D Q_i = -\D\Q_e.
\EE
Also, the electron heat flow must be generalized to
\BE
\D\vq_e = -\ne\k_e k^2\ii\vk\D\Te + \b(nT)_e(\D\vue - \D\vui).
\EE
A decomposition into decoupled longitudinal and transverse components can
be made as before.  The transverse dispersion relation leads to two pairs
of eigenvalues, where each pair obeys
\BE
\l^2 + (\nue + \nui + \e_e + \e_i)\l + \e_i\nue + \e_e\nui + \e_e\e_i,
\EE
where $\nue \doteq \a\taue\m1$, $\nui \doteq (\me/\mi)\nue$, and $\e_s
\doteq \mu_s k^2$.  The
approximate solutions are a \emph{momentum-decay mode} (2--1 balance),
\BE
\l_- = -(\nue + \nui) + \Order{\e},
\EE
and a \emph{momentum-diffusion mode} (1--0 balance),
\BE
\l_+ = -\fR{\e_i\nue + \e_e\nui}{\nue + \nui} + \Order{\e^2}.
\eq{l+}
\EE
If one assumes $\Te = \Ti = T$ and writes $\mu_s = C_sT/(m_s\nu_s)$, where
$C_s$~is a constant, this eigenvalue reduces to
\BE
\l_+ = -T\(\fr{C_i}{\me} + \fr{C_e}{\mi}\)\fR{1}{\nue + \nui}k^2,
\EE
describing diffusion with a hybrid viscosity based on (essentially) the
reduced mass and the total collision frequency.

Remaining are six longitudinal eigenmodes, for which I merely quote the
lowest-order results:
\BI
\item \emph{two plasma oscillations} (6--4 balance):  $\l = \pm \ii \wp$, where
$\wp^2 
\doteq \sum_s 
\wps^2$ and $\wps \doteq [4\pi (n q^2/m)_s]\ehalf$;
\item \emph{two ion sound waves} (4--2 balance):  $\l = \pm \ii k\cs$, where $\cs
\doteq 
(Z\Te/\mi)\ehalf$; 
\item a \emph{temperature-equilibration mode} (2--1 balance):  $\l = -2\g$, where
$\g \doteq 2\nu_{ie}$;
\item a \emph{thermal-diffusion mode} (1--0 balance):  $\l = -\half
  (\Sigma_{Te} 
+ \Sigma_{Ti})$, where $\Sigma_{Ts} \doteq 2k^2 \kappa_s/3 = \kappa_s/\cv$.
\EI

\subsection{Eigenmodes of a two-species \magnetize d plasma for
\texorpdfstring{$\nu/\abso{\wc} \ll 1$}{small collisionality}}
\label{eigen_B}

Whereas for $\vB = \v0$ decomposition of~$\vu$ into longitudinal and
transverse components is natural ($\vk$~being the only vector in the
problem other than~$\vu$), for $\vB \neq \v0$ a more useful and physically
meaningful 
decomposition is into compressional and vortical components:
\BE
\Wpar \doteq \ii \kpar\upar,
\quad
\Wperp \doteq \ii \vkperp\.\vuperp,
\quad
\Wcross \doteq \ii (\vkperp\cross\vu)\.\bhat.
\EE
Expressing the algebra in this way helps one to make contact with
predictions of the \GK\ formalism, in which vorticity plays a prominent
role.  In \GK s,\footnote{For an introductory review of \GK s with many
  references, see 
\Ref{JAK_ARFM}.} the perpendicular dielectric constant $\dielperp \doteq
\wpi^2/\wci^2$ is assumed to be large.\footnote{For some discussion of
  various regimes, see \Ref[\SECTION II]{JAK_es}.}
  To introduce~$\dielperp$ naturally,
it is convenient to 
normalize frequencies to~$\wci$ and \wavenumber s to~$\rs \doteq \cs/\wci$.

I shall assume that $\e \doteq \nue/\abs{\wce}$ is small.  The normalized
collision frequencies are then $\nuebar \doteq \nue/\wci =
(\abs{\wce}/\wci)(\nue/\abs{\wce}) = \e/\mu$, $\nuibar \doteq \nui/\wci
= \mu\,\nuebar = \e$.  For an optimal ordering, I shall take $\e = \mu\,\ebar$,
where $\ebar = \Order{1}$ \wrt~$\mu$.
This makes $\nuebar = \Order{1}$ in the mass-ratio ordering; later, one can do
a subsidiary ordering \wrt~$\ebar$.

When considering various limiting cases in the small parameters~$\mu$,
$\e$, and $\d \doteq \kbar^2 \doteq k^2\rs^2$ (where
$k$~refers to either~$\kpar$ or~$\kperp$), it is
important to keep in mind that the order of limits may matter.  For
example, any effect involving~$k^2$ is small relative to the interspecies
collisional relaxation rates as $k \to 0$.  However, to ensure
proper cancellations relating to momentum conservation when
evaluating $\det(\mK)$, one must first express 
all collision rates in terms of a common collision frequency.  Since I am
treating~$\nuebar$ as $\Order{1}$, it 
is appropriate to replace $\nuibar \to \mu\,\nuebar$.
Although the mass ratio~$\mu$ is very small and
the full determinant contains terms of various orders in~$\mu$, most of
which can be neglected, the ordered limit $\lim_{\d \to 0}\lim_{\mu\to 0}$
will produce unusual answers since it assumes that the $k^2$~effects are
large compared to the interspecies relaxation rates.  The proper
hydrodynamic limit is $\lim_{\mu\to 0}\lim_{\d \to 0}$.

Regarding the hydrodynamic limit, note that classical transport assumes the
ordering $k^2\lmfp^2 \sim (\lmfp/L)^2 \ll 1$, where $\lmfp \doteq \vt/\nu$
and $L$~is a characteristic gradient scale length.  In the
large-$\dielperp$ limit, the natural 
dimensionless \wavenumber\ that appears is~$k\rs$. Since $\lmfp \gg \rs$ in
a hot plasma and
$\kperp\lmfp =
(\lmfp/\rs)(\kperp\rs) \gg \kperp\rs$, the requirement $\kperp\lmfp \ll 1$
does not inevitably require $\kperp\rs \ll 1$.  However, that limit is
implied by the assumption that $\kperp L = \Order{1}$ provided that $\rs/L
\ll 1$. 

In general, I construct from the linearized
Braginskii equations a $10\times 10$ matrix that acts on the column vector 
$(\D\ne/\ne,\,
\D\ni/\ni,\,
\Omega_{\parallel e}/\wci,\,
\Omega_{\parallel i}/\wci,\,
\Omega_{\perp e}/\wci,\,
\Omega_{\perp i}/\wci,\,
\allowbreak
\Omega_{\times e}/\wci,\,
\Omega_{\times i}/\wci,\,
\D\Te/\Te,\,
\D\Ti/\Ti)\Tr$.  A complete description of the eigenvalues and eigenvectors
of that matrix for all possible limits in the multidimensional
space of small parameters is beyond the scope of this paper.  Below I shall
merely consider a few illustrative special cases.

\subsubsection{\Magnetize d eigenvalues in the limit of zero dissipation}

When all of the dissipation parameters as well as the gyroviscous stresses
are set to zero (as discussed above, this is 
\emph{not} the hydrodynamic limit because I hold~$\kpar$ and~$\kperp$
finite), and with $\lbar \doteq \l/\wci$, one finds
\BE
\mK =
\begin{pmatrix}
-\lbar & 0 & -1 & 0 & -1 & 0 & 0 & 0 & 0 & 0\\
0 & -\lbar & 0 & -1 & 0 & -1 & 0 & 0 & 0 & 0 \\
\Ehatpar + \kparbar^2 &
 -\Ehatpar & -\mu\lbar & 0 &
0 & 0 & 0 & 0 & \kparbar^2 & 0\\
-\Ehatpar & 
\Ehatpar + \tau\kparbar^2 & 0 & -\lbar &
0 & 0 & 0 & 0 & 0 & \tau\kparbar^2\\
\Ehatperp + \kperpbar^2 &
-\Ehatperp & 0 & 0 & 
-\mu\lbar & 0 & -1 & 0 & \kperpbar^2 & 0\\
-\Ehatperp & 
\Ehatperp + \tau\kperpbar^2 & 0 & 0 &
-0 & -\lbar & 0 & 1 & 0 & \tau\kperpbar^2\\
0 & 0 & 0 & 0 & 1 & 0 & -\mu\lbar & 0 & 0 & 0\\
0 & 0 & 0 & 0 & 0 & -1 & 0 & -\lbar & 0 & 0\\
0 & 0 & -2/3 & 0 & -2/3 & 0 & 0 & 0 & -\lbar
& 0\\ 
0 & 0 & 0 & -2/3 & 0 & -2/3 & 0 & 0 & 0 & -\lbar
\end{pmatrix},
\eq{K0}
\EE
where $\tau \doteq \Ti/\Te = 1$ (the perturbations are around an absolute
equilibrium with a common temperature), $\Ehatpar \doteq
\dielperp(\kpar^2/k^2)$, 
and $\Ehatperp \doteq \dielperp(\kperp^2/k^2)$.  For small~$\mu$,
large~$\dielperp$, and small~$\kperp^2$, the characteristic polynomial is
dominantly 
\BE
(10/3)[\dielperp(\kperp^2/k^2)]\kparbar^2 \lbar^2 +
\dielperp(\kpar^2/k^2)\lbar^4 + \dielperp(\kpar^2/k^2)\lbar^6
+ \mu\l^8 + \mu^3\lbar^{10} = 0.
\EE
Clearly, two eigenvalues vanish; those will be resolved when dissipation is
included.  For the remaining eigenvalues, 
first balance \wrt\ small~$\mu$.  The 10--8 balance gives $\lbar^2 = -\mu\m2$,
which when the normalizations are unwrapped gives $\w^2 = \wce^2$, the
low-density limit of the \emph{upper hybrid wave}.  The 8--6 balance gives
$\lbar^2 = 
-\dielperp(\kpar^2/k^2)$.  These modes are the strongly \magnetize d limit of
the plasma oscillations; in \GK s, they are known as the \emph{$\w_{\rm
  H}$~modes}. 
For the remaining balances, which involve~$\mu^0$, use a subsidiary
ordering \wrt\ small~$\kparbar$.  The 6--4 balance gives $\lbar^2 = -1$,
the ion 
cyclotron wave.  The 4--2 balance gives $\lbar^2 = -(10/3)\kparbar^2$.
These are 
the \emph{ion sound waves}, but with a thermodynamic coefficient that
differs from 
the value of~1 that follows from collisionless kinetic theory.  The issue
here is the same one that was raised in the discussion of the incorrect
thermal correction to the un\magnetize d plasma oscillations [\Eq{53}]; the
Braginskii equations are correct only for the collisional limit $\w \ll \nu$.

\subsubsection{\Magnetize d eigenvalues for purely perpendicular propagation
with no dissipation}

For purely perpendicular propagation, the characteristic
polynomial derived from \Eq{K0} changes to
$
\dielperp\mu\,\lbar^6 + \mu\,\lbar^8 + \mu^3\lbar^{10} = 0
$.
The sound waves have disappeared; four vanishing eigenvalues will be
resolved by dissipation.  Furthermore,
the 8--6 balance is now changed to $\lbar^2 = -\dielperp$ (the order of the
limits $\mu \to 0$ and $\kparbar \to 0$ matters), or in dimensional
variables $\w^2 = \wpi^2$; this is the low-density limit of the
\emph{lower-hybrid waves}.


\subsubsection{\Magnetize d eigenvalues for purely perpendicular propagation
with dissipation}

With all terms included, one finds
$
\mK = (\mK\up{1\dots4}_{10\times 4},\, \mK\up{5,6}_{10\times 2}
,\, \mK\up{7\dots10}_{10\times 4})
$,
where
\BM
\BAams
&\mK\up{1\dots4} \doteq
\NN\\
&\ \begin{bmatrix}
-\lbar & 0 & -1 & 0 
\\
0 & -\lbar & 0 & -1 
\\ 
\Ehatpar + \kparbar^2 &  -\Ehatpar & -\mu(\lbar + \a\nuebar +
\case43\Sigmabar_{0\parallel e}^u + \Sigmabar_{2\perp e}^u) & \mu\a \nuebar
\\
-\Ehatpar &\Ehatpar + \tau\kparbar^2 & \a\nuibar & -(\lbar + \a\nuibar +
 \case43\Sigmabar_{0\parallel i}^u + \Sigmabar_{2\perp i}) 
\\
\Ehatperp + \kperpbar^2 & -\Ehatperp & \mu(\case23\Sigmabar_{0\perp e}^u -
 \Sigmabar_{2\perp e}^u) & 0 
\\
-\Ehatperp &\Ehatperp+\tau\kperpbar^2 & 0 & \case23\Sigmabar_{0\perp i}^u -
 \Sigmabar_{2\perp i}^u 
\\ 
0 & 0 & 0 & 0 
\\
0 & 0 & 0 & 0 
\\
0 & 0 & -\case23(1+\b) & \case23\b 
\\ 
0 & 0 & 0 & -\case23 
\end{bmatrix},
\EAams
\BAams
&\mK\up{5,6} \doteq 
\NN\\
&\ \begin{bmatrix}
 -1 & 0 
\\
 0 & -1 
\\
\mu(\case23\Sigmabar_{0\parallel e}^u - \Sigmabar_{2\parallel e}^u) & 0 
\\
0 &  \case23\Sigmabar_{0\parallel i}^u - \Sigmabar_{2\parallel i}^u 
\\
-\mu(\lbar +\nuebar + \third\Sigmabar_{0\perp e}^u + \Sigmabar_{1\perp e}^u +
\Sigmabar_{2\parallel e}^u) & \mu\nuebar
\\
\nuibar & -(\lbar+\nuibar + \third\Sigmabar_{0\perp i}^u +
\Sigmabar_{1\perp i}^u + 
\Sigmabar_{2\parallel i}^u) 
\\
 1 & 0 
\\
 0 & -1 
\\
 -\case23 & 0 
\\
 0 & -\case23 
\end{bmatrix},
\EAams
\BE
\mK\up{7\dots10} \doteq
\begin{bmatrix}
 0 & 0 & 0 & 0 
\\
 0 & 0 & 0 & 0 
\\
 0 & 0 & (1+\b)\kparbar^2 & 0 
\\
 0 & 0 & -\b \kparbar^2 & \tau\kparbar^2
\\
 -1 & 0 & \kperpbar^2 & 0 
\\
 0 & 1 & 0 &\tau\kperpbar^2
\\ 
 -\mu(\l+\nuebar) & \mu\nuebar & \case32\kperpbar^2\mu\nuebar & 0
\\
\nuibar & -(\l+\nuibar) & 0 & -\case32\kperpbar^2\mu\nuebar
\\
\mu\nuebar & -\mu\nuebar & -(\lbar+\gbar + \Sigmabar_e^T) & \gbar 
\\
0 & 0 &\gbar & -(\lbar + \gbar + \Sigmabar_i^T) 
\end{bmatrix}.
\EE
Here $\nue \equiv \nu_{ei} = \taue\m1$ and $\nui \equiv \nu_{ie}= \mu\nue$;
overlines denote normalizations \wrt~$\wci$;
$\Sigmabar^u$~denotes quantities related to the stress tensor, with the 
numerical subscripts following Braginskii's notation (\App{Viscosity}) and
$\parallel$ or~$\perp$ denoting~$\kpar^2$ or~$\kperp^2$, \eg,
$\Sigmabar_{2\perp e}^u \doteq \kperp^2\mu_{2e}/\wci$ (distinguish the mass
ratio~$\mu$, which is unsubscripted, from the classical
viscosities~$\mu_{ps}$, which are subscripted);  $\Sigmabar^T$~denotes the full
thermal-diffusion coefficient [\eg, $\Sigmabar^T_e \doteq
(2/3)(\kpar^2\k_{\parallel e} + \kperp^2\k_{\perp e})/\wci$]; and $\gbar \doteq
2\nuibar$.

The general dispersion relation for arbitrary~$\kpar/\kperp$ is
complicated, even in the hydrodynamic limit, and will not be discussed
here.  The case of purely perpendicular propagation, however, is
analytically tractable.
With $\kpar = 0$, small~$\mu$, and large~$\Dperp$, one finds 
that the characteristic polynomial is dominantly
\BAams
&2\d^3\a\dielperp\e^3\Sigmabar_{2\perp i}^u(\Sigmabar_e^T +
\Sigmabar_i^T)\kperpbar^2(1+\t)\lbar
+ 2\d^2\a\dielperp\e^2\Sigmabar_{2\perp i}^u(\Sigmabar_e^T +
\Sigmabar_i^T)\lbar^2
\NN\\
&\quad + 2\d\a\dielperp\e^2(2\Sigmabar_{2\perp i}^u + \Sigmabar_e^T +
\Sigmabar_i^T)\lbar^3
+ 4\a\dielperp\e^2\lbar^4
\NN\\
&\quad+ \a\dielperp\e\lbar^5
+ \dielperp\mu\lbar^6
+ \a\e\lbar^7
+ \mu\lbar^8
+ (2+\a)\mu^2\e\lbar^9
+ \mu^3\lbar^{10},
\eq{lperp}
\EAams
\EM
where I have inserted the hydrodynamic ordering parameter~$\d$ to
explicitly remind one of the order of the transport
terms\footnote{Since~$\d$ has been included explicitly, one must
treat~$\kperpbar^2$ as order unity in \Eq{lperp}.}
 in~$\kperpbar^2$. 
One eigenvalue vanishes; that is resolved by viscous dissipation for small
but nonzero~$\kpar$.

As stated above, the proper order of limits is $\lim_{\mu\to 0}\lim_{\d \to
0}$.  With respect to~$\d$, the dominant balances are to set to zero either
the first two lines of \Eq{lperp} (balances~A) or the last term of the
second line plus the last line (balances~B). For the A~balances, all of the
terms fall on a line in a $\d$-ordered Kruskal diagram, so there is no
asymptotic simplification for small~$\d$.
After one factors out
$2\a\dielperp\e^2$ and sets $\d = 1$, the A~balances become
\BE
\e\Sigmabar_{2\perp i}^u(\Sigmabar_e^T + \Sigmabar_i^T)\kperpbar^2(1+\t)\lbar
+ \Sigmabar_{2\perp i}^u(\Sigmabar_e^T + \Sigmabar_i^T)\lbar^2
+ (2\Sigmabar_{2\perp i}^u + \Sigmabar_e^T +
\Sigmabar_i^T)\lbar^3
+ 2\lbar^4 = 0.
\EE
With respect to~$\e$, the dominant balances are 2--1 and 4--3--2.  
The 2--1 balance is
\BE
\lbar = -(1+\t)\kperpbar^2\e
\quad
\hbox{or}
\quad
\l = -(1+\t)\kperp^2\Dperp,
\EE
where $\Dperp \doteq \ri^2\nui = \re^2\nue$; this is the ambipolar
\emph{cross-field density-diffusion mode}.  The 4--3--2 balance factors
into 
\BE
(\lbar + \Sigmabar_{2\perp i}^u)[\lbar + \half(\Sigmabar_e^T +
\Sigmabar_i^T)] = 0, 
\EE
yielding a \emph{momentum-diffusion mode} and a \emph{thermal-diffusion mode}.

The B~balances do not involve~$\d$, so one is free to consider balances
\wrt~$\mu$.  Upon factoring out~$\lbar^4$ and replacing $\e = \mu\ebar$,
one has
\BE
4\a\dielperp\mu^2\ebar^2 
+ \a\dielperp\mu\ebar\lbar
+ \dielperp\mu\lbar^2
+ \a\mu\ebar\lbar^3
+ \mu\lbar^4
+ (2+\a)\mu^3\ebar\lbar^5
+ \mu^3\lbar^6 = 0,
\EE
which contains the balances 1--0, 4--3--2--1, and 6--4.  The 1--0 balance
yields
\BE
\lbar = -4\mu\ebar = -4\e
\quad
\hbox{or}
\quad
\l = -4\nui;
\EE
this is an interspecies \emph{heat-relaxation mode}.  The 4--3--2--1
balance can be further 
ordered \wrt\ large~$\dielperp$, yielding (i)~the 2--1 balance
\BE
\lbar = -\a\ebar
\quad
\hbox{or}
\quad
\l = -\a\nue,
\EE
which is an interspecies \emph{momentum-relaxation mode}; and (ii)~the 4--2
balance 
\BE
\lbar^2 = -\dielperp
\quad
\hbox{or}
\quad
\l^2 = -\wpi^2,
\EE
the \emph{lower-hybrid waves}.  Finally, the 6--4 balance yields
\BE
\lbar^2 = -\mu\m2
\quad
\hbox{or}
\quad
\l^2 = -\wce^2,
\EE
the \emph{upper-hybrid waves}.

For the particular orderings I chose,
the eigenmodes obtained for perpendicular propagation are physically
reasonable; clearly, though, other orderings will lead to different
results.  In the general case, numerical work is essential.  But the
analytical exploration of the linearized Braginskii equations provides a
good example of several important lessons from asymptotic analysis, namely
that the order of limits can matter and that dominant balances in
polynomial equations can be usefully \analyze d in terms of Kruskal diagrams.

\section{Projection operators:  Caveats and further examples}
\label{Caveats}

As shown in the main text, an appropriate choice of projection operator
leads to a systematic derivation of multispecies transport theory, at least
to first order in the gradients.  (See Part~II for a discussion of
second-order effects.)
However, although the projection method is quite powerful, it can be
misused, as I shall show 
in \Sec{nonlocality}--\Sec{Plateau} with several simple examples.
Finally, in \Sec{Brownian} I shall discuss the Brownian test particle in
terms of 
two possible projections.  In this case, there is no misuse of the
formalism; both projections are viable.  Understanding why that is so, in
the face of the caveats discussed in the next three subsections,
should lead one to a deeper appreciation for the overall content and
consistency of the formalism.

\subsection{Projection-operator methods and nonlocality}
\label{nonlocality}

In a 2D vector space with basis vectors 
$
\ve_x \doteq (1,\;0)\Tr
$
and
$
\ve_y \doteq (0,\;1)\Tr
$,
let
$
\vpsi \doteq (\psix,\;\psiy)\Tr
$
obey
\BE
\delt\vpsi + \ii\mL\.\vpsi = 0,
\eq{vpsi_dot}
\EE
where
\BE
\mL = \begin{pmatrix}0 & \ii\\-\ii & 0\end{pmatrix}.
\EE
(Note that the eigenvalues of~$\mL$ are~$\pm 1$, indicating oscillation.)
The resulting dynamical system is
\BE
\dot\psi_x = \psiy,
\quad
\dot\psi_y = -\psix.
\EE
These can be combined into the wave equation
\BE
\ddot\psi_x + \psix = 0,
\eq{ddot}
\EE
so the variables oscillate sinusoidally with unit frequency.

Let us try to recover
\Eq{ddot} by projecting \Eq{vpsi_dot} into the $x$~direction.  That can be
accomplished by introducing the projection operator
\BE
\P \doteq \ve_x\otimes\ve_x\Tr = 
\begin{pmatrix}1 & 0\\0 & 0\end{pmatrix}.
\EE
The standard equations apply:
\BALams
\delt\P\vpsi + (\P\ii\L\P)\P\vpsi &= -(\P\ii\L\Q)\Q\vpsi,
\eq{P_psi_dot}
\\
\delt\Q\vpsi + (\Q\ii\L\Q)\Q\vpsi &= -(\Q\ii\L\P)\P\vpsi.
\eq{Qvpsi_dot}
\EALams
Simple
calculation shows that the frequency operator $\P\L\P$ vanishes identically
for the present problem, so in spite of its name
it is unrelated to the natural oscillation.
Furthermore, although the standard procedure is to eliminate~$\Q\vpsi$ by
introducing Green's function $\UQ(\t) \doteq H(\t)\exp(-\Q\ii\L\Q\t)$, so
\BE
\Q\vpsi(t) = -\I0t \dd\t\,\UQ(\t)\Q\ii\L\P\vpsi(t-\t),
\EE
it is easy to
show that in the present problem $\Q\L\Q \equiv 0$.  Thus, $\UQ(\t)$~does not
decay in time, 
precluding the possibility of a Markovian description.  Indeed, upon noting
that
\BE
\Q\L\P = (1 - \P)\L\P = \L\P = \begin{pmatrix}0 & 0\\-\ii & 0\end{pmatrix},
\EE
one finds that
\BE
\Q\vpsi(t) = -\I0{t}\dd\tbar\,\begin{pmatrix}0\\\psix(\tbar)\end{pmatrix};
\EE
then, with
\BE
\P\L\Q = \P\L(1 - \P) = \P\L = 
\begin{pmatrix}0 & \ii\\
0 & 0\end{pmatrix},
\EE
one finds that \Eq{P_psi_dot} becomes
\BE
0 = \Partial{}{t}\begin{pmatrix}\psix(t)\\0\end{pmatrix}
+ \I0t \dd\tbar\,\begin{pmatrix}\psix(\tbar)\\0\end{pmatrix}.
\EE
The desired $x$~component can be extracted by dotting with~$\ve_x\Tr$.  Of
course, the 
nonlocal integro-differential equation that results is
equivalent to the second-order differential equation \EQ{ddot}. 

This trivial example shows that it is not inevitable that a projected
description must
be time-local or Markovian.  Note that the original vector system
\EQ{vpsi_dot} of 
coupled ordinary differential equations (ODEs) is local in time;
time-history integration is introduced as a consequence 
of the projection.\footnote{This phenomenon is well known in the theory of
  classical Brownian motion\cite{Wang-Uhlenbeck}.}
This is actually typical \behavior\ when the system supports linear waves.
Very special conditions must hold in order that a projected system is
Markovian.  At least, what is required is that Green's function $\UQ(\t)$ (an
operator) 
decays sufficiently rapidly in time when acting on the orthogonal subspace
[as is required for \Eq{Qvpsi_dot}],
a property that is related to the 
spectrum of $\Q\L\Q$.  Note that $\Q\L\Q$ will always possess a null space
that is at least $p$-dimensional, where $p$~is the dimension of the
projected subspace, since $(\Q\L\Q)\P = 0$.  But if it also possesses a
zero or very small eigenvalue in the $\Q$~direction
(as in the present example), some sort of nonlocality must ensue.  It is
for this 
reason that for applications to fluid equations $\Q$~is chosen to project
into the directions orthogonal to the null space of the collision
operator.\footnote{Projection into the directions orthogonal to the null
  space of the collision operator is necessary but not sufficient.  When
  that operator is the linearized Landau operator (or the linearized
  Boltzmann operator for neutral fluids), it is clear from the
  discussion in \Sec{spectrum} that perturbations in the orthogonal
  directions decay rapidly.  But the famous example of the long-time
  algebraic tails 
  of correlation functions [$C(\t) \sim \t\m{d/2}$, where $d$~is the
  dimension of space] discovered by \Ref{Alder-Wainwright} shows that
  those familiar operators omit essential physics.  Specifically,
  collisional processes at the microscopic level excite long-lived
  hydrodynamic excitations, and the nonlinear mode coupling of those
  fluctuations leads to a slow component that does not lie in the null
  space of the standard operators.  In \magnetize d plasmas, the phenomenon
  is called the generation of convective cells and was treated by
  \Ref{JAK_cells}.  [For a discussion of the neutral-fluid problem, see
 \Ref[Sec.~21.5]{Balescu}, \Ref[Sec.~S11.A]{Reichl2},
\Ref[Chap.~9]{Zwanzig}, and the references listed in \Ref{JAK_cells}.] 
One method of attack is to augment the dimensionality
  of the standard hydrodynamic projector~$\P$ to include the extra slow
  directions, as was done by \Ref{JAK_cells}.  The resulting effect on the
  transport coefficients is 
  particularly severe in 2D, where the transport is distinctly nonlocal.
  (An attempt at a Markovian description leads to divergent transport
  coefficients, $\eta \sim \InT \dd\t\,\t\m1 \propto
  \lim_{\t\to\infty}\ln\t$.)}

\subsection{Projection operators and the dispersion relation of Langmuir
  oscillations}
\label{Langmuir}

One of the wonderful yet dangerous features of the
projection-operator 
formalism is that one can project onto virtually anything.  For the case of
collisional transport, the choice of projection operator is relatively
obvious.  In general, however, the formalism can be
confusing because the zero-frequency limit is not allowed for
high-frequency modes and because of the appearance of~$\Q$ in the relevant
Green's function.

I remarked in \Sec{eigen_OCP} that the Braginskii equations predict an
incorrect dispersion relation for
the Langmuir oscillations $\w = \Omegak \approx \pm\wp$ as a consequence of
a violation of the Markovian approximation by those high-frequency modes.
Before the Markovian approximation is made, however, a
projection-operator formalism must produce a formally correct result, no matter
into what space the dynamics are projected.  The proper dispersion relation,
including both reactive and dissipative parts, must follow from a correct
evaluation of the frequency-dependent $\mSigmahat(\w)$.  I shall
demonstrate this by \analyzing\ the projection into the density subspace for
the well-known model problem of linearized, collisionless Vlasov response.
Note that for a high-frequency mode the null eigenspace of the collision
operator is no longer relevant and cannot be used to motivate a useful
projection operator.  Projecting into the 1D density subspace is the
simplest operation that can be done, yet it demonstrates some nontrivial
manipulations. 

Thus, consider\footnote{A closely related version of this calculation is the
  discussion of \Ref[App.~F]{JAK_thesis}, which treats the short-time limit
  of the two-time correlation function of the many-body plasma.}
$
\P = \ket 1>\bra 1|
$
and project the perturbed kinetic equation \EQ{Df_dot}, assuming a
collisionless electron plasma with neutralizing ion background.  The
frequency matrix 
involves the single matrix element $\<1 | \vv | 1> = \v0$.  The
electric-field term vanishes under~$\P$ because it is a perfect derivative
in velocity space.  Thus, the perturbed density
evolves according to
\BE
-\ii\w\fR{\D \nhat(\vkw)}{n} + \Sigmahat(\vkw)\(1 +
\fr{1}{k^2\lDe^2}\)\fR{\D\nhat(\vkw)}{n} = 
\fr{\D n(0)}{n}, 
\eq{Dnhat_dot}
\EE
where\footnote{In this discussion I shall use the notation $\one$ instead
  of~$1$ for the identity operator in order to avoid confusion with scalar
  functions of velocity.}
\BE
\Sigmahat(\vkw) \doteq \vk\.\<\vv \Q | [-\ii(\w\one - \Q\vk\.\vv\Q +
  \ii\e\one)]\m1 | 
\Q\vv>\.\vk.
\eq{Sigmahat_Losc}
\EE
The term in $(k\lDe)\m2$ arises from the $\Q$~projection of the $\D\vE$
term in \Eq{Df_dot} upon using Poisson's 
equation to express $\D\vE$ in terms of the electron charge density.
In \Eq{Sigmahat_Losc}, 
the left-most and right-most $\Q$'s may be omitted if desired because $\Q =
\one - \P$,
$\P\ket\vv> = \v0$, and $\bra\vv|\P = \v0$; for the same reason, it is
sufficient to write $\Q\vk\.\vv\Q = \Q\vk\.\vv$.  The matrix element, a
second-rank tensor, depends only on~$\vk$, so it must have
the form
\BE
\<\vv | [-\ii(\w\one - \Q\vk\.\vv\Q + \ii\e\one)]\m1 | \vv> = a(k)\khat\khat +
b(k)(\mone - 
\khat\khat).
\EE
Only the $\khat\,\khat$ term contributes to \Eq{Sigmahat_Losc}, so one must
evaluate 
\BE
\Sigmahat \doteq \<\vk\.\vv | [-\ii(\w\one - \Q\vk\.\vv + \ii\e\one)\m1 |
  \Q\vk\.\vv>. 
\EE
Addition and subtraction of~$\w$ to the right-most ket leads to
\BE
\Sigmahat = -\ii\w\<\vk\.\vv | (\w\one - \Q\vk\.\vv + \ii\e\one)\m1>.
\eq{Sigmahat}
\EE
The term in angular brackets is similar, though not identical, to the
integral that defines the electrostatic susceptibility:
\BE
\chio(\vkw) \doteq -(k\lDe)\m2 J(\vkw),
\EE
where
\BE
J(\vkw) \doteq \BIG\<\fr{\vk\.\vv}{\w - \vk\.\vv+\ii\e}>.
\EE
To simplify \Eq{Sigmahat}, use $\Q = \one - \P$ and define the operators
\BE
\A \doteq (\w - \vk\.\vv)\one,
\quad
\B \doteq \P\vk\.\vv
\EE
so that
\BE
J = \<\B\A\m1>.
\EE
Then
\BE
\Sigmahat/(-\ii\w) = \<\B(\A+\B)\m1>.
\EE
With the aid of the operator identity \EQ{ABm1}, one finds
  $\Sigmahat = -\ii\w I$, where
\BE
I \doteq \<\B(\A+\B)\m1> = \<\B\A\m1> -
\<\B\A\m1\B(\A+\B)\m1>. 
\eq{I_def}
\EE
Since $\B = \ket 1>\bra\vk\.\vv|$, the last term factors:
\BE
\<\B\A\m1\B(\A+\B)\m1> = \<\B\A\m1>\<\B(\A+\B)\m1> = JI.
\EE
Thus, \Eq{I_def} becomes
\BE
I = J - JI,
\EE
the solution of which is
\BE
I = J/(1+J).
\eq{I}
\EE
Upon defining the electrostatic dielectric function as
\BE
\diel(\vkw) \doteq 1 +\chio(\vkw) = 1 - J(\vkw)/(k^2\lDe^2),
\EE
one readily obtains the solution to \Eq{Dnhat_dot} with the aid of \Eq{I}:
\BE
\fr{\D\nhat(\vkw)}{n} = \fR{1+J}{-\ii\w\diel}\fr{\D n(0)}{n}.
\EE
Since one has
\BE
1 + J = 1 + \BIG\<\fr{\vk\.\vv}{\w - \vk\.\vv + \ii\e}> = \w\BIG\<\fr{1}{\w -
  \vk\.\vv + \ii\e}>,
\EE
the final result is
\BE
\fr{\D\nhat(\vkw)}{n} = \fr{1}{\diel(\vkw)}\(\Int \dd\vv\,\fr{\fM(\vv)}{-\ii(\w -
  \vk\.\vv + 
  \ii\e)}\)\fr{\D \nhat(\vk,t=0)}{n}.
\EE
This is to be compared with the exact solution of the linearized Vlasov
problem:
\BE
\D\nhat(\vkw) = \fr{1}{\diel(\vkw)}\nbar\Int \dd\vv\,\fr{\D
  \fhat(\vv,\vk,t=0)}{-\ii(\w - 
  \vk\.\vv + \ii\e)}.
\EE
As must be so, the results agree when the initial perturbation is chosen to
have the form 
\BE
\D \fhat(\vv,\vk,t=0) = \fR{\D \nhat(\vk,t=0)}{\nbar}\fM(\vv)
\EE
(\ie, when only a density perturbation is imposed initially).  The
contribution from the vertical part of~$\D f(0)$ is contained in the
propagated initial-condition term that was ignored in the elimination of
$\Q\ket\chi>$.  That is, the system was prepared to lie in the density
subspace.

Results such as \Eqs{ABm1} or \EQ{I} show that one must be very careful to
not confuse 
unmodified propagators with ones modified with the orthogonal
projector~$\Q$.  Although here I was working with the single-particle
propagator, similar relations arise in many-body physics, where the
$N$-particle Liouville propagator arises; that was the original situation
discussed by \Ref{Mori}.  Some further interpretations and generalizations
of the formalism are discussed in appendixes~D and~E of \Ref{JAK_thesis}.

\subsection{The plateau phenomenon; unmodified \vs\ modified propagators}
\label{Plateau}

According to \Ref{Kubo57}, linear response functions can be couched
as two-time correlations of time-dependent currents for which the time
dependence is 
induced by the Liouville propagator $\exp(-\ii\cL t)$.  However, the transport
coefficients discussed in the present article involve currents defined with
the modified propagator\footnote{I did not work with the Liouville
  operator in the present paper.  That is done in Part~II, where the
  salient results of Part~I are recovered from the \gspace\ formalism.}
 $\exp(-\ii\Q\cL\Q t)$.  The subtle but crucial
difference relates to the distinction between `internal' and `external'
response functions and to a \behavior\ known as the \emph{plateau
  phenomenon}\cite[and references therein]{Berne,Mazenko73}.  Here I
provide a brief introduction. 

The simplest illustration involves the familiar stochastic Langevin equation
for the random momentum~$\Tilde{p}$ of an un\magnetize d test particle of
mass~$M$: 
\BE
\Total{\Tilde{p}}{t} + \nu\Tilde{p} = \Tilde{f}\Ext(t),
\eq{L0}
\EE
where 
$\Tilde{f}\Ext$~is taken to be \centre d Gaussian white noise with
covariance\footnote{The 
abbreviation ext stands for external.  It refers to the fact that
physically $\Tilde{f}\Ext$  arises from the random motions of the bath
particles, which are external relative to the identity of a test particle.}
\BE
F\Ext(t,t') \doteq \<\Tilde{f}\Ext(t)\Tilde{f}\Ext(t')> = 2\Dp\Dirac{t-t'};
\EE
the constant~$\Dp$~is the momentum-space diffusion coefficient [such that
at short times $\nu t \ll 1$ the mean-square momentum
fluctuations obey $\<\d p^2>(t) = 2\Dp t$].  It is well
known and can be easily proven from \Eq{L0} that on the collisional time
scale the fluctuation level saturates at 
the level $\<\d p^2> = \Dp/\nu$.  The \sstate\ 
balance between kinetic energy and the thermal energy of a bath at
temperature~$T$ then
leads to the Einstein relation $\Dp/M\nu = T$. 

The existence of that \Sstate\ leads to a peculiar but important property
of the two-time correlation function~$F(t,t')$ of the total force
\BE
\Tilde{f} \doteq -\nu\Tilde{p} + \Tilde{f}\Ext.
\EE
From
\BE
\<\d p^2>(t) = \I0t \dd\tbar\I0t \dd\tbar'\,F(\tbar,\tbar'),
\EE
one can calculate a running `total diffusion coefficient' according to
\BE
D\Tot_p(t) \doteq \Half\Total{\<\d p^2>}{t} = \I0t \dd\taubar\,F(\taubar).
\EE
The fact that $\dd\<\d p^2>/\dd t \to 0$ as $t \to \infty$ implies that
rigorously 
\BE
\InT \dd\taubar\,F(\taubar) = 0.
\eq{InT_F}
\EE
Since it is easy to see that $F(\t) \to F\Ext(\t)$ as $\t \to 0$,
\Eq{InT_F} implies that $F(\t)$~possesses a \emph{long, negative tail}
whose integrated contribution exactly cancels the diffusive contribution
embodied in~$F\Ext$.  This can be seen explicitly by direct calculation,
which shows that
\BE
F(\t) = 2\Dp\(\Dirac{\t} - \Half\nu \ee^{-\nu\t}\).
\eq{F_tail}
\EE
This function and its time integral are illustrated in \Fig{neg_tail3}.

\FIGURE[0.8]{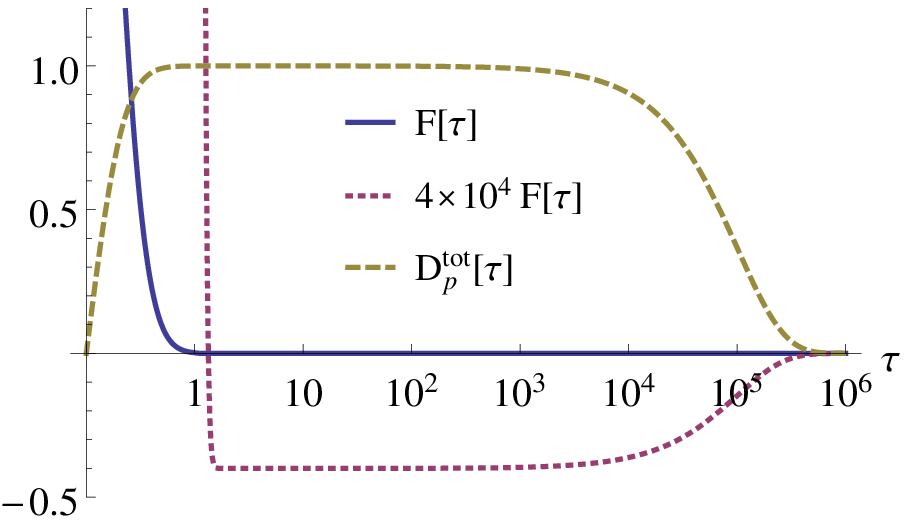}{(colour online)  Solid curve:  the function $F(\t)$
  [\protect\Eq{F_tail}] for 
  $\Dp = 1$; dotted curve:  $4\times 10^4F(\t)$ (amplified so as to make
  the shape of the negative tail visible); dashed curve:  the running
  diffusion coefficient $\Dp\Tot(\t)$, showing that a plateau forms
  after a few autocorrelation times and that the total area under~$F(\t)$
  goes to zero after a few collision times.    The delta function in
  \Eq{F_tail} 
has been opened up to be a Gaussian with standard deviation $\s =
  (2/\pi)\ehalf/3$, so that one unit in~$\t$ corresponds to 3 microscopic
  correlation times~$\tac$.  In these units, $\nu$~is chosen to
  be~$10\m5$.  The $\t$~axis is linear for $\t \le 1$ and logarithmic for
  $\t > 1$.}

The integral \EQ{InT_F} is the one-sided Fourier transform of~$F(\t)$
evaluated at $\w = 0$.  The one-sided transform of \Eq{F_tail} is
\BE
\Fhat(\w) = \Dp - \Dp\fr{\nu}{-\ii(\w + \ii\nu)},
\EE
or, upon dividing by~$MT$ and invoking the Einstein relation,
\BE
\Phat(\w) \doteq \Fhat(\w)/(MT) = \nu - \nu[-\ii(\w + \ii\nu)]\m1\nu,
\eq{Phat_nu}
\EE
where the last term has been written in a way that will be easy to compare
with the
more general formula \EQ{Phat_gen} derived later. Another interesting
form is 
\BE
\Phat\m1(\w) = \nu\m1 + (-\ii\w)\m1.
\eq{Phat_inverse}
\EE
From either \Eq{Phat_nu} or \Eq{Phat_inverse}, one finds
\BE
\Phat(0) = 0
\eq{Phat(0)}
\EE
as a signature of the long-time tail.

More generally, the constant~$\nu$ could be replaced by a
frequency-dependent relaxation rate~$\nuhat(\w)$ with $\nuhat(0) = \nu$.
Then the message is that whereas $\nuhat(\w)$ approaches a nonzero limit as
$\w \to 0$, $\Phat(\w)$ vanishes in that limit.

The physical difference between~$\nuhat(\w)$ and $\Phat(\w)$ is that
$\nuhat(\w)$ describes an \emph{internal polarization} process whereas
$\Phat(\w)$ describes the \emph{total response} to an external
perturbation. 
This distinction is crucial to maintain for all response processes; a
lengthy pedagogical article that includes a discussion of conductivity is
by \Ref{Kubo_response}. 

The representations \EQ{Phat_nu} or \EQ{Phat_inverse} might suggest that
it is possible to extract the 
relaxation coefficient~$\nu$ (or~$\Dp$, \via\ the Einstein relation) from
the \emph{high}-frequency limit of~$\Phat(\w)$:  
\BE
\nu = \lim_{\w\to\hbox{\small`}\infty\hbox{\small'}}\Phat(\w),
\EE
where the inverted commas remind one that $\w\m1 > \tac$,
where $\tac$~is the microscopic autocorrelation
time [the physical width of the delta function in \Eq{F_tail}], since the
Langevin model is not valid for $\t < \tac$.
In the time domain, the statement is that $\Dp\Tot(t) \approx \Dp$ for
$\tac < t \ll \nu\m1$.   This is a
consequence of the plateau \behavior\ in which $\Dp\Tot(t)$ quickly rises to
an essentially constant value, then slowly falls off as the effects of the
negative tail 
manifest.  However, this works only for frequency-independent~$\nu$.  More
generally, it is better to find $\nu = \lim_{\w\to 0}\nu(\w)$ from the
subtracted form
\BE
\nu\m1 = \lim_{\w \to 0}[\Phat\m1(\w) - (-\ii\w)\m1].
\eq{nu_m1}
\EE
A generalization of this result is useful for multispecies plasmas, which
contain interspecies equilibration phenomena that occur on the collisional
timescale.

Now consider the general projection-operator result for the
frequency-dependent hydrodynamic transport matrix:\footnote{In the
  following manipulations, it is easiest to use~$\cL$ everywhere, although
  the outermost~$\cL$'s in the scalar products may be replaced by~$\L$
  [recall the definitions of~$\cL$ and~$\L$ given in \Eq{LL}].}
\BE
\mSigmahat(\w) \doteq \< \vA|\cL\Q\GQhat(\w)\Q\cL|\vA>\.\mM\m1
\EE
and the corresponding function $\Phat(\w)$ defined with~$\U(\w)$ instead
of~$\UQ(\w)$.  These matrices are related algebraically in a
natural generalization of \Eq{Phat_nu}.  To show this, note that
\BE
\Ghat(\w) = [-\ii(\w -\cL + \ii\e)]\m1,
\EE
write $\cL = \P\cL + \Q\cL$, and invoke the identity \EQ{ABm1} with $\A \doteq
-\ii(\w - \Q\cL)$ and $\B \doteq \ii\P\cL$ to find
\BE
\vPhat = \mSigmahat - \<\vA|\cL\Q\GQhat\ii\P\cL \Ghat\Q\cL|\vA>\.\mM\m1.
\eq{Phat_Sigmahat}
\EE
The last term, \sans\ minus sign, is explicitly
\BE
\<\vA|\cL\Q\GQhat\ii|\vA\Tr>\.\mM\m1\.\<\vA|\cL\Ghat\Q\cL|\vA\Tr>\.\mM\m1.
\eq{last_term}
\EE
In the first matrix element of \Eq{last_term}, write
\BE
[-\ii(\w - \Q\cL)]\m1 = (-\ii\w)\m1\mone - (-\ii\w)\m1[-\ii(\w -
  \Q\cL)]\m1\ii\Q\cL. 
\EE
The first term does not contribute because $\Q\ket\vA> = 0$; the second
term reproduces~$\mSigmahat$.  Thus,
\BE
\vPhat = \mSigmahat - \mSigmahat\.\mChat,
\EE
where
\BE
\mChat \doteq (-\ii\w)\m1\<\vA|\cL\Ghat\Q\cL|\vA\Tr>\.\mM\m1.
\EE
Insert the identity $\P + \Q = 1$ after the first~$\cL$.  The $\P$~term
introduces the frequency matrix, and the $\Q$ term reproduces~$\vPhat$:
\BE
\mChat = (-\ii\w)\m1\mOmega\.\<\vA|\Ghat\Q\cL|\vA\Tr>\.\mM\m1
+ (-\ii\w)\m1\vPhat.
\eq{mChat2}
\EE
Write
\BE
\Ghat = (-\ii\w)\m1 - (-\ii\w)\m1\Ghat\ii\cL.
\EE
Again the first term does not contribute; the second term
reproduces~$\mChat$.  Upon solving \Eq{mChat2} for~$\mC$, one finds
\BE
\mChat = [-\ii(\w\mone - \mOmega)]\m1\.\vPhat;
\EE
then solving \Eq{Phat_Sigmahat} for~$\vPhat$ leads to 
\BE
\vPhat  = \mSigmahat - \mSigmahat\.[-\ii(\w\mone - \mOmega +
  \ii\mSigmahat)]\m1\.\mSigmahat.
\eq{Phat_gen}
\EE
This generalizes \Eq{Phat_nu}.  
If the matrices are invertible, \Eq{Phat_gen} can be written as
\BE
\vPhat\m1 = \mSigmahat\m1 + [-\ii(\w\mone - \mOmega)]\m1,
\eq{Phat_m1}
\EE
which generalizes \Eq{Phat_inverse}.

An \alternate\ way of writing \Eq{Phat_gen} is
\BE
\vPhat = \mSigmahat\.[-\ii(\w\mone - \mOmega +
\ii\mSigmahat)]\m1\.[-\ii(\w\mone - \mOmega)]. 
\eq{Phat_alt}
\EE
Upon using the property that the determinant of a product is the product of
the determinants, one finds
\BE
\det(\vPhat) = [\det(\mSigmahat)][\det\oF{-\ii(\w\mone - \mOmega +
    \ii\mSigmahat)}]\m1 [\det\oF{-\ii(\w\mone - \mOmega)}].
\EE
Thus, $\det\oF{\vPhat(\w_i)} = 0$ for any eigenvalue~$\w_i$ of~$\mOmega$.  
This generalizes \Eq{Phat(0)}.  (For the classical Langevin problem, the
frequency vanishes,\footnote{A well-known generalization is the
  harmonically bound Brownian particle, which does have a
  nonzero~$\Omega$.} 
 so $\vPhat$~vanishes at $\w = 0$.)  

Although the transport processes are naturally represented by the
long-wavelength, low-frequency limit of~$\mSigmahat$,
it is also possible to extract them from~$\vPhat$ if one is careful.
The
Langevin example shows that $\vPhat(0)$~is unrelated to~$\mSigmahat(0)$ for
the special case of vanishing frequency matrix, and the scalar version of
\Eq{Phat_alt} for nonzero~$\Omega$ shows that $\vPhat(0)$~has at best a
peculiar relation to~$\mSigmahat(0)$.  To understand how to proceed, first
consider the classical neutral-fluid or OCP cases, for which $\mSigmahat(0) = 
k^2\mD$ (where $\mD$~is the matrix of transport coefficients) and
$\mOmega \propto k$.  It follows that
\BE
\mD = \lim_{k,\w\to 0}k\m2\mSigmahat_k(\w) = \lim_{\w\to 0}\lim_{k\to
0}k\m2\vPhat_k(\w). 
\eq{ordered}
\EE
The order of limits is immaterial for~$\mSigmahat$, but the $k \to 0$ limit
must be taken first for~$\vPhat$.  [In \Eq{Phat_alt}, the $k\to 0$ limit
removes the~$\mOmega$ and the second~$\mSigmahat$.  The~$\w$'s then cancel,
after which the $\w \to 0$ limit evaluates $k\m2\mSigmahat(\v0,\w)$ at $\w =
0$.]
That the ordered limit must be taken for~$\vPhat$ agrees with
the results of \Ref{Kadanoff-Martin}.

This discussion clarifies an argument used by \Ref{Brey}.  They
work in the space--time domain rather than with Fourier transforms.  They
derive an expression equivalent to \Eq{Phat_Sigmahat}, then
assert\footnote{Actually, they expand~$\mSigmahat$ in terms of~$\vPhat$, so
the roles of~$\GQ$ and~$\G$ are reversed.}
 that the
term $\bra \vA |\L\Q\GQ\ii\P$ is negligible through second order in the
gradients because (when $\L$~is proportional to a gradient) it
involves~$\GQ$ evaluated to zeroth order in the 
gradients (namely the identity operator) and $\Q\P = 0$.  Thus, they are
first taking the limit of small gradients (\ie, $k \to 0$).  Then they assert
that `the time integral can be extended to infinity' (\ie, they take the
$\w \to 0$ after passing to the limit of infinite system size).

The arguments of Brey \etal\ fail in the multispecies case where
$\L$~contains the 
collision operator, first because $\Chat$~does not tend to zero with the
gradients, second because $\Q \Chat\Q \neq \Chat$.  Thus, the
un\magnetize d Braginskii 
transport coefficients involve $(\Q\Chat\Q)\m1$ rather than $\Chat\m1 = \InT
\dd \taubar\,\ee^{-\Chat \taubar}$.   

In the multispecies case, the frequency matrix contains terms that
are of the order of 
the collision frequency and do not vanish with~$k$. That would not be a
problem if $\mSigmahat_\vk$ would vanish with~$k$, in which case the
ordered limit \EQ {ordered} would still work.  However, we know from the
discussion at the end of \Sec{multispecies} that the physics of the
high-energy tail is contained in~$\mSigmahat_\vk$, and the contribution of
that tail to the effective collision frequency does not vanish with~$k$.
Thus, one has a
situation analogous to the Langevin model discussed at the beginning of the
section.  One could then resort to the generalization of
\Eq{nu_m1}, which from \Eq{Phat_m1} is
\BE
\mD\m1 = \lim_{\w,k\to 0}k^2\{\vPhat\m1 - [-\ii(\w\mone - \mOmega)]\m1\},
\EE
provided that the matrices are invertible.

\subsection{Projection-operator analysis of the Brownian test particle}
\label{Brownian}

Analysis of the long-time statistical dynamics of the classical Brownian
test particle by projection-operator techniques provides probably the
simplest nontrivial example of the methodology and also provides further
insights into the interactions between null spaces and multidimensional
projections.  To the stochastic momentum equation \EQ{L0}, adjoin
\BE
\Total{\xt}{t} = \Tilde{v} = \Tilde{p}/M.
\EE
It is then well known that for times much longer than the collision
time~$\nu\m1$ \xspace\ diffusion ensues:  
\BE
\<\d x^2> \to 2Dt,
\quad
\hbox{where}
\quad
D = \vT^2/\nu.
\eq{dx2}
\EE
The exact solution of the (linear) Langevin system can also be obtained for
all times from one of two equivalent methods:  (i)~Recognize that~$\xt$
and~$\Tilde{v}$ are jointly Gaussian because they solve linear ODEs driven
by Gaussian noise, then calculate the conditional means and
variances by directly solving the ODEs and appropriately averaging the
solutions; (ii)~solve the equivalent \FPE
\BE
\Partial{f(x,v,t)}{t} + v\Partial{f}{x} = -\Chat f,
\eq{Brownian_FPE}
\EE
where $\Chat$~is given by \Eq{Chat_i}.\footnote{\Equation{Brownian_FPE} is
a special case of a class of PDEs that can be solved exactly, yielding a
joint Gaussian, as discussed by \Ref[\SECTION VIII.6]{van_Kampen}.  The key
features are (i)~the 
first-order terms are linear in the independent variables, and (ii)~the
second-order terms involve only constant coefficients.}

\subsubsection{Projection into the density subspace}

First consider how the \xspace\ diffusion result \EQ{dx2} emerges from the
natural hydrodynamic projection-operator formalism.  It is not difficult to
show that $\Chat$~possesses a single null eigenvalue, with eigenvector
$\ket\He_0> = \ket 
1>$ associated with probability (or density) conservation.  This motivates
the projection of 
\Eq{Brownian_FPE} onto the 1D density subspace, namely $\P = \ket 1>\bra
1|$.  The frequency `matrix' $\Omega = k\<1 | v | 1>$ vanishes by
symmetry (or equivalently by the orthogonality of~$\He_0$ and~$\He_1$).
Because the 
eigenvalues of~$\Chat$ are the Hermite polynomials and the $\Q$~projection
excludes the null space, the modified propagator $\exp(-\Q\Chat\Q t)$
decays on the collisional timescale, so for times longer than~$\nu\m1$ the
Markovian approximation is valid and one may use the specialization of the
result \EQ{meta_cov} to a single density diffusion coefficient.  One
obtains
\BE
\delt n + k^2 D n = 0,
\EE
where
\BE
D = \<v \Q|(\Q\Chat\Q)\m1|\Q v> = \<v | \Chat\m1 | v>
\eq{D_model}
\EE
and I used the facts that $\P\Chat = \Chat\P = 0$ and $\Q\ket v> = (1 -
\P)\ket v> = \ket v>$.\footnote{Note that the last form of \Eq{D_model}
  does not follow from the previous one by setting $\Q\m1\Q = 1$, since
  projection operators other than the identity are not invertible (as
  follows from the property $\Q^2 = \Q$).}
  The solution of the differential equation
\BE
\Chat\ket\psi> = \ket v>
\EE
is easy to obtain and is just $\ket\psi> = \ket v>/\nu$; then
\BE
D = \<v | v>/\nu = \vT^2/\nu,
\eq{D}
\EE
in agreement with the known result \EQ{dx2}.  

\subsubsection{Projection into a multidimensional subspace}

Suppose that one does not recognize the one-dimensional nature of the null
space.  What would happen if one would use a higher-dimensional projection?
As an 
example, I shall consider a 2D projection into the density and velocity
subspaces.  Namely, with velocities normalized to~$\vT$, project with $\P =
\ket 1>\bra 1| + \ket v>\bra v|$.  The frequency matrix becomes
\BE
\mOmega = k\begin{pmatrix}
\< 1 | v | 1> & \< 1 | v | v>\\
\< v | v | 1> & \< v | v | v>
\end{pmatrix}
=
k
\begin{pmatrix}
0 & 1\\
1 & 0
\end{pmatrix}.
\EE
The Markovian approximation is even better justified (the $\Q$~projection
embraces Hermite polynomials of order~2 and above).  The only surviving
transport coefficient can easily be shown to be~$\eta^v_v/k^2 \equiv \mu$;
it is easy to 
calculate, but its value will not be needed to lowest order in the
hydrodynamic limit.  The projected system of equations takes the form
\BALams
\delt n + \ii k u &= 0,
\\
\delt u + \ii k n + k^2\mu u &= -\nu u,
\EALams
where the \rhs\ of the last equation arises because the collision operator
does not conserve momentum.  The dispersion relation is
\BE
\l^2 + (\nu + k^2\mu)\l + k^2 = 0.
\EE
For $k^2 \to 0$, the 1--0 balance is, upon undoing the velocity
normalization, $\l = -k^2D$ with $D$~as in \Eq{D}; this is the density
diffusion mode.  The 2--1 balance is $\l = -\nu$, which describes momentum
relaxation.
\comment
\footnote{Note that it would not be correct in detail to attempt
to refine the momentum eigenmode by retaining the $k^2\mu$ correction
because the Markovian approximation that was employed is not valid for
relaxation on the collisional timescale.}
\endcomment

Thus, although in this example it is natural and optimally efficient to
project into the 1D density 
subspace, no harm is done by employing a higher-dimensional
projection\footnote{Note that the exact solution of the \FPE\ corresponds
to an infinite-dimensional projection.}
 as long as part of it projects into the complete null space of the
 collision operator.  If
such a projection is treated in a mathematically consistent way, the
long-time transport 
must still emerge, though it is likely to arise in ways that are
mathematically 
different depending on the projection.  For example, in the 1D
projection diffusion is represented by the dissipative $\eta$~coefficient,
while in the density plus momentum projection the $\eta$~coefficient
vanishes in the density equation and is subdominant in the momentum
equation; density diffusion arises from the coupling between density and the
momentum-dissipation effect represented by the explicit momentum projection
of the collision operator, which does not conserve momentum.

If one combines the insights of the present example with the caveats from
the earlier examples in this appendix, which show that it is problematical
to project into a subspace of dimension lower than that of the null space,
one deduces that for a Markovian treatment of long-time transport one must
use a projection that at least spans all of the null space of the collision
operator but may be of higher dimensionality if additional information is
desired.  This is a satisfying consistency.  However, although the
projection method is flexible and intuitive, a solid physical understanding
of the effects to be described is essential to a successful exploitation of
the techniques.


\addcontentsline{toc}{section}{References}

\def\ {\unskip\hskip3.33333pt plus 1.66666pt minus 1.11111pt}

\def\pages#1#2{#1 (#2 pages)}
\def\range#1{\rangeo#1.}
\def\rangeo#1-#2.{#1-#2}
\def\press{in press}
\def\semi{}

\def\addperiod#1{#1}
\def\killcomma#1{<KC>}
\def\commatosemicolon{<CS>}

\long\def\pg#1#2#3{#1%
\ifx#2,%
 \ifx#3[%
  \ #3
 \else
  \ifx#3;%
   ; %
  \else
   #2 #3%
  \fi
 \fi
\else
 #2 #3%
\fi
}

\bibliographystyle{jpp-JAK}


\end{document}